\documentclass[aps]{revtex4}
\usepackage{amsfonts}
\usepackage{amsmath}
\usepackage{amssymb,epsf}

\begin{document}

\title{Einstein-Born-Infeld-Massive Gravity:\\
adS-Black Hole Solutions and their Thermodynamical properties}
\author{S. H. Hendi$^{1,2}$\footnote{
email address: hendi@shirazu.ac.ir}, B. Eslam Panah$^{1}$\footnote{%
email address: behzad$_{-}$eslampanah@yahoo.com}, and S. Panahiyan$^{1}$%
\footnote{ email address: ziexify@gmail.com}} \affiliation{$^1$
Physics Department and Biruni Observatory, College of Sciences,
Shiraz
University, Shiraz 71454, Iran\\
$^2$ Research Institute for Astronomy and Astrophysics of Maragha
(RIAAM), Maragha, Iran}

\begin{abstract}
In this paper, we study massive gravity in the presence of
Born-Infeld nonlinear electrodynamics. First, we obtain metric
function related to this gravity and investigate the geometry of
the solutions and find that there is an essential singularity at
the origin ($r=0$). It will be shown that due to contribution of
the massive part, the number, types and places of horizons may be
changed. Next, we calculate the conserved and thermodynamic
quantities and check the validation of the first law of
thermodynamics. We also investigate thermal stability of these
black holes in context of canonical ensemble. It will be shown
that number, type and place of phase transition points are
functions of different parameters which lead to dependency of
stability conditions to these parameters. Also, it will be shown
how the behavior of temperature is modified due to extension of
massive gravity and strong nonlinearity parameter. Next, critical
behavior of the system in extended phase space by considering
cosmological constant as pressure is investigated. A study
regarding neutral Einstein-massive gravity in context of extended
phase space is done. Geometrical approach is employed to study the
thermodynamical behavior of the system in context of heat capacity
and extended phase space. It will be shown that GTs, heat capacity
and extended phase space have consistent results. Finally,
critical behavior of the system is investigated through use of
another method. It will be pointed out that the results of this
method is in agreement with other methods and follow the concepts
of ordinary thermodynamics.
\end{abstract}

\maketitle

\section{Introduction}

Regarding experimental agreements of Einstein gravity (EN) in various area
of astrophysics and cosmology, motivates one to consider it as an acceptable
theory. In addition, adding a constant term $\Lambda$ in the EN-Hilbert
action may lead to agreement between the results of EN-$\Lambda$ gravity
with dark energy prediction.

On the other hand, general relativity is consistent with interaction of
massless spin $2$ fields, in which related gravitons are massless particles
with two degrees of freedom. Since the quantum theory of massless gravitons
is non-renormalizable \cite{Deser84}, one may be motivated for modifying
general relativity to massive gravity. In order to build up a massive theory
with a massive spin $2$ particle propagation, one can add a mass term to the
EN-Hilbert action. This will result into graviton having a mass of $m$ which
in case of $m\rightarrow 0$, the effect of massive gravity will be vanished.
A class of massive gravity theory in flat and curved background which leads
to absence \cite{Fierz} and existence \cite{Boulware} of ghost, have been
investigated. Also, the quantum aspects of the massive gravity and a
nonlinear class of massive gravity in ghost-free field \cite{Hassan} have
been explored in Refs. \cite{Minjoon,Rham}. Generalization to nonlinearly
charged massive black holes was done in Ref. \cite{Saridakis}. More details
regarding the motivations of massive gravity is given in Ref. \cite{Babichev}%
.

In this paper, we are interested in studying the nontrivial adS massive
theory that was investigated in \cite{Vegh,Hassan2011}. The motivation for
this consideration is due to fact that graviton shows similar behavior as
lattice in holographic conductor \cite{Davison}. In other words, a Drude
like behavior is observed for the case of massless graviton in this theory
which makes the role of graviton similar to lattice. Another interesting
subject for study in this theory is metal-insulator transition \cite%
{MassiveADSCFT}. Recently, charged massive black holes with consideration of
this theory have been investigated in \cite{Cai2015}. The $P-V$\ criticality
of these solutions and their geometrical thermodynamic aspects have been
studied \cite{Xu2015,HendiPEM}. Also, the generalization to
Gauss-Bonnet-Maxwell-massive gravity and its stability, geometrical
thermodynamics and $P-V$ criticality have been investigated \cite{HPEmassive}%
.

One of the main problems of Maxwell's electromagnetic field theory for a
point-like charge is that there is a singularity at the charge position and
hence, it has infinite self-energy. To overcome this problem in classical
electrodynamics, Born and Infeld in Ref. \cite{BI} introduced a nonlinear
electromagnetic field, with main motivation, to solve infinite self-energy
problem by imposing a maximum strength of the electromagnetic field. Then,
Hoffmann in Ref. \cite{Hoffmann} investigated EN gravity in the presence of
Born-Infeld (BI) electrodynamics. In recent two decades, exact solutions of
gravitating black objects in the presence of BI theory have been vastly
investigated \cite{BIpapers,BIpapers2}. Another interesting property of BI
is that, BI type effective action arises in an open superstring theory and
D-branes are free of physical singularities \cite{Fradkin}. For a review of
aspects of BI theory in the context of string theory see Ref. \cite{Gibbons}%
. Recently, there has been growing interest in Eddington-inspired BI gravity
in context of black holes and cosmology \cite{Eddington}. Also, it was
proposed that one can consider BI theory as a gravitational theory \cite%
{Born-G}. Dualization of the BI theory and some of the special properties of
this theory have been investigated in Ref. \cite{Dulaization}.

There are several approaches for studying and obtaining critical behavior
and phase transition points of black holes: First method is based on
studying heat capacity. It was pointed out that roots and divergencies of
the heat capacity are representing phase transition points. In other words,
in place of roots and divergencies of the heat capacity system may go under
phase transition. Another important property of the heat capacity is
investigation of the thermal stability. Systems with positive heat capacity
are denoted to be in thermally stable states. Therefore, the stability
conditions are indicated by changes in sign of heat capacity \cite{Myung}.
This is known as canonical ensemble.

In the second method, by using the renewed interpretation of cosmological
constant as thermodynamical variable, one can modify the thermodynamical
structure of the phase space \cite{cosmological}. One of the most important
property of this method is the similarity of critical behavior of the black
holes and ordinary thermodynamical Van der Waals liquid/gas systems \cite%
{Vander}. Recently, it was pointed out that the extended phase space should
be interpreted as an RG-flow in the space of field theories, where isotherm
curves codify how the number of degrees of freedom $N$ (or the central
charge $c$) runs with the energy scale \cite{Pedraza}. On the other hand, it
was shown that variation of cosmological constant could be corresponded to
variation of number of the colors in Yang-Mills theory residing on the
boundary spacetime \cite{Yang}.

The third method is using geometrical concept for studying critical
behavior. In other words, by employing a thermodynamical potential and its
corresponding extensive parameters, one can construct phase space. The
divergencies of Ricci scalar in constructed metric are denoted as phase
transition points. There are several metrics for this method that one can
name: Weinhold \cite{Weinhold}, Quevedo \cite{Quevedo} and HPEM \cite{HPEM}
which has mass as thermodynamical potential and Ruppeiner \cite{Ruppeiner}
in which entropy is considered as thermodynamical potential. These metrics
are used in context of heat capacity. Another set of metrics was introduced
in Ref. \cite{PVEinstein} which can be used in context of extended phase
space.

Finally, a fourth method was introduced in Ref. \cite{PVEinstein} which is
based on denominator of the heat capacity. In this method by replacing
cosmological constant with pressure in denominator of the heat capacity and
solving it with respect to pressure, a new relation is obtained for
pressure. The existence of maximum for obtained relation, represents the
critical pressure and volume in which phase transition takes place. The
behavior of system in case of this method is consistent with ordinary
thermodynamical concepts \cite{HendiPEM,PVEinstein}.

The outline of the paper will be as follow. In Sec. \ref{FieldEq}, we
introduce action and basic equations related to EN-BI-massive gravity. Sec. %
\ref{MassiveSol} is devoted to obtain the black hole solutions of this
gravity and investigation of the geometrical structure of them. In the next
section, we calculate conserved and thermodynamic quantities related to
obtained solutions and check the validation of the first law of
thermodynamics. In section \ref{Stability}, we study thermal stability of
the EN-BI-massive black hole solutions in canonical ensemble. Next, we
consider cosmological constant as pressure and study the critical behavior
of the system. Then we employ the geometrical methods for investigating
thermodynamical behavior of the system and extend this study by another
method. In the last section we present our conclusions.

\section{Basic Equations \label{FieldEq}}

The $d$-dimensional action of EN-massive gravity with negative cosmological
constant and a nonlinear electrodynamics is
\begin{equation}
\mathcal{I}=-\frac{1}{16\pi }\int d^{d}x\sqrt{-g}\left[ \mathcal{R}-2\Lambda
+L(\mathcal{F})+m^{2}\sum_{i}^{4}c_{i}\mathcal{U}_{i}(g,f)\right] ,
\label{Action}
\end{equation}%
where $\mathcal{R}$ is the scalar curvature, $\Lambda =-\frac{\left(
d-1\right) \left( d-2\right) }{2l^{2}}$ is the negative cosmological
constant and $f$ \ is a fixed symmetric tensor. In Eq. (\ref{Action}), $%
c_{i} $ are constants and $\mathcal{U}_{i}$ are symmetric polynomials of the
eigenvalues of the $d\times d$ matrix $\mathcal{K}_{\nu }^{\mu }=\sqrt{%
g^{\mu \alpha }f_{\alpha \nu }}$ which can be written as follows
\begin{eqnarray}
\mathcal{U}_{1} &=&\left[ \mathcal{K}\right], \;\;\;\;\; \mathcal{U}_{2} =%
\left[ \mathcal{K}\right] ^{2}-\left[ \mathcal{K}^{2}\right], \;\;\;\;\;
\mathcal{U}_{3} =\left[ \mathcal{K}\right] ^{3}-3\left[ \mathcal{K}\right] %
\left[ \mathcal{K}^{2}\right] +2\left[ \mathcal{K}^{3}\right] ,  \notag \\
&& \mathcal{U}_{4} =\left[ \mathcal{K}\right] ^{4}-6\left[ \mathcal{K}^{2}%
\right] \left[ \mathcal{K}\right] ^{2}+8\left[ \mathcal{K}^{3}\right] \left[
\mathcal{K}\right] +3\left[ \mathcal{K}^{2}\right] ^{2}-6\left[ \mathcal{K}%
^{4}\right].  \notag
\end{eqnarray}

Here, we want to study a particular model of nonlinear electrodynamics
called BI theory which has attracted lots of attentions due to its relation
to effective string actions. The function $L(\mathcal{F})$ for BI theory is
given as
\begin{equation}
L(\mathcal{F})=4\beta ^{2}\left( 1-\sqrt{1+\frac{\mathcal{F}}{2\beta ^{2}}}%
\right) ,  \label{LBI}
\end{equation}%
where $\beta $ is the BI parameter, the Maxwell invariant is $\mathcal{F}%
=F_{\mu \nu }F^{\mu \nu }$ in which $F_{\mu \nu }=\partial _{\mu }A_{\nu
}-\partial _{\nu }A_{\mu }$ is the electromagnetic field tensor and $A_{\mu
} $ is the gauge potential.

Variation of the action (\ref{Action}) with respect to the metric tensor $%
g_{\mu \nu }$ and the Faraday tensor $F_{\mu \nu }$, leads to
\begin{equation}
G_{\mu \nu }+\Lambda g_{\mu \nu }-\frac{1}{2}g_{\mu \nu }L(\mathcal{F})-%
\frac{2F_{\mu \lambda }F_{\nu }^{\lambda }}{\sqrt{1+\frac{\mathcal{F}}{%
2\beta ^{2}}}}+m^{2}\chi _{\mu \nu }=0,  \label{Field equation}
\end{equation}%
\begin{equation}
\partial _{\mu }\left( \frac{\sqrt{-g}F^{\mu \nu }}{\sqrt{1+\frac{\mathcal{F}%
}{2\beta ^{2}}}}\right) =0,  \label{Maxwell equation}
\end{equation}%
where $G_{\mu \nu }$ is the EN tensor and $\chi _{\mu \nu }$ is the massive
term with the following form
\begin{eqnarray}
\chi _{\mu \nu } &=&-\frac{c_{1}}{2}\left( \mathcal{U}_{1}g_{\mu \nu }-%
\mathcal{K}_{\mu \nu }\right) -\frac{c_{2}}{2}\left( \mathcal{U}_{2}g_{\mu
\nu }-2\mathcal{U}_{1}\mathcal{K}_{\mu \nu }+2\mathcal{K}_{\mu \nu
}^{2}\right) -\frac{c_{3}}{2}(\mathcal{U}_{3}g_{\mu \nu }-3\mathcal{U}_{2}%
\mathcal{K}_{\mu \nu }+  \notag \\
&&6\mathcal{U}_{1}\mathcal{K}_{\mu \nu }^{2}-6\mathcal{K}_{\mu \nu }^{3})-%
\frac{c_{4}}{2}(\mathcal{U}_{4}g_{\mu \nu }-4\mathcal{U}_{3}\mathcal{K}_{\mu
\nu }+12\mathcal{U}_{2}\mathcal{K}_{\mu \nu }^{2}-24\mathcal{U}_{1}\mathcal{K%
}_{\mu \nu }^{3}+24\mathcal{K}_{\mu \nu }^{4}).  \label{massiveTerm}
\end{eqnarray}

\section{Black hole solutions in EN-BI-Massive gravity \label{MassiveSol}}

In this section, we obtain static nonlinearly charged black holes in context
of massive gravity with adS asymptotes. For this purpose we consider the
metric of $d$-dimensional spacetime in the following form
\begin{equation}
ds^{2}=-f(r)dt^{2}+f^{-1}(r)dr^{2}+r^{2}h_{ij}dx_{i}dx_{j},\
i,j=1,2,3,...,n~,  \label{Metric}
\end{equation}%
where $h_{ij}dx_{i}dx_{j}$ is a $(d-2)$ dimension line element for an
Euclidian space with constant curvature $\left( d-2\right) (d-3)k$ and
volume $V_{d-2}$. We should note that the constant $k$, which indicates that
the boundary of $t=constant$ and $r=constant$, can be a negative
(hyperbolic), zero (flat) or positive (elliptic) constant curvature
hypersurface.

We consider the ansatz metric \cite{Cai2015}
\begin{equation}
f_{\mu \nu }=diag(0,0,c^{2}h_{ij}),  \label{f11}
\end{equation}%
where $c$ is a positive constant. Using the metric ansatz (\ref{f11}), $%
\mathcal{U}_{i}$'s are in the following forms \cite{Cai2015}
\begin{equation}
\mathcal{U}_{1}=\frac{d_{2}c}{r},\text{ \ \ }\mathcal{U}_{2}=\frac{%
d_{2}d_{3}c^{2}}{r^{2}},\text{ \ }\mathcal{U}_{3}=\frac{d_{2}d_{3}d_{4}c^{3}%
}{r^{3}},\text{ \ }\mathcal{U}_{4}=\frac{d_{2}d_{3}d_{4}d_{5}c^{4}}{r^{4}},
\notag
\end{equation}%
in which $d_{i}=d-i$. Using the gauge potential ansatz $A_{\mu }=h(r)\delta
_{\mu }^{0}$ in electromagnetic equation (\ref{Maxwell equation}) and
considering the metric (\ref{Metric}), we obtain
\begin{equation}
h(r)=-\sqrt{\frac{d_{2}}{d_{3}}}\frac{q}{r^{d_{3}}}\ \mathcal{H},
\label{h(r)}
\end{equation}%
in which $\mathcal{H}$ is the following hypergeometric function
\begin{equation}
\mathcal{H}=\ _{2}\mathcal{F}_{1}\left( \left[ \frac{1}{2},\frac{d_{3}}{%
2d_{2}}\right] ,\left[ \frac{3d_{7/3}}{2d_{2}}\right] ,-\Gamma \right) ,
\label{H}
\end{equation}%
where $\Gamma =\frac{d_{2}d_{3}q^{2}}{\beta ^{2}r^{2d_{2}}}$ and $q$ is an
integration constant which is related to the electric charge. Also, the
electromagnetic field tensor in $d$-dimensions is given by
\begin{equation}
F_{tr}=\frac{\sqrt{d_{2}d_{3}}q}{r^{d_{2}}\sqrt{1+\Gamma }}.  \label{Ftr}
\end{equation}

Now, we are interested in obtaining the static black hole solutions. One may
use components of Eq. (\ref{Field equation}) and obtain metric function $%
f(r) $. We use the $tt$ and $x_{1}x_{1}$ components of the Eq. (\ref{Field
equation}), which can be written as
\begin{eqnarray}
e_{tt} &=&\left\{ d_{2}m^{2}c\left[
c_{1}r^{3}+d_{3}c_{2}cr^{2}+d_{3}d_{4}c_{3}c^{2}r+d_{3}d_{4}d_{5}c_{4}c^{3}%
\right] -2\Lambda r^{4}\right.  \notag \\
&&\left. -d_{2}d_{3}r^{2}f-d_{2}r^{3}f^{\prime }+4\beta
^{2}r^{4}+d_{2}d_{3}r^{2}k\right\} \sqrt{1-\left( \frac{h^{\prime }}{\beta }%
\right) ^{2}}-4\beta ^{2}r^{4}=0,  \label{tteq} \\
&&  \notag \\
e_{x_{1}x_{1}} &=&d_{3}m^{2}c\left[
c_{1}r^{3}+d_{3}d_{4}c_{2}cr^{2}+d_{3}d_{4}d_{5}c_{3}c^{2}r+d_{3}d_{4}d_{5}d_{6}c_{4}c^{3}%
\right] -2\Lambda r^{4}  \notag \\
&&-2d_{3}r^{3}f^{\prime }-d_{3}d_{4}r^{2}f-r^{4}f^{\prime \prime }+4\beta
^{2}r^{4}-4\beta r^{4}\sqrt{\beta ^{2}-h^{\prime 2}}+d_{3}d_{4}r^{2}k=0.
\label{x1x1eq}
\end{eqnarray}

We can obtain the metric function $f(r)$, by using the Eqs. (\ref{tteq}) and
(\ref{x1x1eq}) with the following form
\begin{eqnarray}
f\left( r\right) &=&k-\frac{m_{0}}{r^{d_{3}}}+\left( \frac{4\beta
^{2}-2\Lambda }{d_{1}d_{2}}\right) r^{2}-\frac{4\beta ^{2}r^{2}}{d_{1}d_{2}}%
\sqrt{1+\Gamma }+\frac{4d_{2}q^{2}\mathcal{H}}{d_{1}r^{2d_{3}}}  \notag \\
&&+m^{2}\left\{ \frac{cc_{1}}{d_{2}}r+c^{2}c_{2}+\frac{d_{3}c^{3}c_{3}}{r}+%
\frac{d_{3}d_{4}c^{4}c_{4}}{r^{2}}\right\} ,  \label{f(r)}
\end{eqnarray}%
where $m_{0}$ is an integration constant which is related to the total mass
of the black hole. It should be noted that, obtained metric function (\ref%
{f(r)}), satisfy all components of the Eq. (\ref{Field equation}),
simultaneously.

Now, we are in a position to review the geometrical structure of this
solution, briefly. We first look for the essential singularity(ies). The
Ricci scalar and the Kretschmann scalar are
\begin{eqnarray}
\lim_{r\longrightarrow 0}R &\longrightarrow &\infty , \\
\lim_{r\longrightarrow 0}R_{\alpha \beta \gamma \delta }R^{\alpha \beta
\gamma \delta } &\longrightarrow &\infty ,
\end{eqnarray}%
and so confirm that there is a curvature singularity at $r=0$. The Ricci and
Kretschmann scalars are $\frac{2d}{d_{2}}\Lambda $ and $\frac{8d}{%
d_{1}d_{2}^{2}}\Lambda ^{2}$ at $r\longrightarrow \infty $. Therefore, the
asymptotic behavior of these solutions are (a)dS for ($\Lambda <0$ ).

On the other hand, in the absence of massive parameter ($m=0$), the solution
(\ref{f(r)}) reduces to an $d$-dimensional asymptotically adS topological
black hole with a negative, zero or positive constant curvature hypersurface
in the following form%
\begin{equation}
f\left( r\right) =k-\frac{m_{0}}{r^{d_{3}}}-\frac{4\beta ^{2}r^{2}}{%
d_{1}d_{2}}\sqrt{1+\Gamma }+\left( \frac{4\beta ^{2}}{d_{1}d_{2}}+\frac{1}{%
l^{2}}\right) r^{2}+\frac{4d_{2}q^{2}\mathcal{H}}{d_{1}r^{2d_{3}}}.
\label{fEBI}
\end{equation}

\begin{figure}[tbp]
$%
\begin{array}{cc}
\epsfxsize=7cm \epsffile{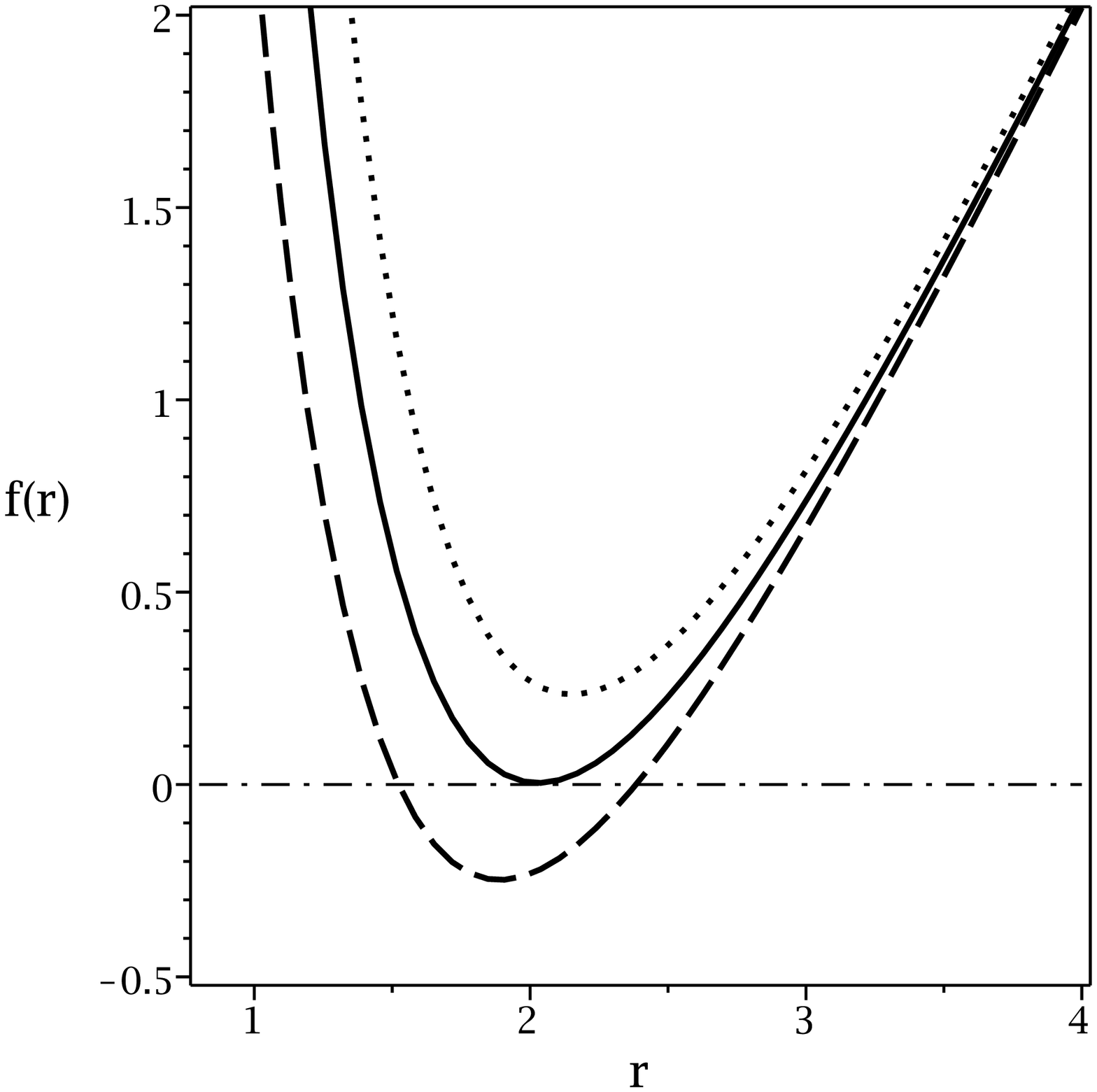} & \epsfxsize=7cm %
\epsffile{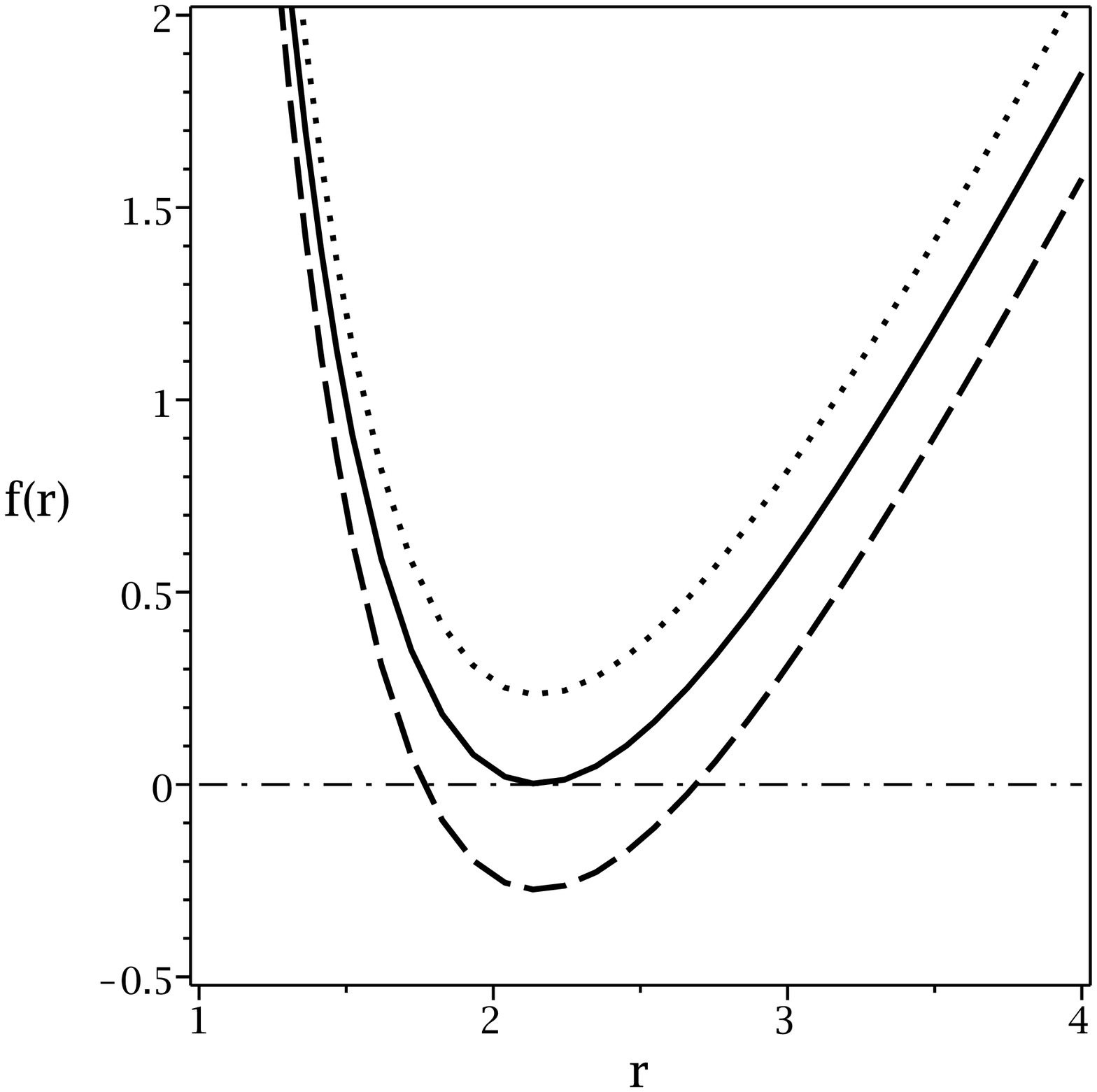}%
\end{array}
$%
\caption{$f(r)$ versus $r$ for $\Lambda=-0.8$, $q=1$, $\protect\beta=0.6$, $%
m=1.4$, $c=-0.8$, $c_{1}=2$, $c_{3}=-4$, $c_{4}=0$, $k=1$, and $d=4$.
\newline
Left diagram for $m_{0}=5$, $c_{2}=1.00$ (dashed line), $c_{2}=1.22$
(continues line) and $c_{2}=1.35$ (dotted line). \newline
Right diagram for $c_{2}=1.40$, $m_{0}=5.80$ (dashed line), $m_{0}=5.64$
(continues line) and $m_{0}=5.5$ (dotted line).}
\label{Figfr1}
\end{figure}

\begin{figure}[tbp]
\epsfxsize=8cm \centerline{\epsffile{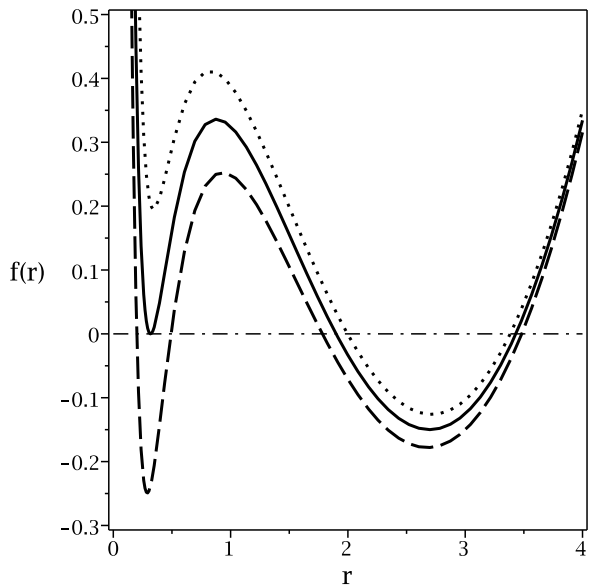}}
\caption{$f(r)$ versus $r$ for $\Lambda=-1$, $q=0.5$, $\protect\beta=7$, $%
m=0.5$, $c=0.4$, $c_{1}=-40$, $c_{2}=60$, $c_{3}=1$, $c_{4}=0$,
$k=1$ and $d=6$.
\newline
diagrams for $m_{0}=1.75$ (dashed line), $m_{0}=1.67$ (continues line) and $%
m_{0}=1.61$ (dotted line).}
\label{Figfr2}
\end{figure}

\begin{figure}[tbp]
$%
\begin{array}{cc}
\epsfxsize=8cm \epsffile{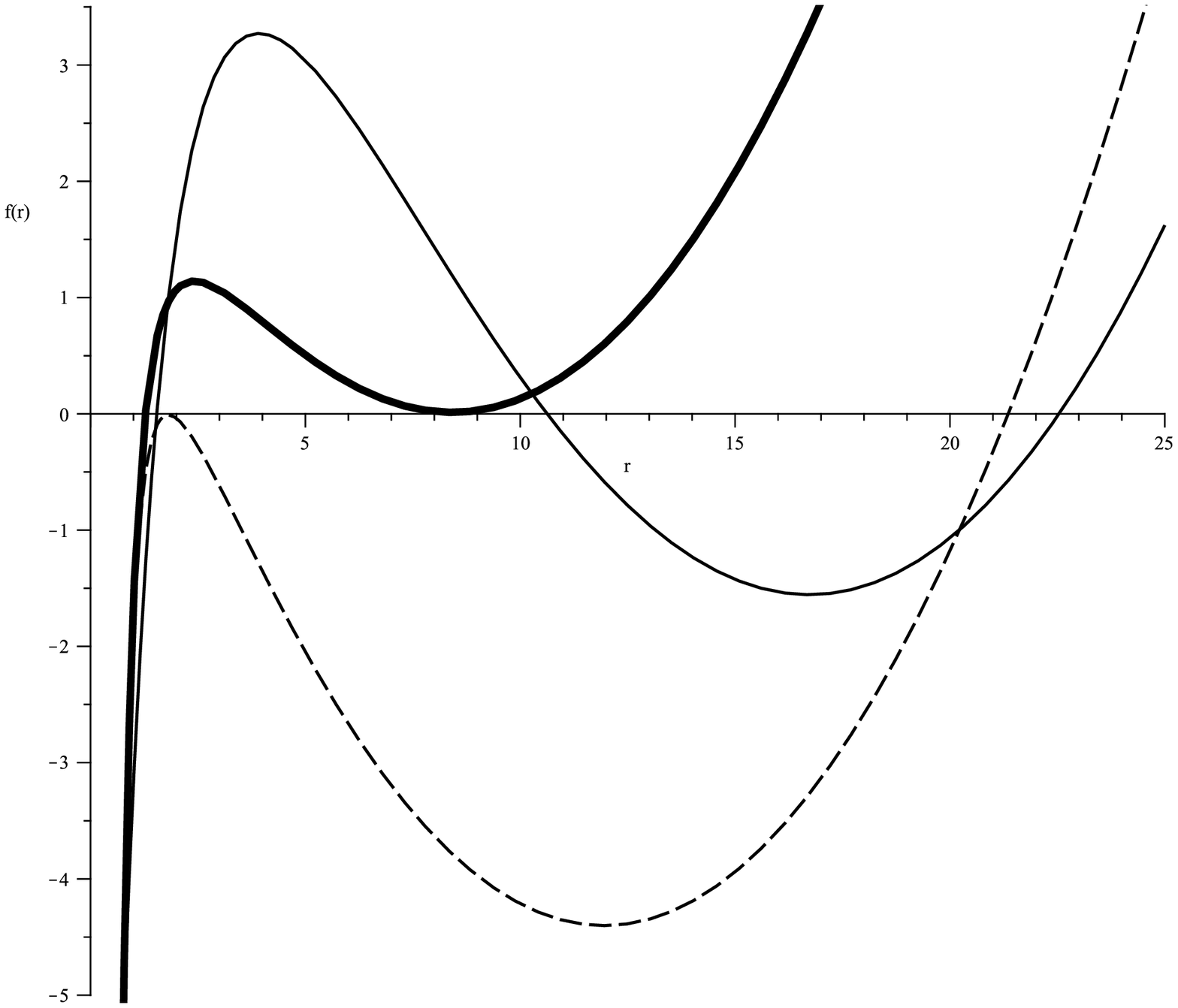} & \epsfxsize=5.5cm %
\epsffile{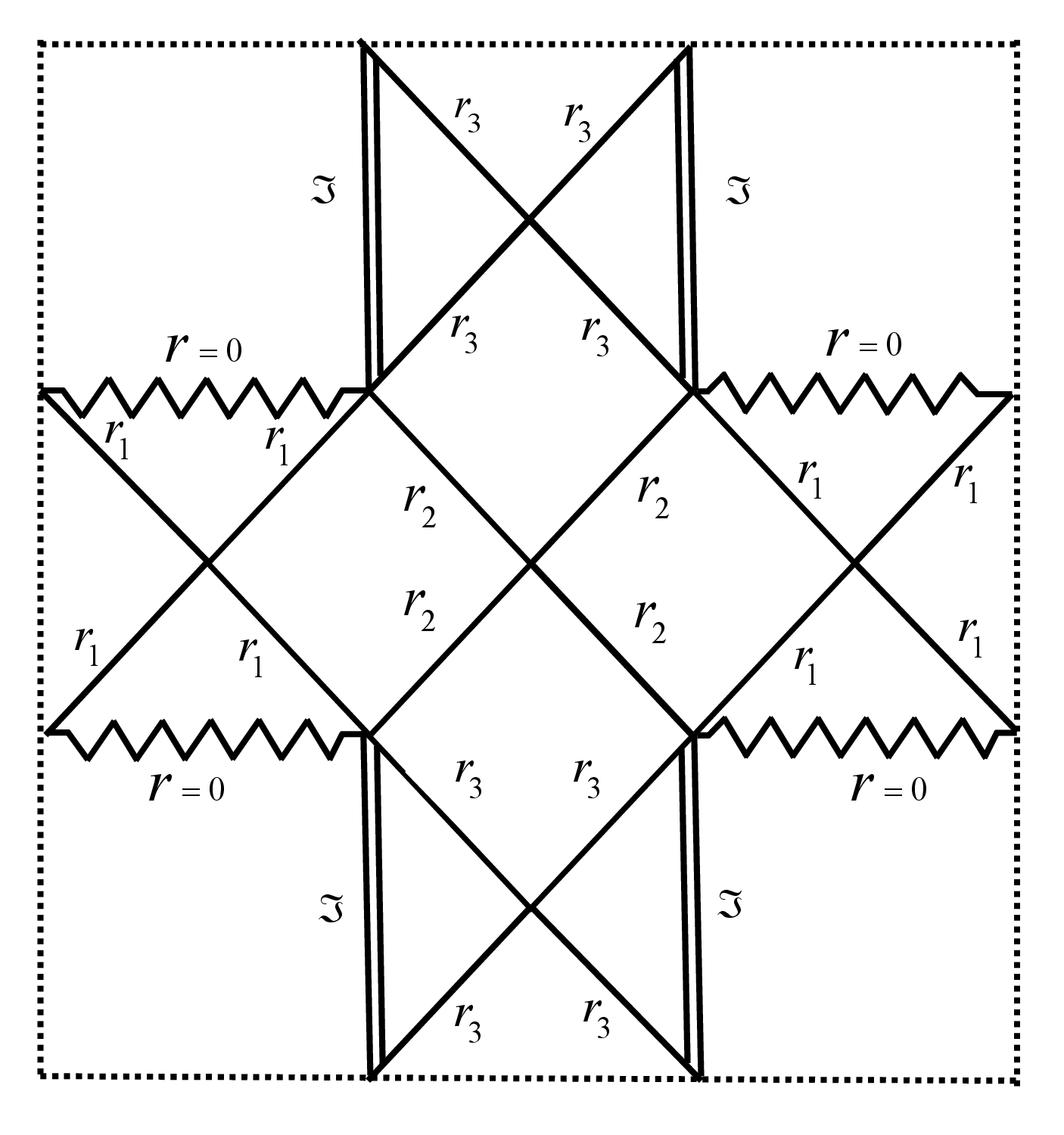} \\
\epsfxsize=5.5cm \epsffile{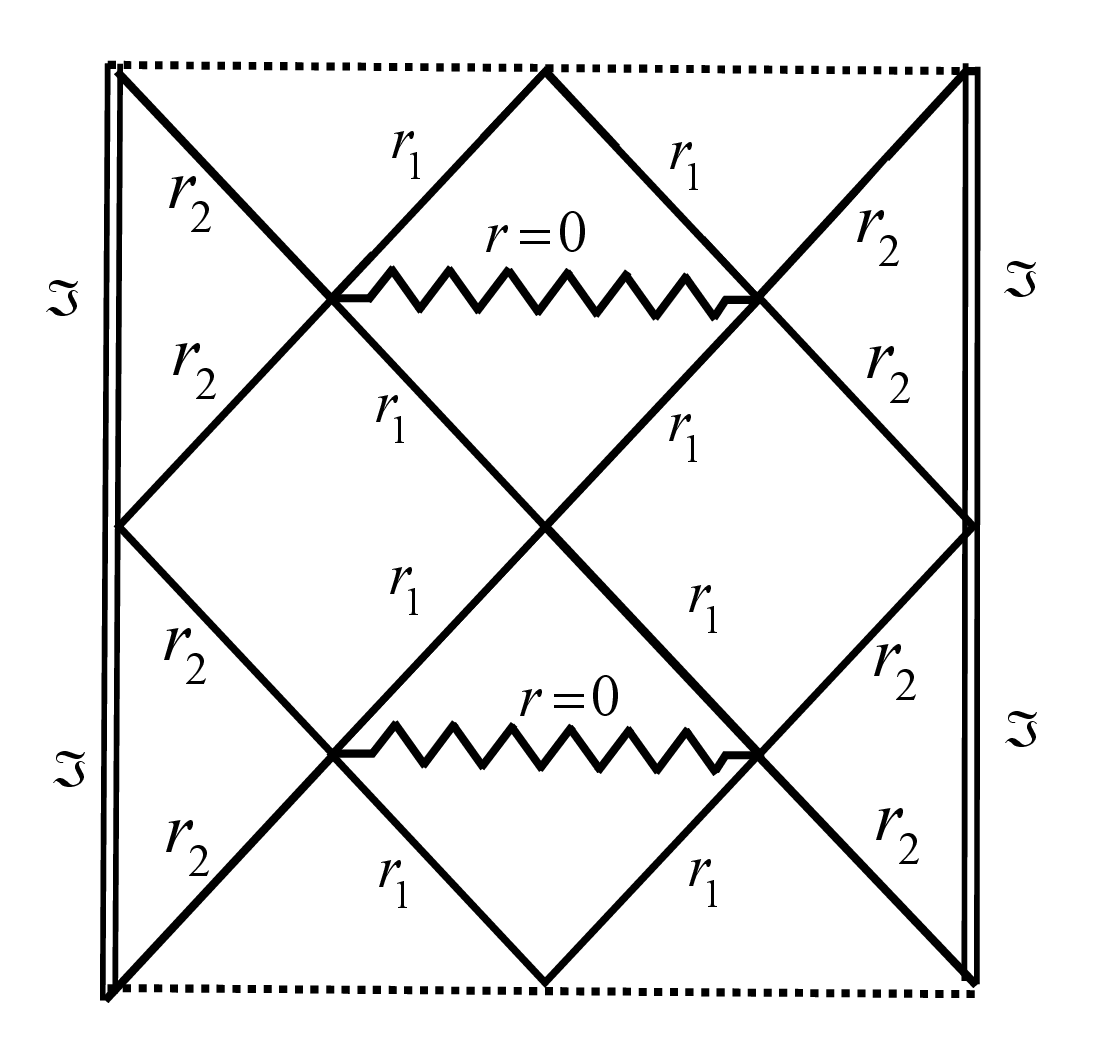} & \epsfxsize=4.7cm %
\epsffile{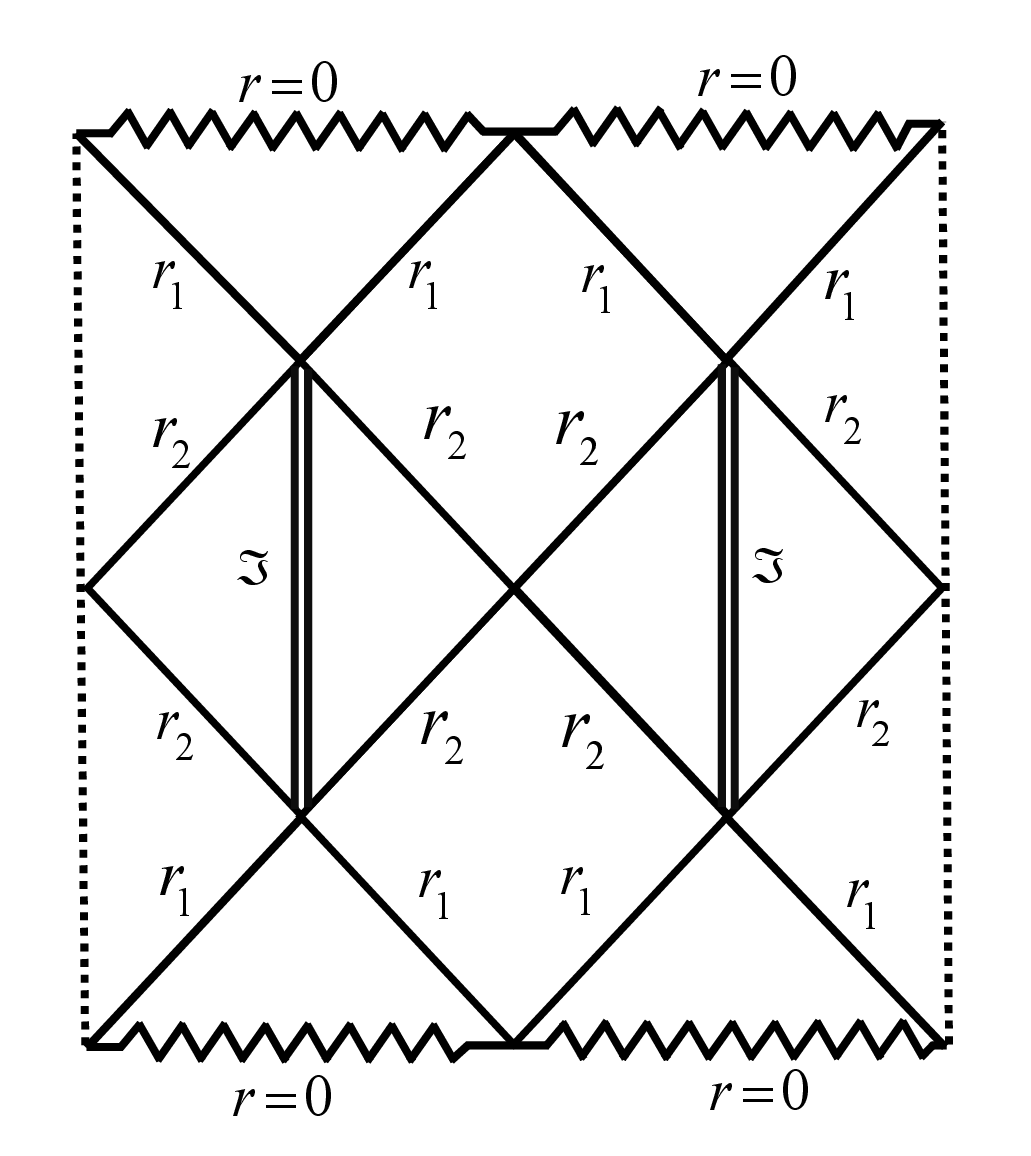}%
\end{array}
$%
\caption{Metric functions and Carter--Penrose diagrams for the
asymptotically adS black holes with spacelike singularity. Three horizons
(continuous line of metric function and related Carter--Penrose diagram in
right-up panel), two horizons which inner one is extreme (dotted line of
metric function and related Carter--Penrose diagram in left-down panel) and
two horizons which outer one is extreme (dashed line of metric function and
related Carter--Penrose diagram in right-down panel).}
\label{Pen}
\end{figure}

In order to study the effects of the EN-BI-massive gravity on metric
function, we have plotted various diagrams (Figs. \ref{Figfr1} -- \ref{Pen}
). 

By considering specific values for the parameters, metric function has
different behaviors. Depending on the choices of the parameters,
EN-BI-massive black holes can behave like Reissner--Nordstr\"{o}m black
holes. In other words, these black holes may have two horizons, one extreme
horizon and without horizon (naked singularity) (see Fig. \ref{Figfr1} for
more details). On the other hand, by adjusting some of the parameters of
EN-BI-massive black holes, we encounter with interesting behaviors. The
solutions may have three or higher horizons (Figs. \ref{Figfr2} and \ref{Pen}%
). The existence of three or higher horizons for black holes is due to the
presence of massive gravity \cite{HendiPEM,HPEmassive}. In addition to the
significant effects of massive term, we should note that the nonlinearity
parameter can affect on the number of horizons. In addition, $\beta$ can
change the type of singularity. In other words, depending on the parameters,
one can find a $\beta_{c}$ in which singularity is spacelike for $%
\beta<\beta_{c}$, and it would be timelike for $\beta>\beta_{c}$ (see \cite%
{BIpapers2} for more details).

Now, we give a brief discussion regarding Carter--Penrose diagrams. In order
to study the conformal structure of the solutions, one may use the conformal
compactification method through plotting the Carter--Penrose (conformal)
diagrams. As we mentioned before, depending on the value of nonlinearity
parameter, $\beta$, one may encounter with timelike or spacelike
singularity. Penrose diagrams regarding to timelike singularity was
discussed in \cite{HPEmassive} (see conformal diagrams in \cite{HPEmassive}%
). Here we focus on special case in which singularity is spacelike ($%
\beta<\beta_{c}$). In other words, the singularity of this nonlinearly
charged black holes behaves like uncharged Schwarzschild solutions (see Fig. %
\ref{Pen}). This means that, although massive and nonlinearity parts of the
metric function can change the type of singularity and horizon structure of
black holes, they does not affect asymptotical behavior of the solutions.
Drawing the Carter--Penrose diagrams, we find the causal structure of the
solutions are asymptotically well behaved.

\section{Thermodynamics}

In this section, we calculate the conserved and thermodynamic quantities of
the static EN-BI-massive black hole solutions in $d$-dimensions and then
check the first law of thermodynamics.

By using the definition of Hawking temperature which is related to the
definition of surface gravity on the outer horizon $r_{+}$, one can find%
\begin{equation}
T=\frac{m^{2}c}{4\pi r_{+}^{3}}\left[
c_{1}r_{+}^{3}+d_{3}c_{2}cr_{+}^{2}+d_{3}d_{4}c_{3}c^{2}r_{+}+d_{3}d_{4}d_{5}c_{4}c^{3}%
\right] +\frac{\left( 2\beta ^{2}-\Lambda \right) r_{+}}{2\pi d_{2}}+\frac{%
d_{3}k}{4\pi r_{+}}-\frac{\beta ^{2}r_{+}}{\pi d_{2}}\sqrt{1+\Gamma _{+}},
\label{TotalTT}
\end{equation}%
where $\Gamma _{+}=\frac{d_{2}d_{3}q^{2}}{\beta ^{2}r_{+}^{2d_{2}}}$. The
electric charge, $Q$, can be found by calculating the flux of the electric
field at infinity, yielding%
\begin{equation}
Q=\frac{V_{d_{2}}\sqrt{d_{2}d_{3}}}{4\pi }q.  \label{TotalQ}
\end{equation}

In order to obtain the entropy of the black holes, one can employ the area
law of the black holes. It is a matter of calculation to show that entropy
has the following form \cite{Beckenstein}
\begin{equation}
S=\frac{V_{d_{2}}}{4}r_{+}^{d_{2}},  \label{TotalS}
\end{equation}

It was shown that by using the Hamiltonian approach or counterterm method,
one can find the mass $M$ of the black hole for massive gravity as
\begin{equation}
M=\frac{d_{2}V_{d_{2}}}{16\pi }m_{0},  \label{TotalM}
\end{equation}%
in which by evaluating metric function on the horizon ($f\left(
r=r_{+}\right) =0$), one can obtain
\begin{eqnarray}
M &=&\frac{d_{2}V_{d_{2}}}{16\pi }\left( kr_{+}^{d_{3}}-\frac{2r_{+}^{d_{1}}%
}{d_{1}d_{2}}\Lambda -\frac{4\beta ^{2}r_{+}^{d_{1}}}{d_{1}d_{2}}\left[
\sqrt{1+\Gamma _{+}}-1\right] +\frac{4d_{2}q^{2}}{d_{1}r_{+}^{d_{3}}}%
\mathcal{H}_{+}\right.  \notag \\
&&\left. +\frac{cm^{2}r_{+}^{d_{5}}}{d_{2}}\left[
d_{2}d_{3}d_{4}c_{4}c^{3}+d_{2}d_{3}c_{3}c^{2}r_{+}+d_{2}c_{2}cr_{+}^{2}+c_{1}r_{+}^{2}%
\right] \right) ,  \label{Mass}
\end{eqnarray}%
where $\mathcal{H}_{+}=\ _{2}\mathcal{F}_{1}\left( \left[ \frac{1}{2},\frac{%
d_{3}}{2d_{2}}\right] ,\left[ \frac{3d_{7/3}}{2d_{2}}\right] ,-\Gamma
_{+}\right) $.

It is notable that, $U$ is\ the electric potential, which is defined in the
following form%
\begin{equation}
U=A_{\mu }\chi ^{\mu }\left\vert _{r\rightarrow \infty }\right. -A_{\mu
}\chi ^{\mu }\left\vert _{r\rightarrow r_{+}}\right. =\sqrt{\frac{d_{2}}{%
d_{3}}}\frac{q}{r_{+}^{d_{3}}}\ \mathcal{H}_{+}.  \label{TotalU}
\end{equation}

Having conserved and thermodynamic quantities at hand, we are in a position
to check the first law of thermodynamics for our solutions. It is easy to
show that by using thermodynamic quantities such as charge (\ref{TotalQ}),
entropy (\ref{TotalS}) and mass (\ref{TotalM}), with the first law of black
hole thermodynamics%
\begin{equation}
dM=TdS+UdQ,
\end{equation}%
we define the intensive parameters conjugate to $S$ and $Q$. These
quantities are the temperature and the electric potential
\begin{equation}
T=\left( \frac{\partial M}{\partial S}\right) _{Q}\ \ \ and\ \ \ \ \ \ \ \ \
U=\left( \frac{\partial M}{\partial Q}\right) _{S},  \label{TU}
\end{equation}%
which are the same as those calculated for temperature (\ref{TotalTT}) and
electric potential (\ref{TotalU}).

\section{Heat Capacity and Stability in Canonical Ensemble \label{Stability}}

Here, we study the stability conditions and the effects of different factors
on these conditions. The stability conditions in canonical ensemble are
based on the signature of the heat capacity. The negativity of heat capacity
represents unstable solutions which may lead to following results: unstable
solutions may go under phase transition and acquire stable states. This
phase transition could happen whether when heat capacity meets a root(s) or
has a divergency. Therefore, the roots of regular numerator and denominator
of the heat capacity are phase transition points. In the other scenario, the
heat capacity is always negative. This is known as non-physical case. But
there is a stronger condition which is originated from the temperature. The
positivity of the temperature represents physical solutions whereas its
negativity is denoted as non-physical one. Therefore, in order to getting
better picture and enriching the results of our study, we investigate both
temperature and heat capacity, simultaneously.

The heat capacity is given by
\begin{equation}
C_{Q}=\frac{T}{\left( \frac{\partial ^{2}M}{\partial S^{2}}\right) _{Q}}=%
\frac{T}{{\left( \frac{\partial T}{\partial S}\right) _{Q}}}.  \label{CQ}
\end{equation}

Considering Eqs. (\ref{TotalTT}) and (\ref{TotalS}), it is a matter of
calculation to show that%
\begin{eqnarray}
{\left( \frac{\partial T}{\partial S}\right) _{Q}} &=&-\frac{d_{3}k}{%
d_{2}r_{+}^{d_{1}}}+\frac{2\left( 2\beta ^{2}-\Lambda \right) }{\pi
d_{2}^{2}r_{+}^{d_{3}}}-\frac{d_{3}m^{2}c}{\pi d_{2}r_{+}^{d_{2}}}\left[
3d_{4}d_{5}c_{4}c^{2}+2d_{4}c_{3}cr_{+}+c_{2}cr_{+}^{2}\right]  \notag \\
&&-\frac{4\beta ^{2}}{\pi d_{2}^{2}r_{+}^{d_{3}}}\left( 1+\Gamma _{+}\right)
^{\frac{3}{2}}+\frac{4d_{3}q^{2}\left( d_{1}+\Gamma _{+}\right) }{\pi
d_{2}r_{+}^{3d-7}\sqrt{1+\Gamma _{+}}}.  \label{Heat}
\end{eqnarray}

In order to study the effects of different parameters on stability
conditions and temperature, we have plotted various diagrams (Figs. \ref%
{Fig1} - \ref{Fig5}).
\begin{figure}[tbp]
$%
\begin{array}{cc}
\epsfxsize=7cm \epsffile{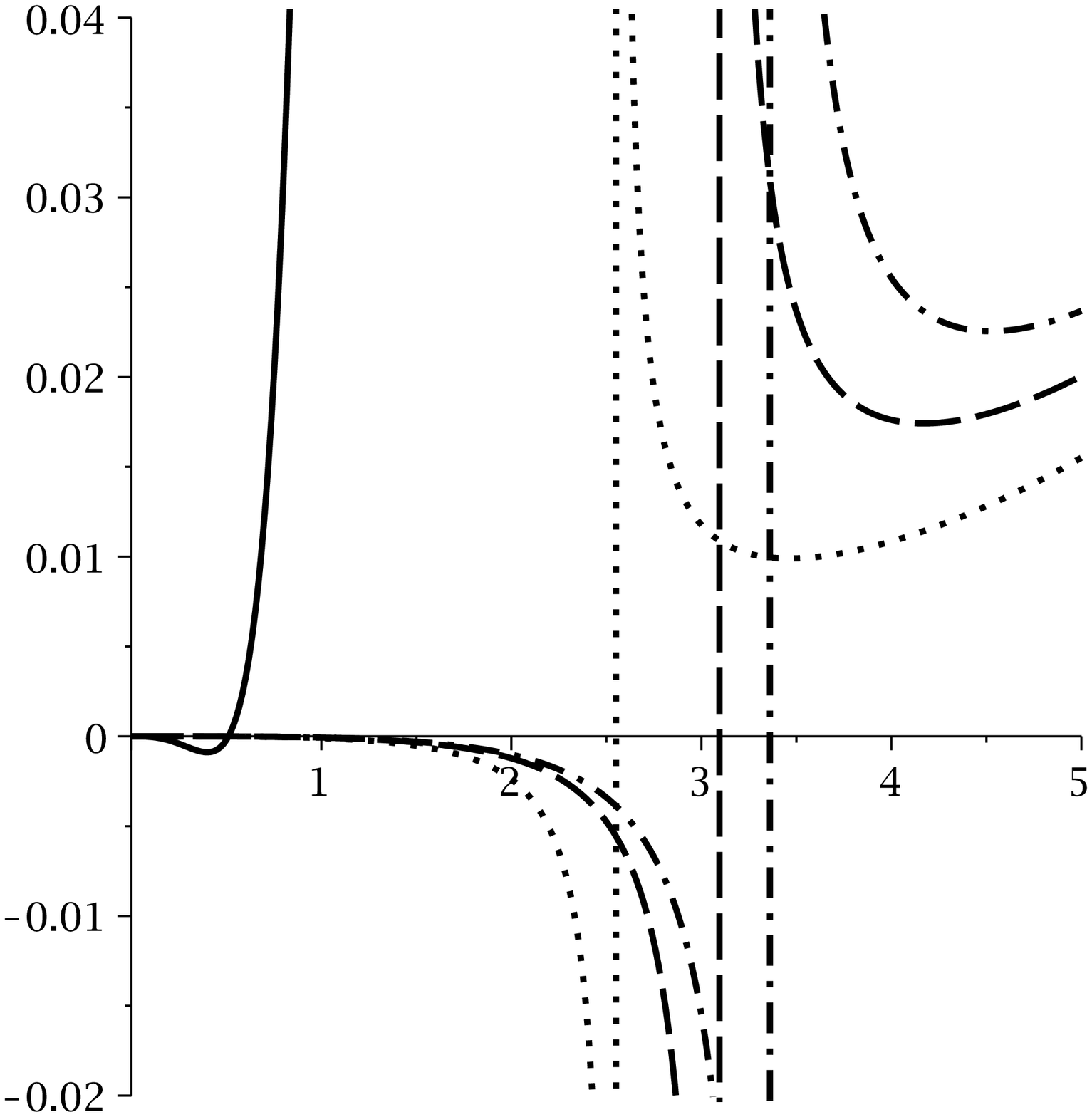} & \epsfxsize=7cm %
\epsffile{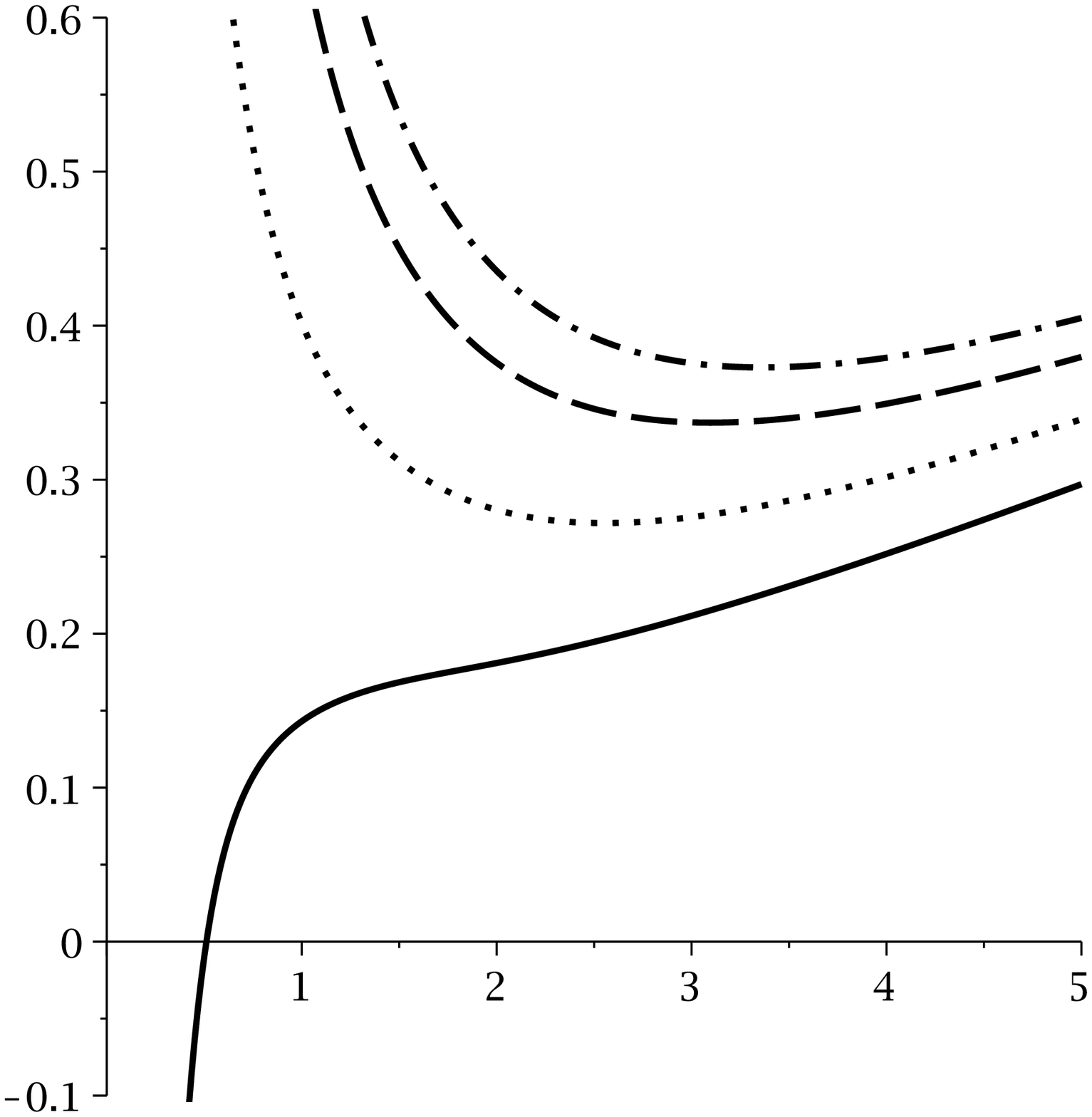}%
\end{array}
$%
\caption{For different scales: $C_{Q}$ (left panel) and $T$ (right panel)
versus $r_{+}$ for $q=1$, $\Lambda =-1$ $c=c_{1}=c_{2}=c_{3}=2$, $c_{4}=0$, $%
\protect\beta =0.5$, $d=5$ and $k=1$; $m=0$ (continues line), $m=0.25$
(dotted line), $m=0.35$ (dashed line) and $m=0.40$ (dashed-dotted line).}
\label{Fig1}
\end{figure}

\begin{figure}[tbp]
$%
\begin{array}{ccc}
\epsfxsize=5.5cm \epsffile{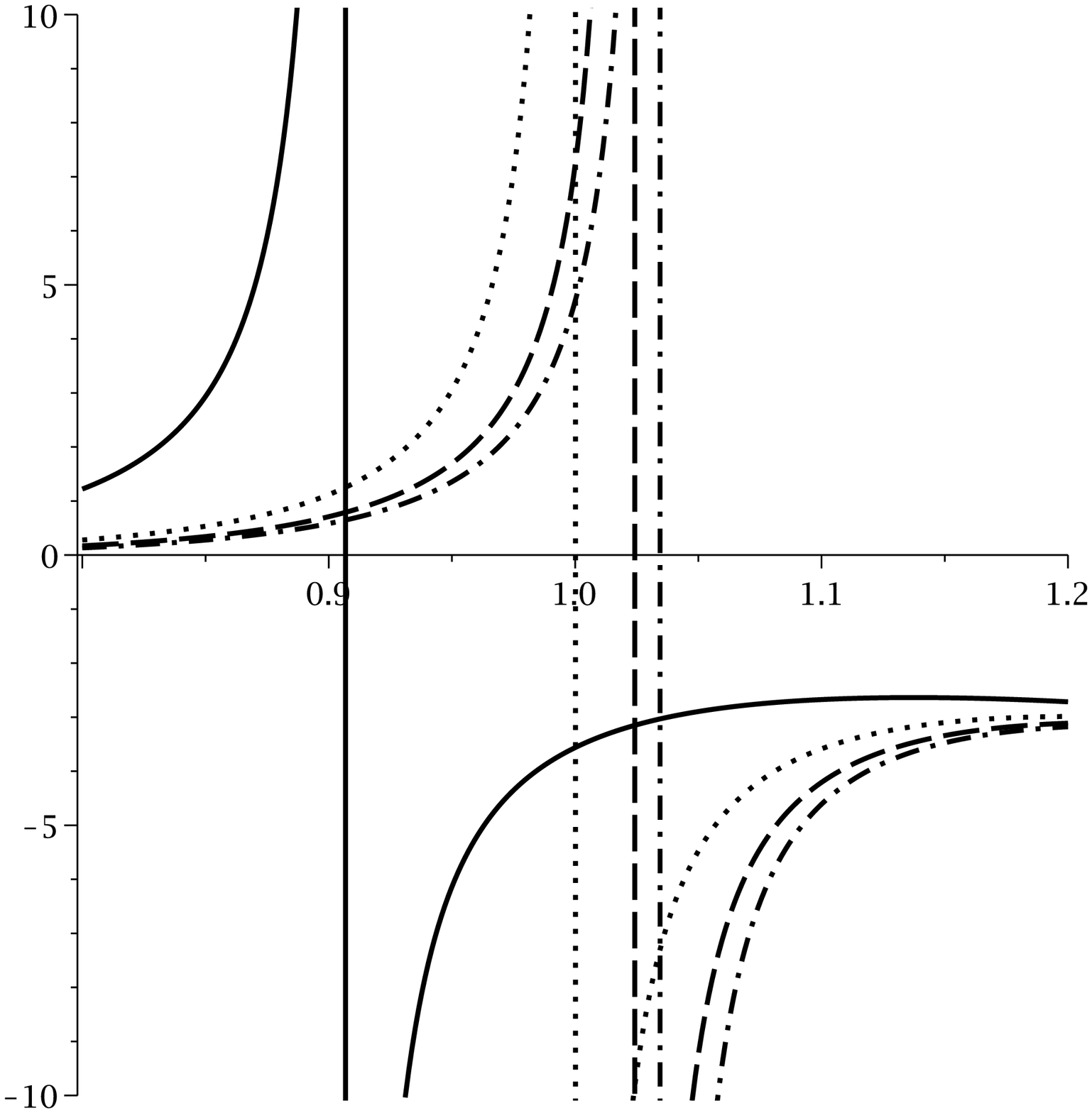} & \epsfxsize=5.5cm %
\epsffile{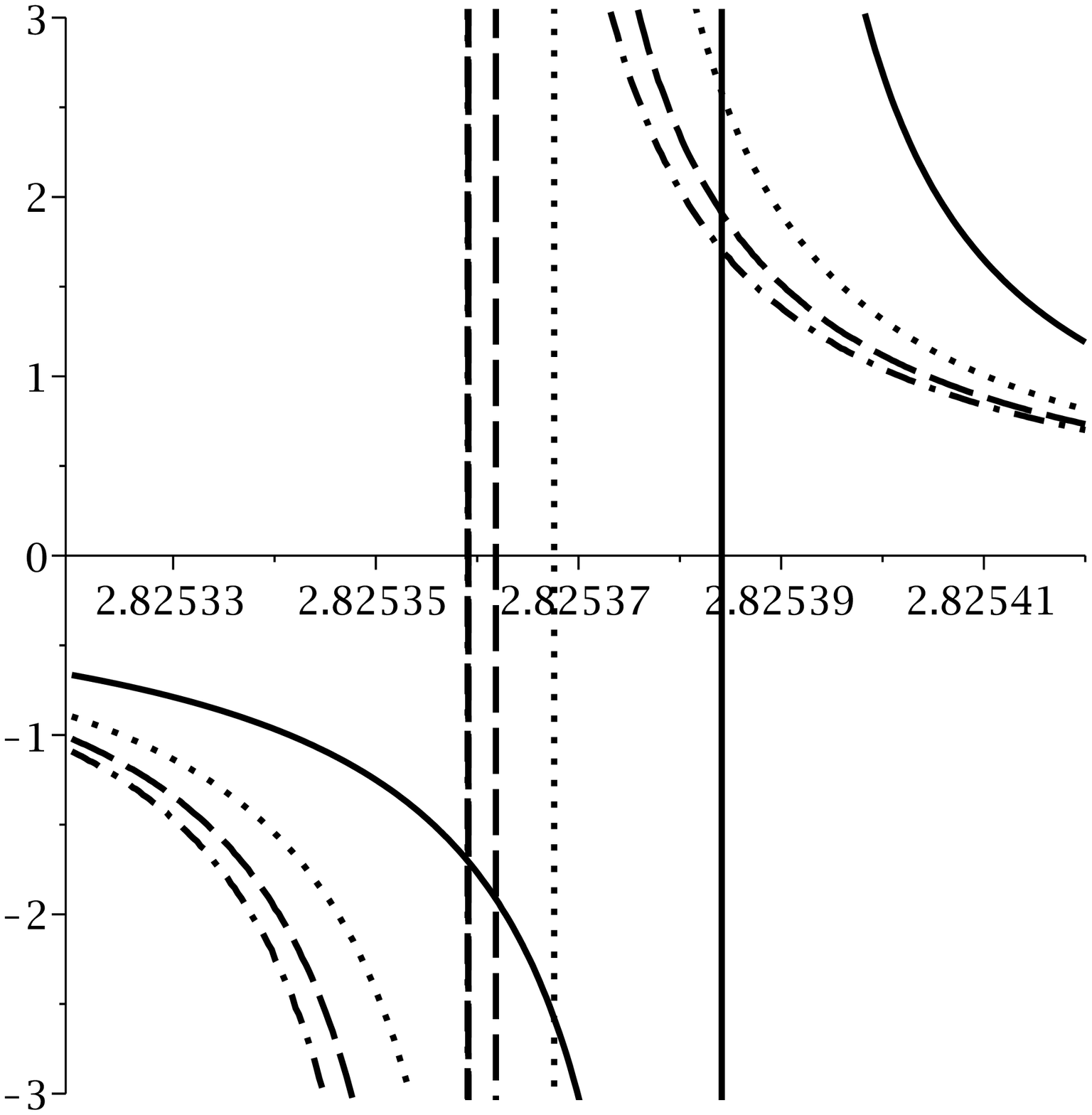} & \epsfxsize=5.5cm \epsffile{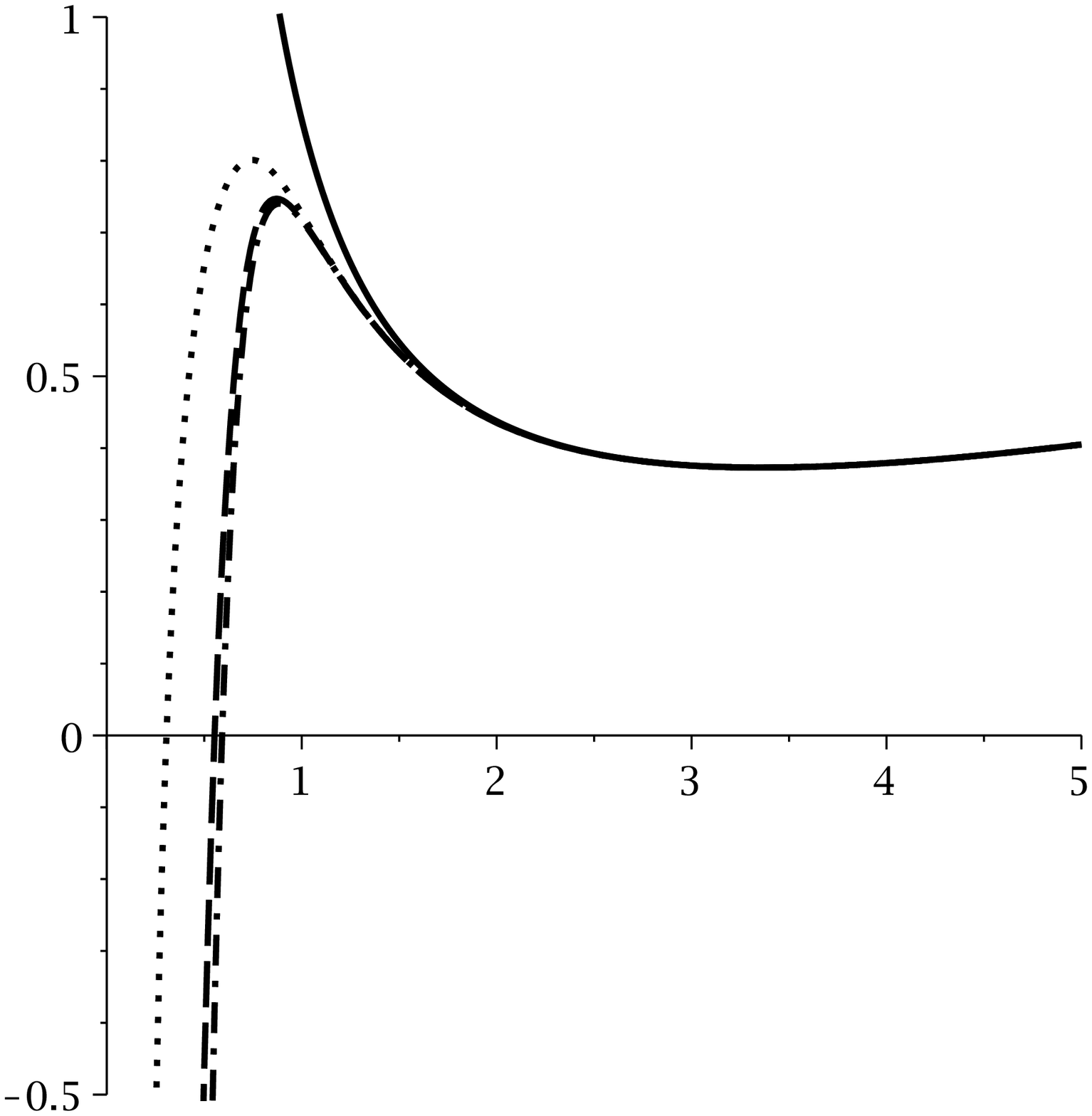}%
\end{array}
$%
\caption{For different scales: $C_{Q}$ (left and middle panels) and $T$
(right panel) versus $r_{+}$ for $q=1$, $\Lambda=-1$, $c=c_{1}=c_{2}=c_{3}=2$%
, $c_{4}=0$, $m=0.30$, $d=5$ and $k=1$; $\protect\beta=2$ (continues line), $%
\protect\beta=3$ (dotted line), $\protect\beta=4$ (dashed line) and $\protect%
\beta=5$ (dashed-dotted line).}
\label{Fig2}
\end{figure}

\begin{figure}[tbp]
$%
\begin{array}{cc}
\epsfxsize=7cm \epsffile{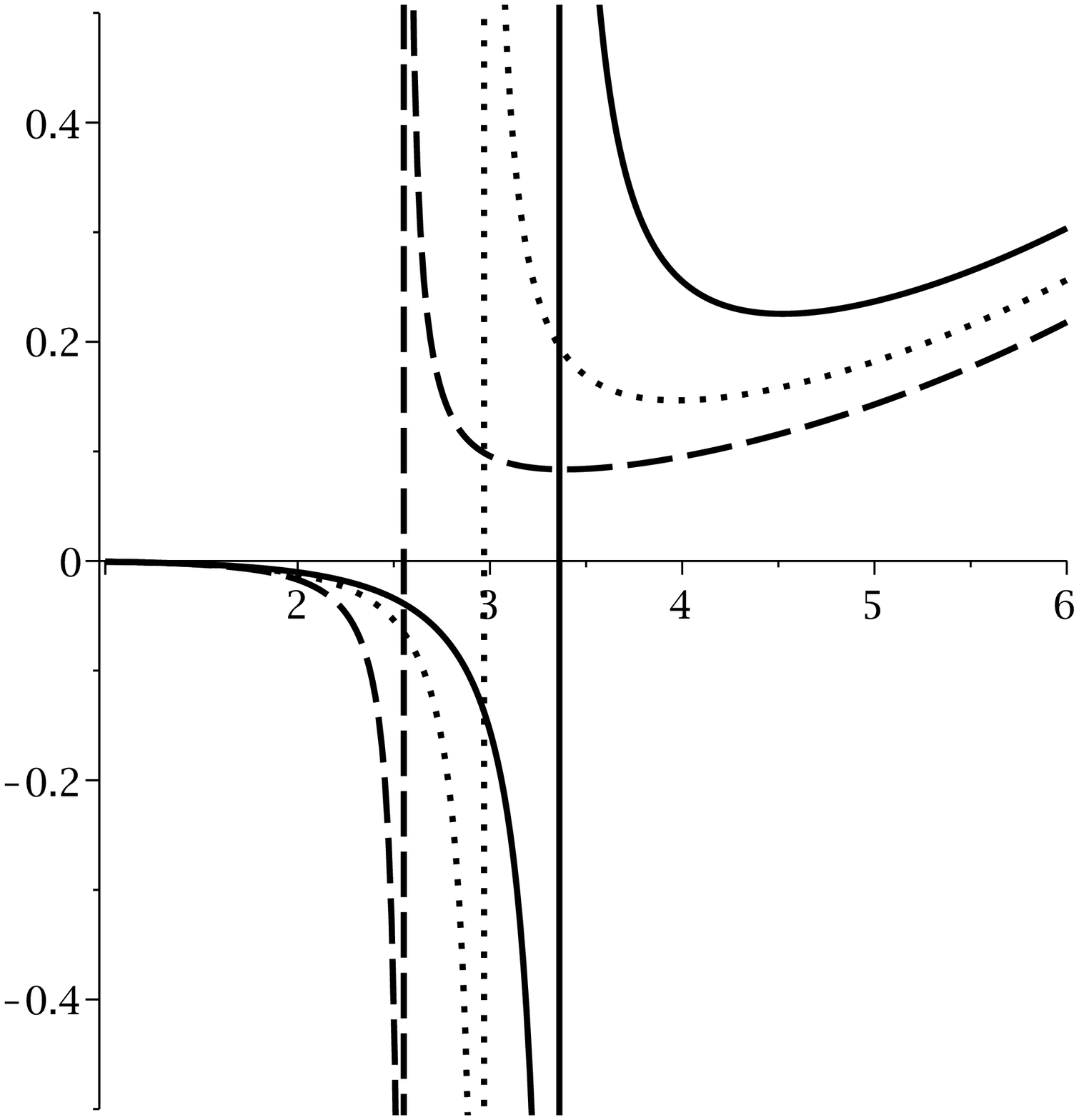} & \epsfxsize=7cm %
\epsffile{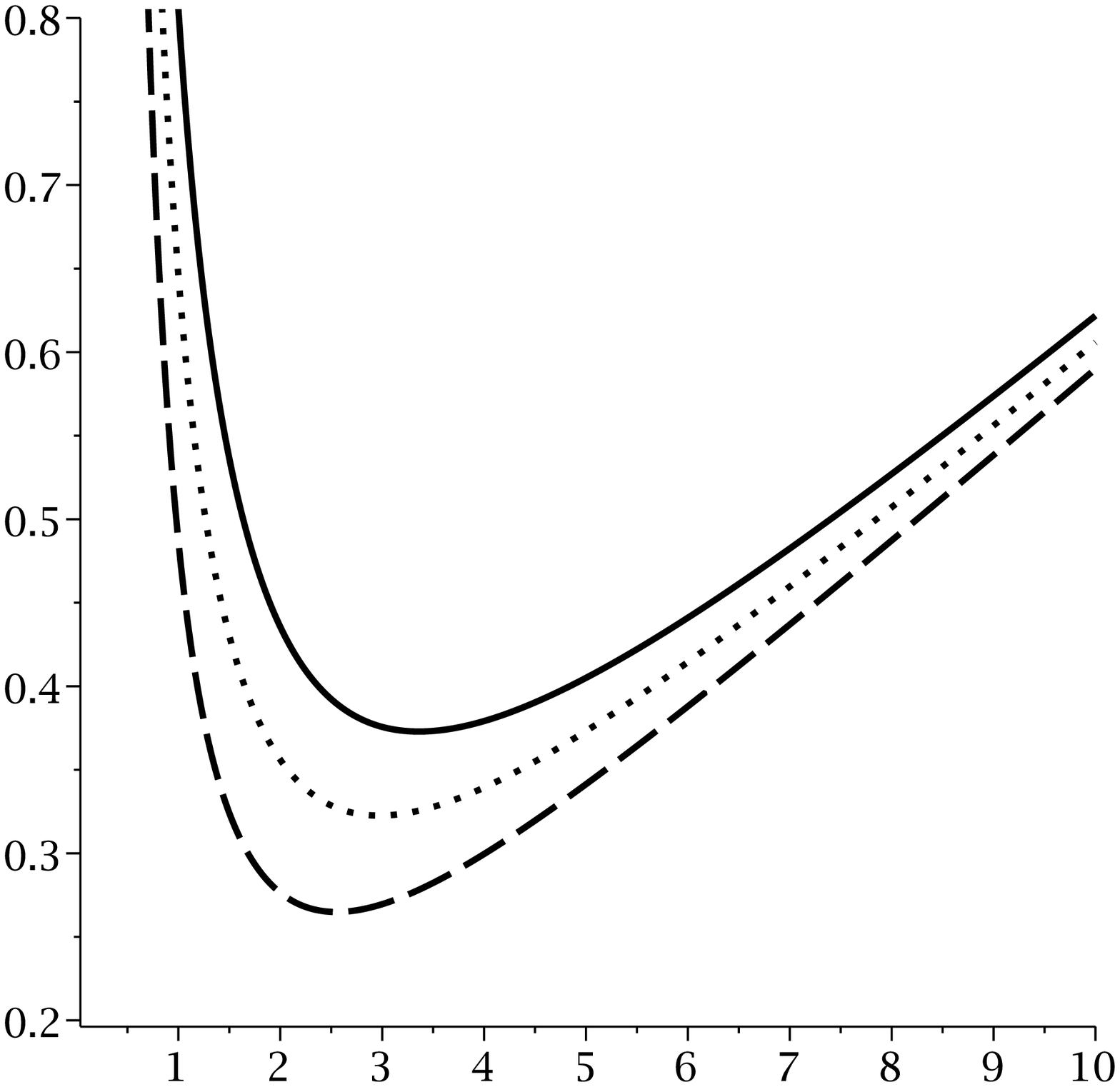}%
\end{array}
$%
\caption{For different scales: $C_{Q}$ (left and middle panels) and $T$
(right panel) versus $r_{+} $ for $q=1$, $\Lambda=-1$, $%
c=c_{1}=c_{2}=c_{3}=2 $, $c_{4}=0$, $m=0.4$, $d=5$ and $\protect\beta=0.5$; $%
k=1$ (continues line), $k=0$ (dotted line) and $k=-1$ (dashed line).}
\label{Fig3}
\end{figure}

\begin{figure}[tbp]
$%
\begin{array}{cc}
\epsfxsize=7cm \epsffile{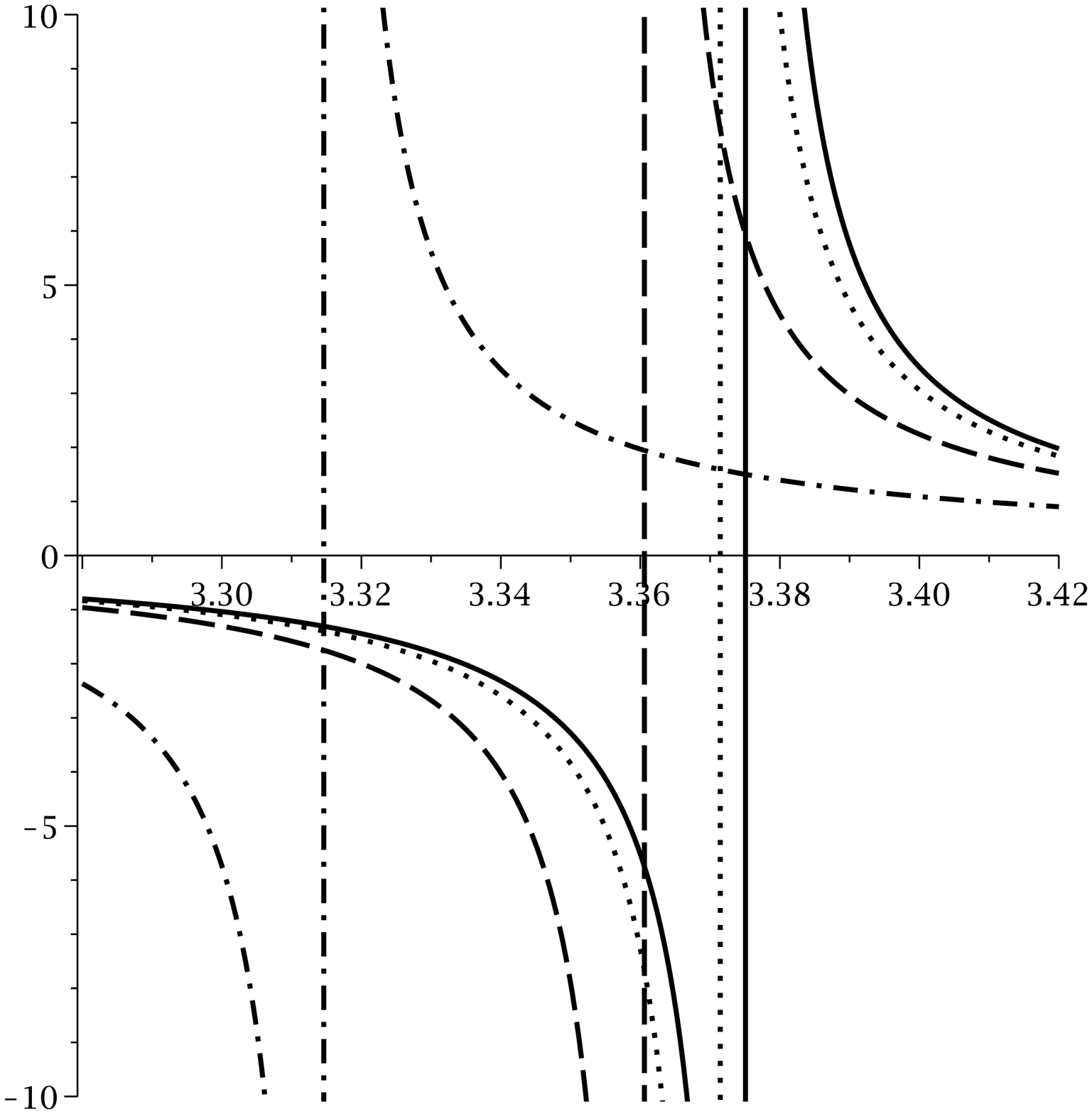} & \epsfxsize=7cm %
\epsffile{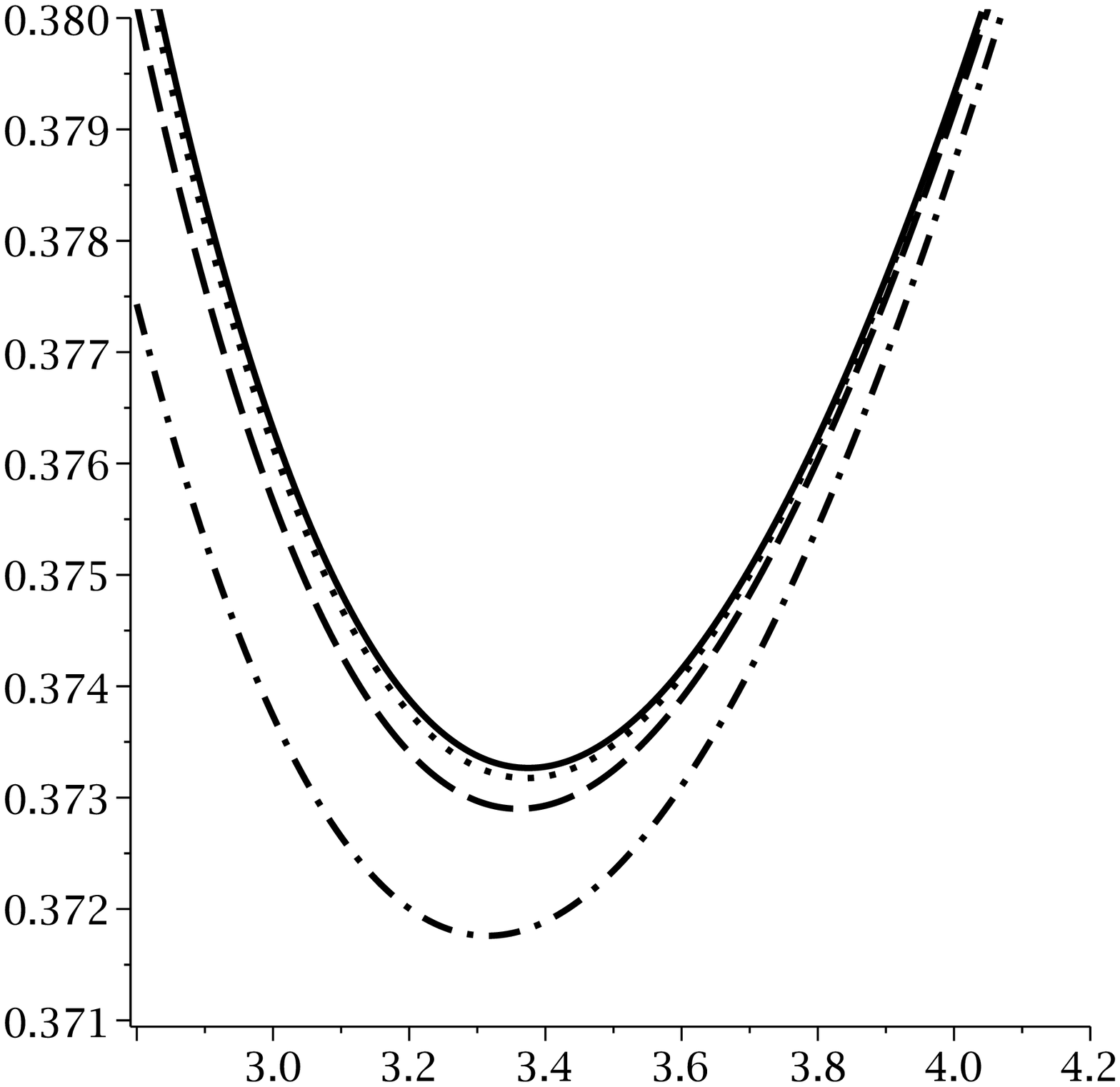}%
\end{array}
$%
\caption{For different scales: $C_{Q}$ (left panel) and $T$ (right panel)
versus $r_{+}$ for $\Lambda=-1$, $c=c_{1}=c_{2}=c_{3}=2$, $c_{4}=0$, $%
\protect\beta=0.5$, $m=0.4$, $d=5 $ and $k=1$; $q=0$ (continues line), $%
q=0.5 $ (dotted line), $q=1$ (dashed line) and $q=2$ (dashed-dotted line).}
\label{Fig4}
\end{figure}

\begin{figure}[tbp]
$%
\begin{array}{ccc}
\epsfxsize=7cm \epsffile{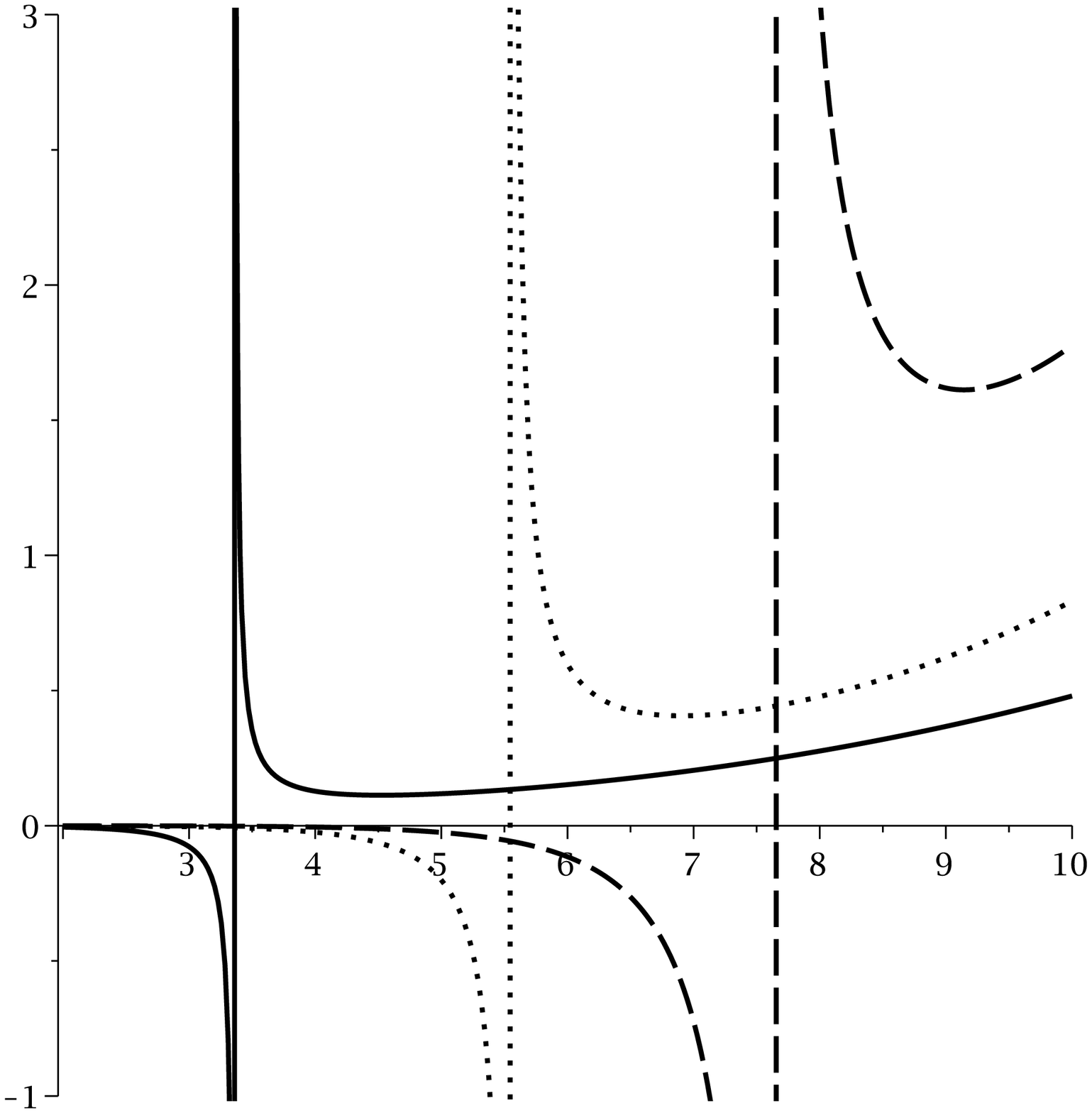} & \epsfxsize=7cm %
\epsffile{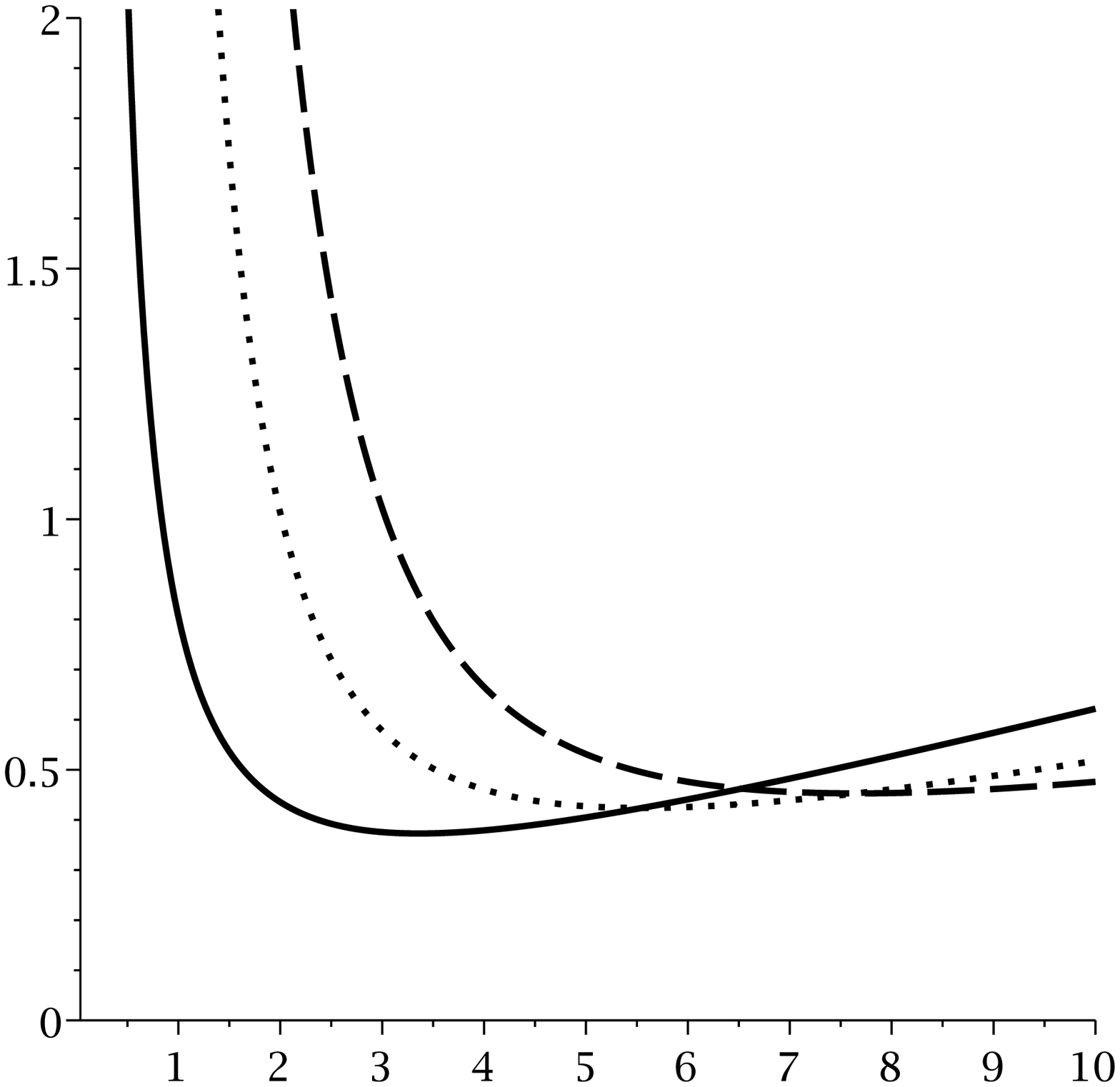} &
\end{array}
$%
\caption{For different scales: $C_{Q}$ (left and middle panels) and $T$
(right panel) versus $r_{+} $ for $q=1$, $\Lambda =-1$, $%
c=c_{1}=c_{2}=c_{3}=2$, $c_{4}=0$, $\protect\beta =0.5$, $m=0.4$ and $k=1$; $%
d=5$ (continues line), $d=6$ (dotted line) and $d=7$ (dashed line).}
\label{Fig5}
\end{figure}


Interestingly, in the absence of the massive parameter (Fig. \ref{Fig1}
right panel), temperature starts from $-\infty $ and it is only an
increasing function of horizon radius with a root. Therefore, we have two
regions of physical and non-physical solutions. Adding massive gravity could
modify the behavior of the temperature into an $U$ shape diagram starting
from $+\infty $ without any root. The extremum is an increasing function of
massive parameter (Fig. \ref{Fig1} right panel) and dimensions (Fig. \ref%
{Fig5} right panel), whereas, it is a decreasing function of $k$ (Fig. \ref%
{Fig3} right panel) and electric charge (Fig. \ref{Fig4} right panel). The
only exception for this behavior is for strong nonlinearity parameter.
Interestingly, for large values of nonlinearity parameter, a massive-less
like behavior is observed (Fig. \ref{Fig2} right panel). In other words, the
temperature starts from $-\infty $ but the effect of the massive could be
seen through two extrema. The root of temperature and smaller extremum are
increasing functions of $\beta $ and larger extremum is a decreasing
function of it. Another interesting property of temperature is the effect of
dimensions. Studying Fig. \ref{Fig5} (right panel) shows that the
temperature for every two sets of dimensions will coincide with each other.
In other words, there are places in which despite differences in dimensions,
two black holes with two different dimensions and same values for other
parameters will have same temperature in a special $r_{+}$.

In absence of massive gravity, black holes could acquire temperature from
zero to $+\infty $, whereas adding massive, will cause the black holes never
acquire some temperature. This effect is vanished in case of large
nonlinearity parameter. In other words, the strength of nonlinearity
parameter has opposing effects to massive's ones. Also, the $U$ shape
diagram indicates that for every temperature that black holes can acquire
two horizons exist except for the extremum. Therefore, considering Hawking
radiation, one is not able to recognize the size of these black holes by
measuring their Hawking radiation. It is worthwhile to mention that extrema
and root(s) of temperature are phase transition points of heat capacity.

Regarding stability, it is evident that in absence of the massive gravity,
there exists a region of the instability which is located where the
temperature is negative. Therefore, this is a non-physical solution (Fig. %
\ref{Fig1} left panel). Interestingly, by adding massive gravity, the
non-physical region is vanished and heat capacity acquires divergence point
without any root. Before divergence point, the heat capacity is negative.
Therefore, in this region black holes are unstable. In divergence point,
black holes go under phase transition of smaller unstable to larger stable
black holes. The divergence point is an increasing function of massive
parameter (Fig. \ref{Fig1} left panel) and dimensions (Fig. \ref{Fig5} left
panel), whereas, it is a decreasing function of $k$ (Fig. \ref{Fig3} left
panel) and electric charge (Fig. \ref{Fig4} left panel).

Interestingly, in strong nonlinearity parameter, the mentioned behavior is
modified. In this case black holes enjoy one root and two divergence points.
Before root and between two divergencies, heat capacity is negative and
between root and smaller divergence point and after larger divergence point,
heat capacity is positive. According to thermodynamical concept, systems go
under phase transition to acquire stable states. Therefore, following phase
transitions take place: non-physical unstable to physical stable (in root),
large unstable to smaller stable (in smaller divergence point) and smaller
unstable to larger stable black holes (in larger divergency). Root and
smaller divergence point are increasing functions of $\beta$ (Fig. \ref{Fig2}
left panel), whereas, larger divergency is a decreasing function of it (Fig. %
\ref{Fig2} middle panel). It is worthwhile to mention that larger divergency
is not highly sensitive to variation of nonlinearity parameter.

Comparing obtained results for heat capacity (regarding phase transitions)
and the behavior of the temperature, one can see that larger to smaller
phase transition takes place at maximum (compare Fig. \ref{Fig3} left
diagram with right) and smaller to larger one happens at minimum (compare
Fig. \ref{Fig3} middle diagram with right) of temperature. Therefore, one is
able to recognize the type and number of phase transition by only studying
temperature's diagrams.

\section{$P-V$ criticality of charged black holes in EN-BI-Massive gravity
\label{PV}}

Now, we are in a position to study the critical behavior of the system
through phase diagrams. Using the renewed interpretation of the cosmological
constant as thermodynamical pressure, one can use following relation to
rewrite thermodynamical relations of the solutions in spherical horizon \cite%
{Vander}
\begin{equation}
P=-\frac{\Lambda }{8\pi },  \label{P}
\end{equation}
which results into following conjugating thermodynamical variable
corresponding to pressure \cite{Vander}
\begin{equation}
V=\left( \frac{\partial H}{\partial P}\right) _{S,Q}.  \label{V}
\end{equation}

Due to existence of the pressure in obtained relation for total mass of the
black holes, one can interpret the total mass as thermodynamical quantity
known as Enthalpy. This interpretation will lead to the following relation
for Gibbs free energy \cite{Vander}
\begin{equation}
G=H-TS=M-TS.  \label{G}
\end{equation}

Now by using Eqs. (\ref{TotalM}) and (\ref{P}) with the relations of volume
and Gibbs free energy (Eqs. (\ref{V}) and (\ref{G})), one finds
\begin{equation}
V=\frac{\omega _{d_{2}}}{d_{1}}r_{+}^{d_{1}},
\end{equation}%
and
\begin{eqnarray}
G &=&\frac{r_{+}^{d_{1}}}{d_{1}d_{2}}P+\frac{m^{2}c^{2}r_{+}^{d_{5}}}{16\pi }%
\left( 3d_{3}d_{4}c_{4}c^{2}+2d_{3}c_{3}cr_{+}+c_{2}r_{+}^{2}\right) +\frac{%
d_{2}^{2}q^{2}\mathcal{H}_{+}}{2\pi d_{1}r_{+}^{d_{3}}}+  \notag \\
&&\frac{\beta ^{2}r_{+}^{d_{1}}}{4\pi d_{1}d_{2}}\sqrt{1+\Gamma _{+}}+\frac{%
r_{+}^{d_{3}}}{16\pi }.
\end{eqnarray}

Obtained relation for volume indicates that volume of the black holes is
only a function of the topology of the solutions and independent of
electrodynamics and gravitational extensions, directly.

In order to obtain critical values, one can use $P-V$ diagrams. In other
words, by studying inflection point properties one can obtain critical
values in which phase transitions may take place. Therefore, one can use
\begin{equation}
\left( \frac{\partial P}{\partial r_{+}}\right) _{T} =\left( \frac{\partial
^{2}P }{\partial r_{+}^{2}}\right) _{T} =0.  \label{infel}
\end{equation}

Considering obtained values for temperature (\ref{TotalTT}) and pressure (%
\ref{P}), one can obtain pressure as
\begin{equation}
P=\frac{d_{2}T}{4r_{+}}-\frac{m^{2}c}{16\pi r_{+}^{4}}\left[
c_{1}r_{+}^{3}+d_{3}c_{2}cr_{+}^{2}+d_{3}d_{4}c_{3}c^{2}r_{+}+d_{3}d_{4}d_{5}c_{4}c^{3}%
\right] -\frac{d_{2}d_{3}}{16\pi r_{+}^{2}}+\frac{\beta ^{2}}{4\pi }\left(
\sqrt{1+\Gamma _{+}}-1\right) .  \label{PP}
\end{equation}

Now, by considering Eq. (\ref{infel}) with obtained relation for pressure (%
\ref{PP}), one can obtain two relations for finding critical quantities. Due
to economical reasons, we will not present them. Regarding the contribution
of electromagnetic part, it is not possible to obtain critical horizon
analytically, and therefore, we use numerical method. Considering the
variation of $\beta$ and massive parameter, one can draw following tables

\begin{center}
\begin{tabular}{ccccc}
\hline\hline
$m$ & $r_{c}$ & $T_{c}$ & $P_{c}$ & $\frac{P_{c}r_{c}}{T_{c}}$ \\
\hline\hline
$0.000000$ & $1.8264628$ & $0.1334354$ & $0.02200146$ & $0.3011558$ \\ \hline
$0.100000$ & $1.8263848$ & $0.1334835$ & $0.02200495$ & $0.3010822$ \\ \hline
$1.000000$ & $1.7953522$ & $0.1382092$ & $0.02233685$ & $0.2901581$ \\ \hline
$5.000000$ & $1.6278897$ & $0.2562956$ & $0.03187871$ & $0.2024811$ \\ \hline
$10.000000$ & $0.7052643$ & $0.7329062$ & $0.13205092$ & $0.1270705$ \\
\hline
\end{tabular}
\\[0pt]
\vspace{0.1cm} Table ($1$): $q=1$, $\beta =0.5$,
$c_{1}=c_{2}=c_{3}=0.2$, $c_{4}=0$ and $d=5$. \vspace{0.5cm}
\end{center}


\begin{center}
\begin{tabular}{ccccc}
\hline\hline
$\beta $ & $r_{c}$ & $T_{c}$ & $P_{c}$ & $\frac{P_{c}r_{c}}{T_{c}}$ \\
\hline\hline
$1.000000$ & $1.7819632$ & $0.1693410$ & $0.0296043$ & $0.3115245$ \\ \hline
$2.100000$ & $1.8174387$ & $0.1677996$ & $0.0290170$ & $0.3142835$ \\ \hline
$3.500000$ & $1.8235114$ & $0.1675323$ & $0.0289161$ & $0.3147386$ \\ \hline
$4.000000$ & $1.8256045$ & $0.1674399$ & $0.0288813$ & $0.3148944$ \\ \hline
$5.000000$ & $1.8265678$ & $0.1673974$ & $0.0288653$ & $0.3149659$ \\ \hline
\end{tabular}
\\[0pt]
\vspace{0.1cm} Table ($2$): $q=1$, $m=0.1$, $c_{1}=c_{2}=c_{3}=2$,
$c_{4}=0$ and $d=5$. \vspace{0.5cm}
\end{center}

\begin{figure}[tbp]
$%
\begin{array}{ccc}
\epsfxsize=5cm \epsffile{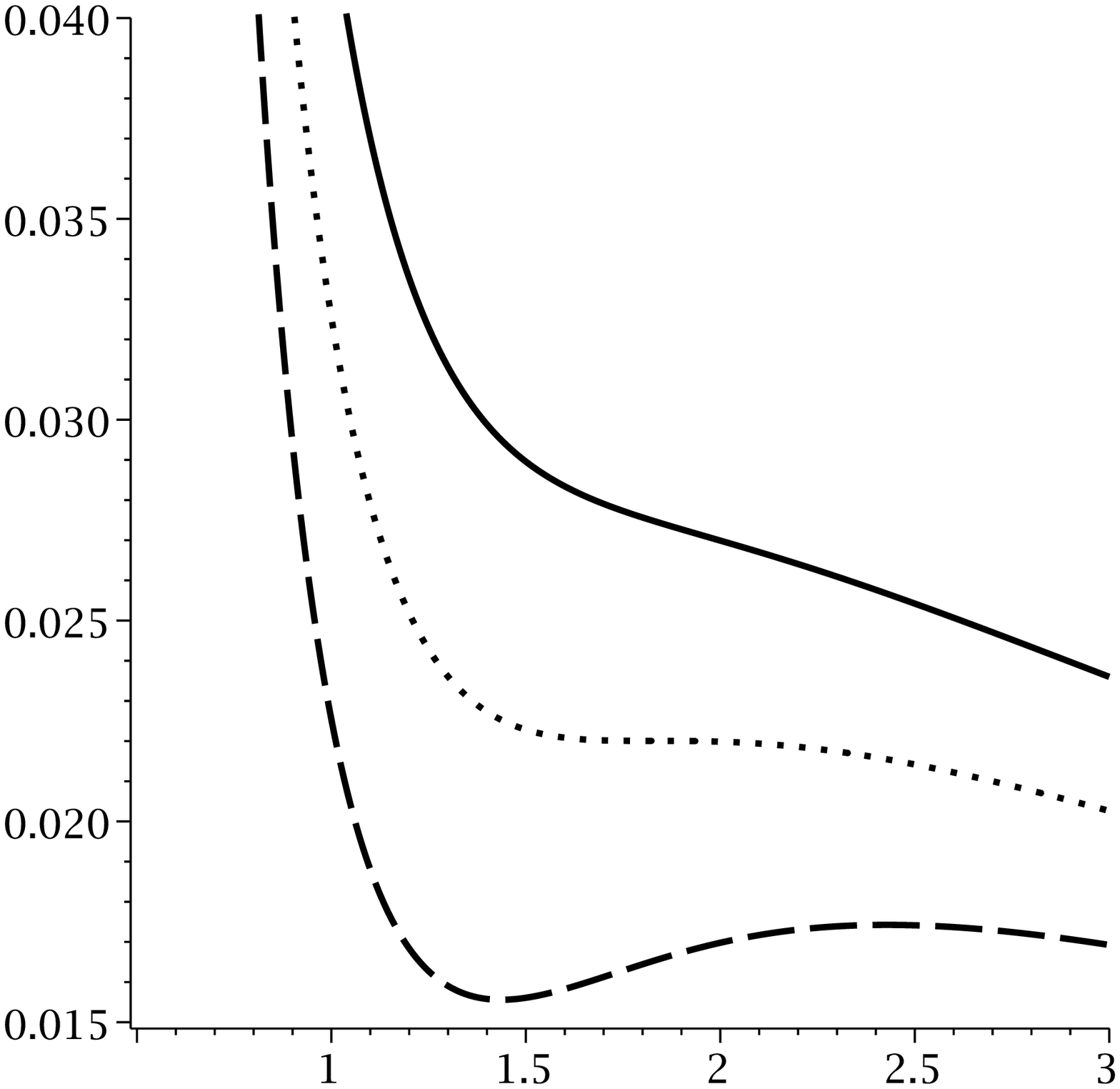} & \epsfxsize=5cm \epsffile{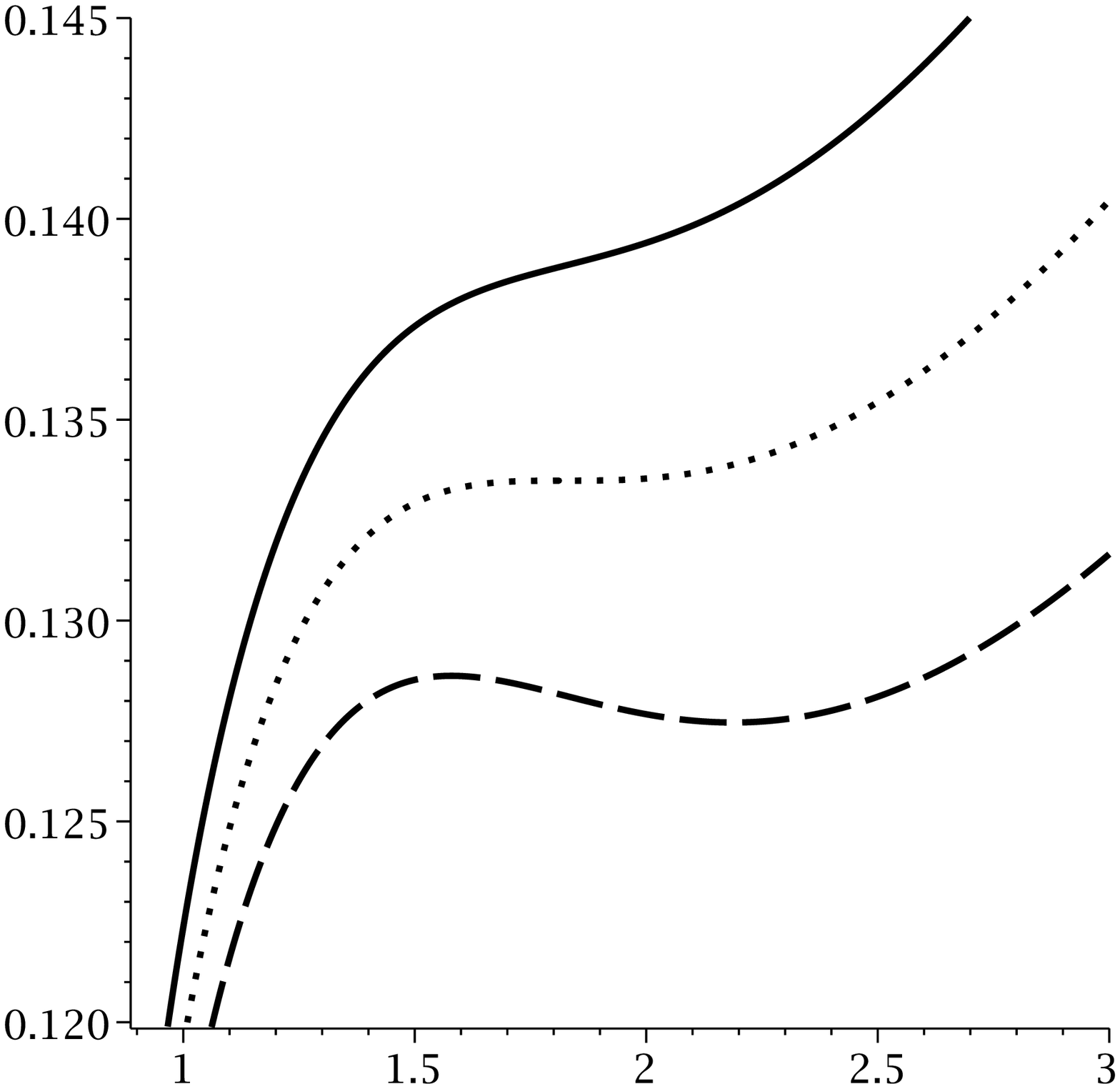}
& \epsfxsize=5cm \epsffile{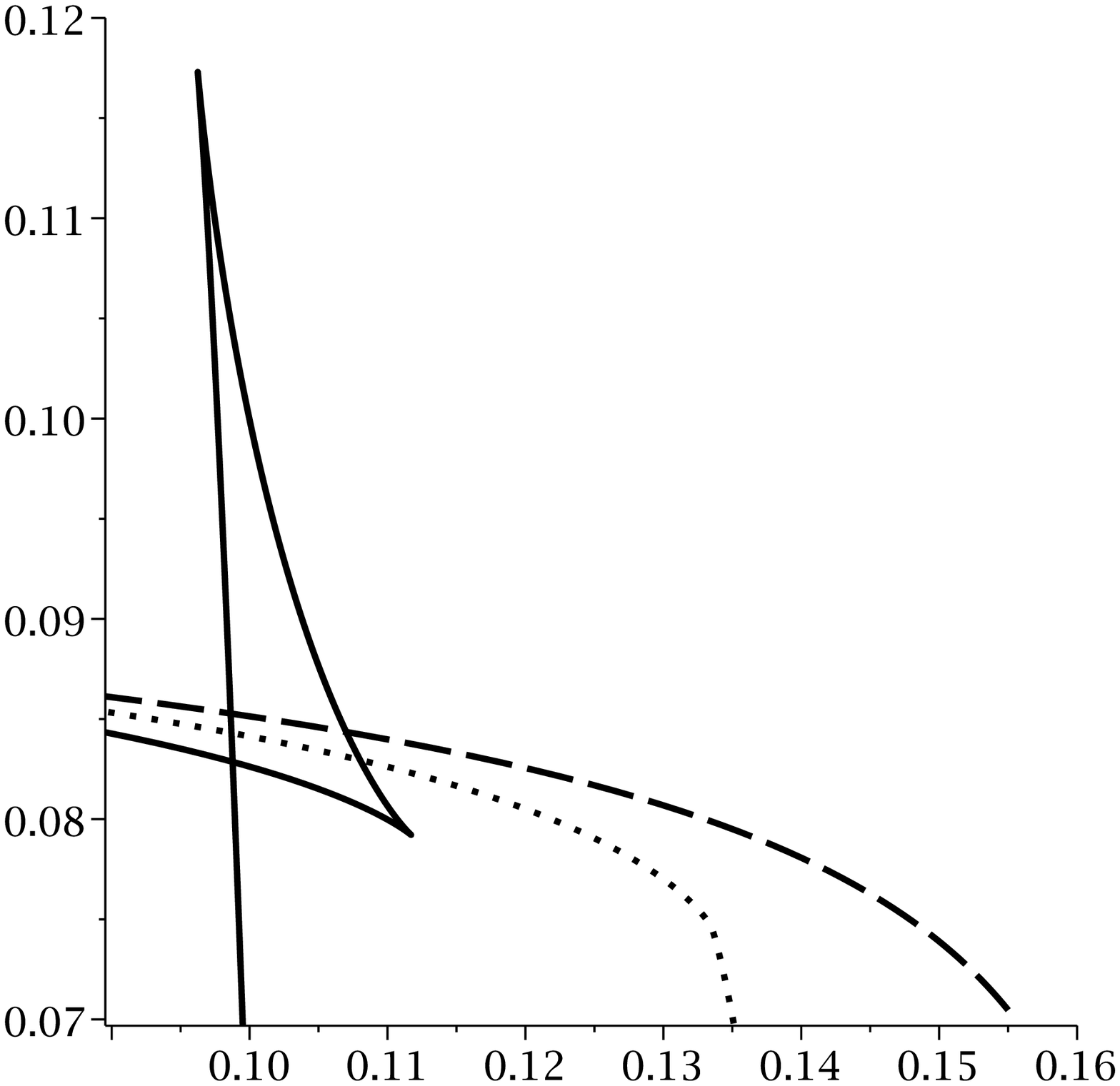}%
\end{array}
$%
\caption{ $P-r_{+}$ (left), $T-r_{+}$ (middle) and $G-T$ (right) diagrams
for $\protect\beta=0.5$, $q=1$, $m=0.1$, $c=c_{1}=c_{2}=c_{3}=0.2$, $c_{4}=0$ and $d=5$%
. \newline
$P-r_{+}$ diagram, from up to bottom $T=1.1T_{c}$, $T=T_{c}$ and $T=0.9T_{c}$%
, respectively. \newline
$T-r_{+}$ diagram, from up to bottom $P=1.1P_{c}$, $P=P_{c}$ and $P=0.9P_{c}$%
, respectively. \newline
$G-T$ diagram for $P=0.5P_{c}$ (continuous line), $P=P_{c}$ (dotted line)
and $P=1.5P_{c}$ (dashed line). }
\label{Fig6}
\end{figure}

\begin{figure}[tbp]
$%
\begin{array}{ccc}
\epsfxsize=5cm \epsffile{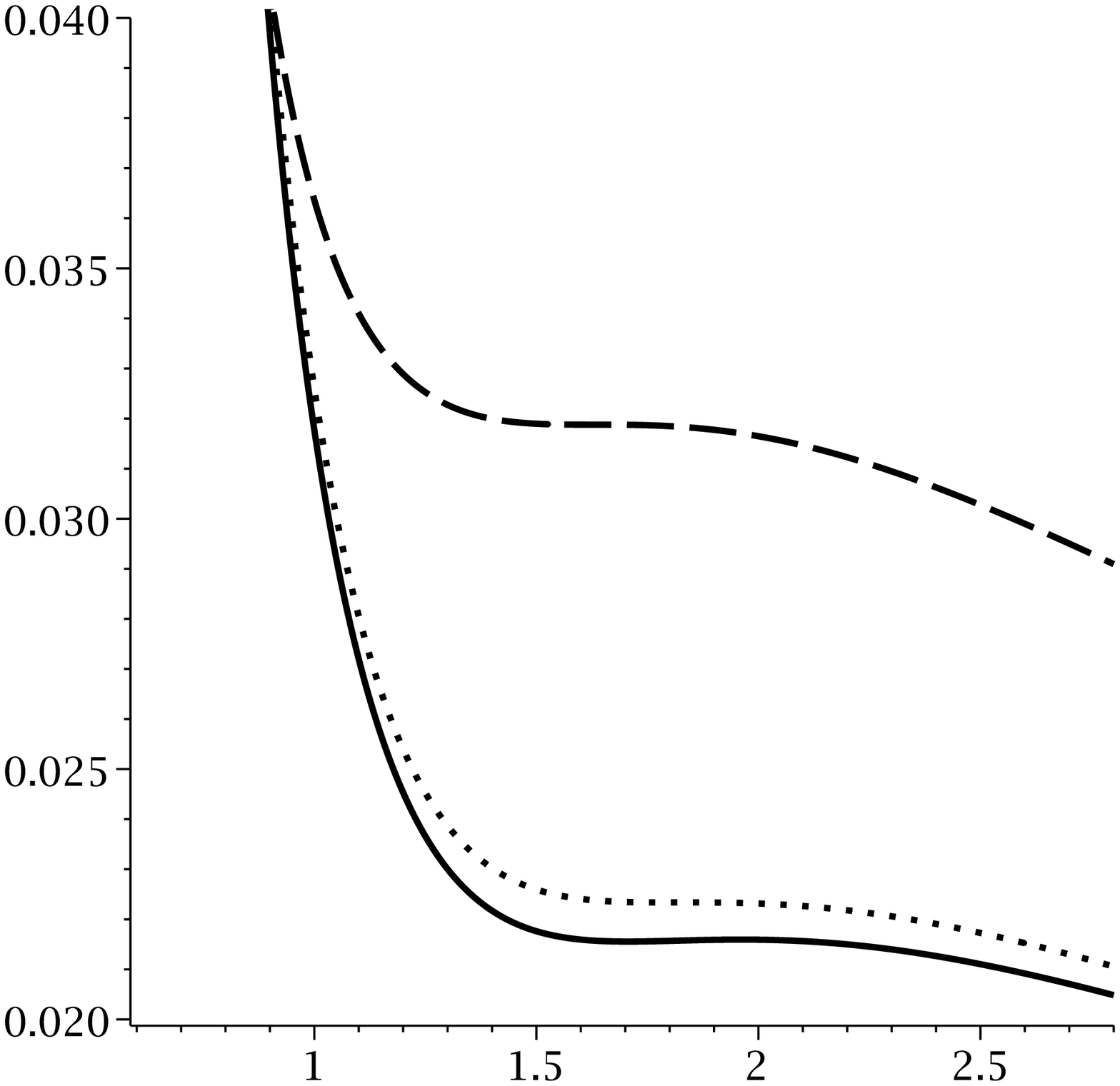} & \epsfxsize=5cm %
\epsffile{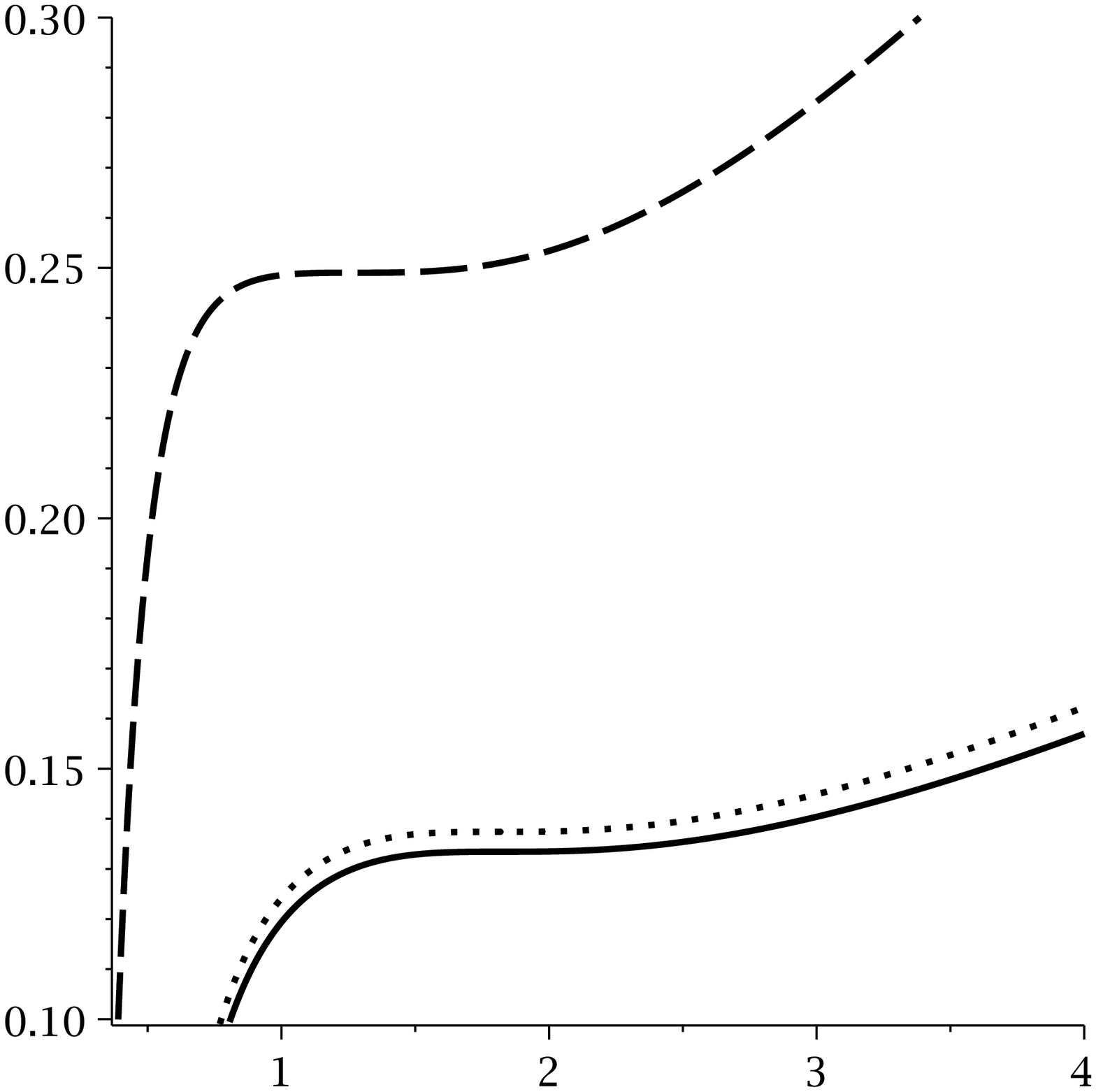} & \epsfxsize=5cm \epsffile{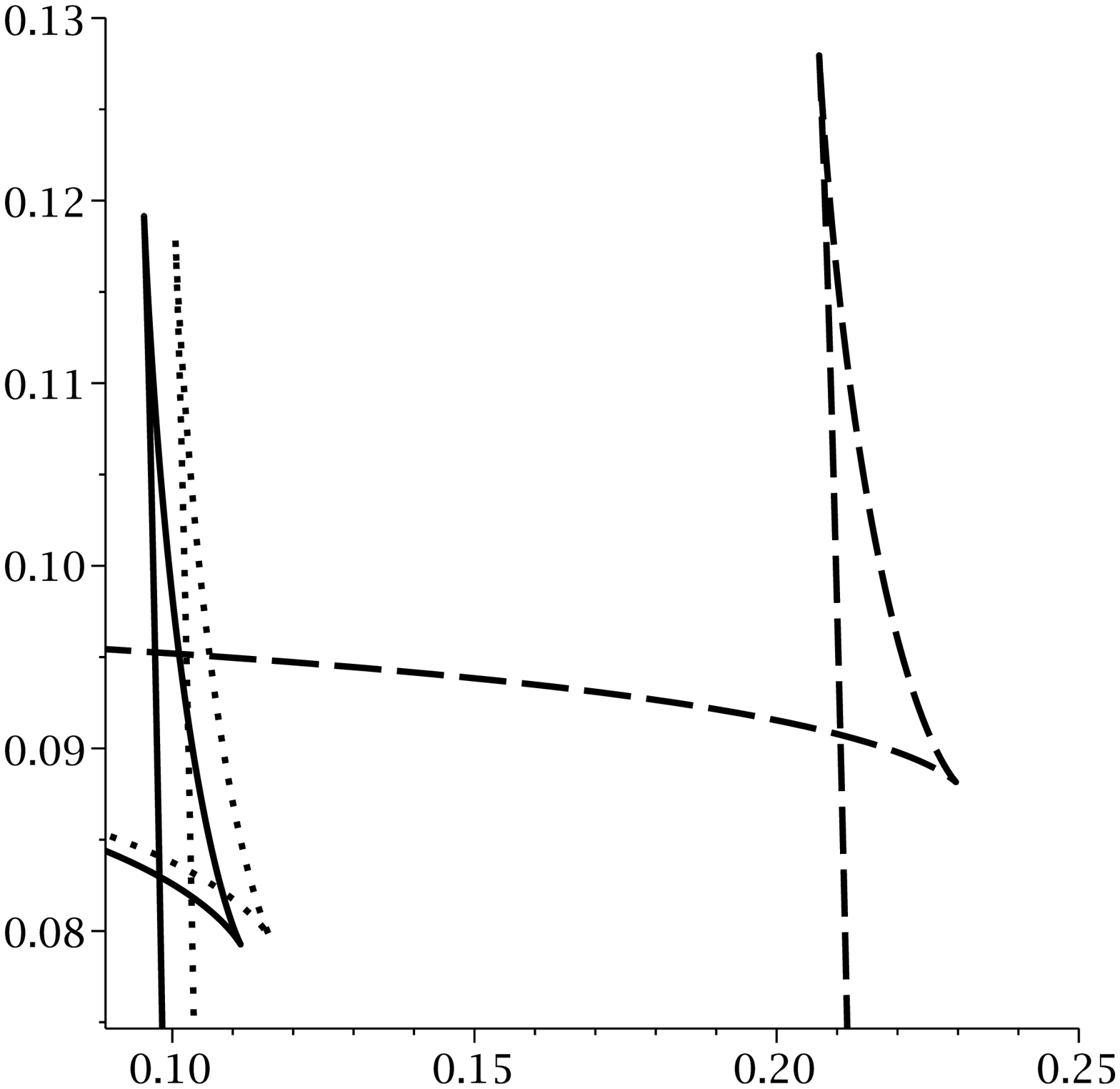}%
\end{array}
$%
\caption{ $P-r_{+}$ (left), $T-r_{+}$ (middle) and $G-T$ (right)
diagrams for $\protect\beta=0.5$, $q=1$,
$c=c_{1}=c_{2}=c_{3}=0.2$, $c_{4}=0$, $d=5$, $m=0$ (continuous
line), $m=1$ (dotted line) and $m=5$ (dashed line). \newline
$P-r_{+}$ diagram for $T=T_{c}$, $T-r_{+}$ diagram for $P=P_{c}$
and $G-T$ diagram for $P=0.5P_{c}$. } \label{Fig7}
\end{figure}

\begin{figure}[tbp]
$%
\begin{array}{ccc}
\epsfxsize=5cm \epsffile{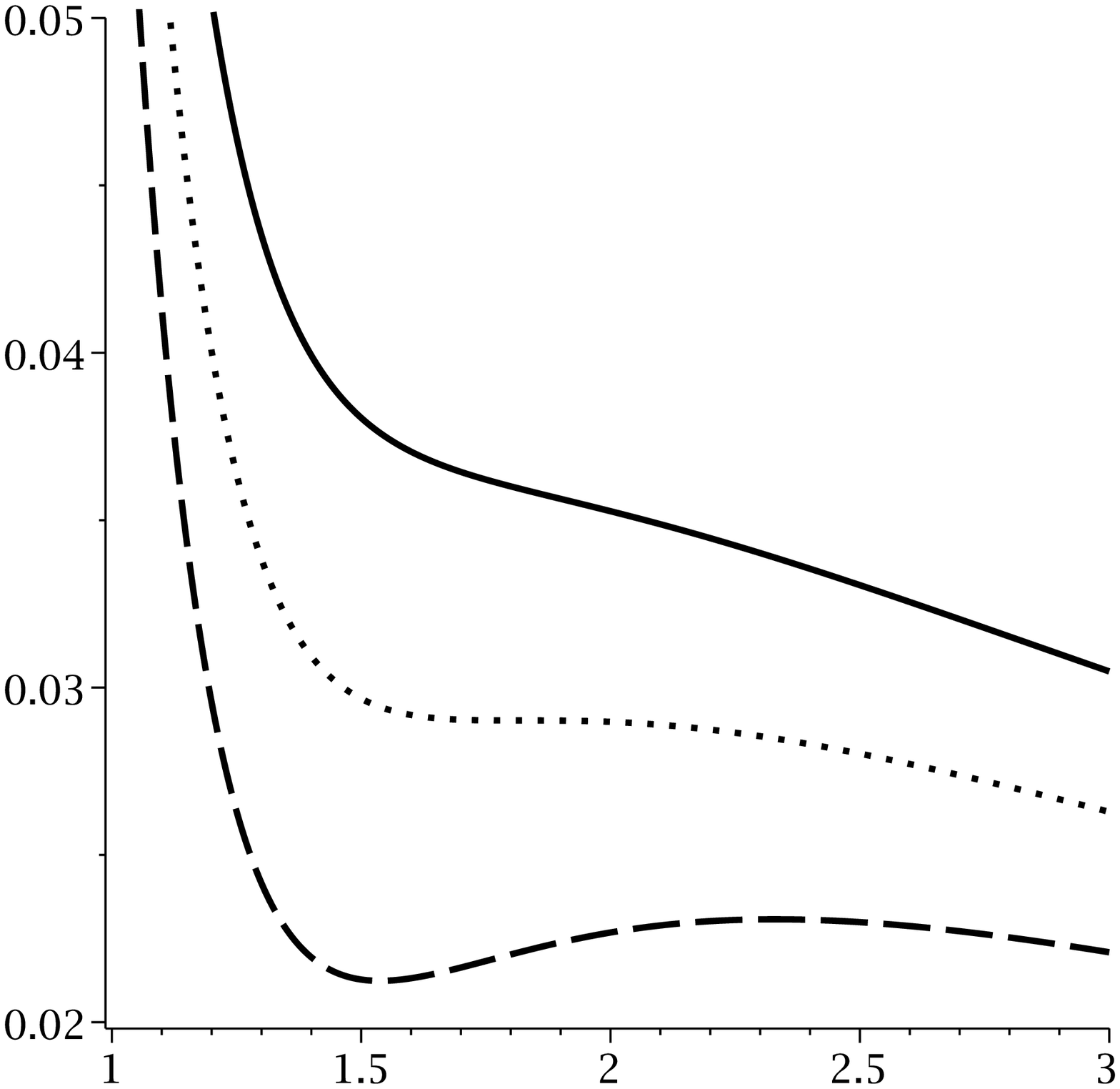} & \epsfxsize=5cm %
\epsffile{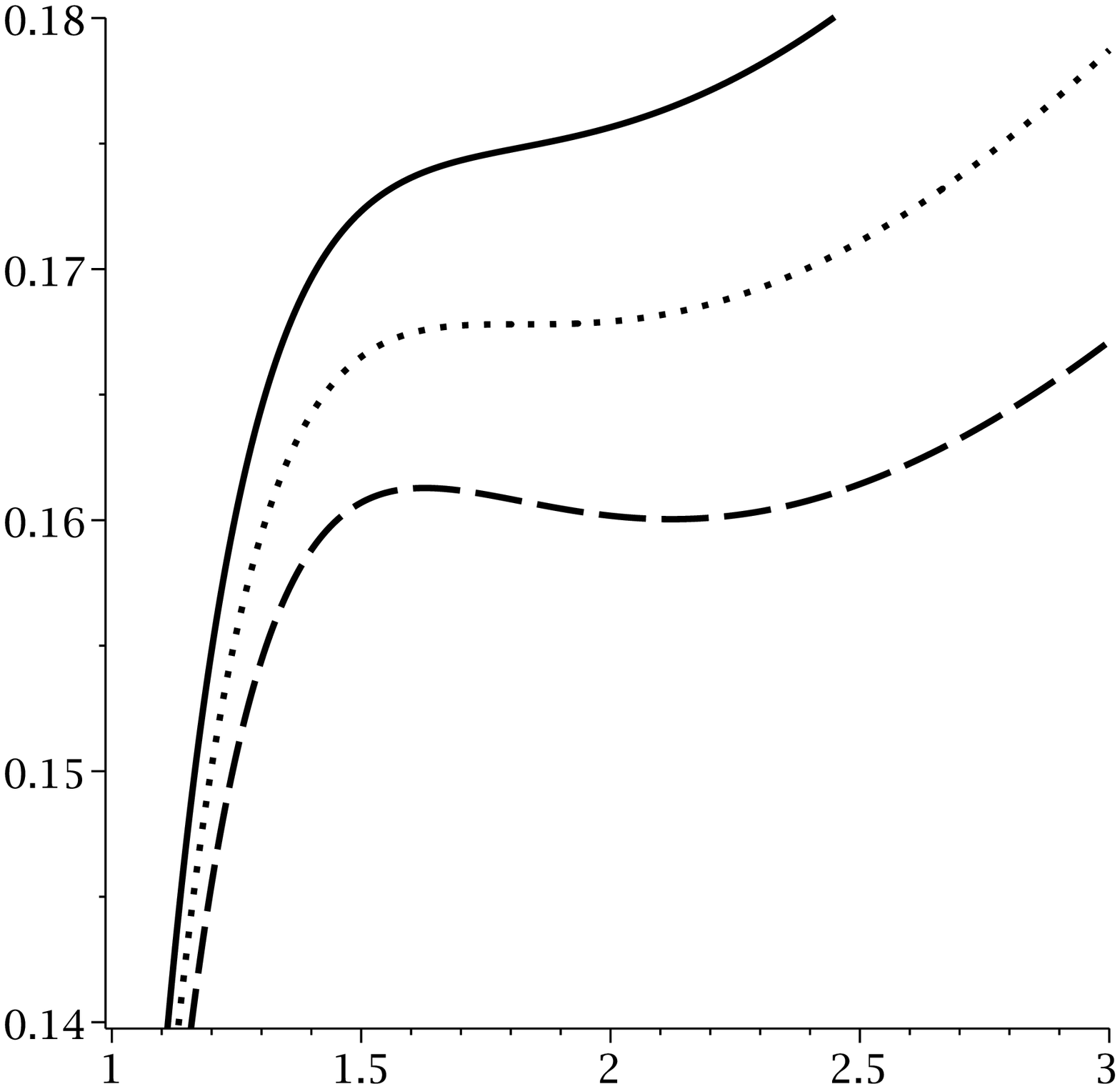} & \epsfxsize=5cm \epsffile{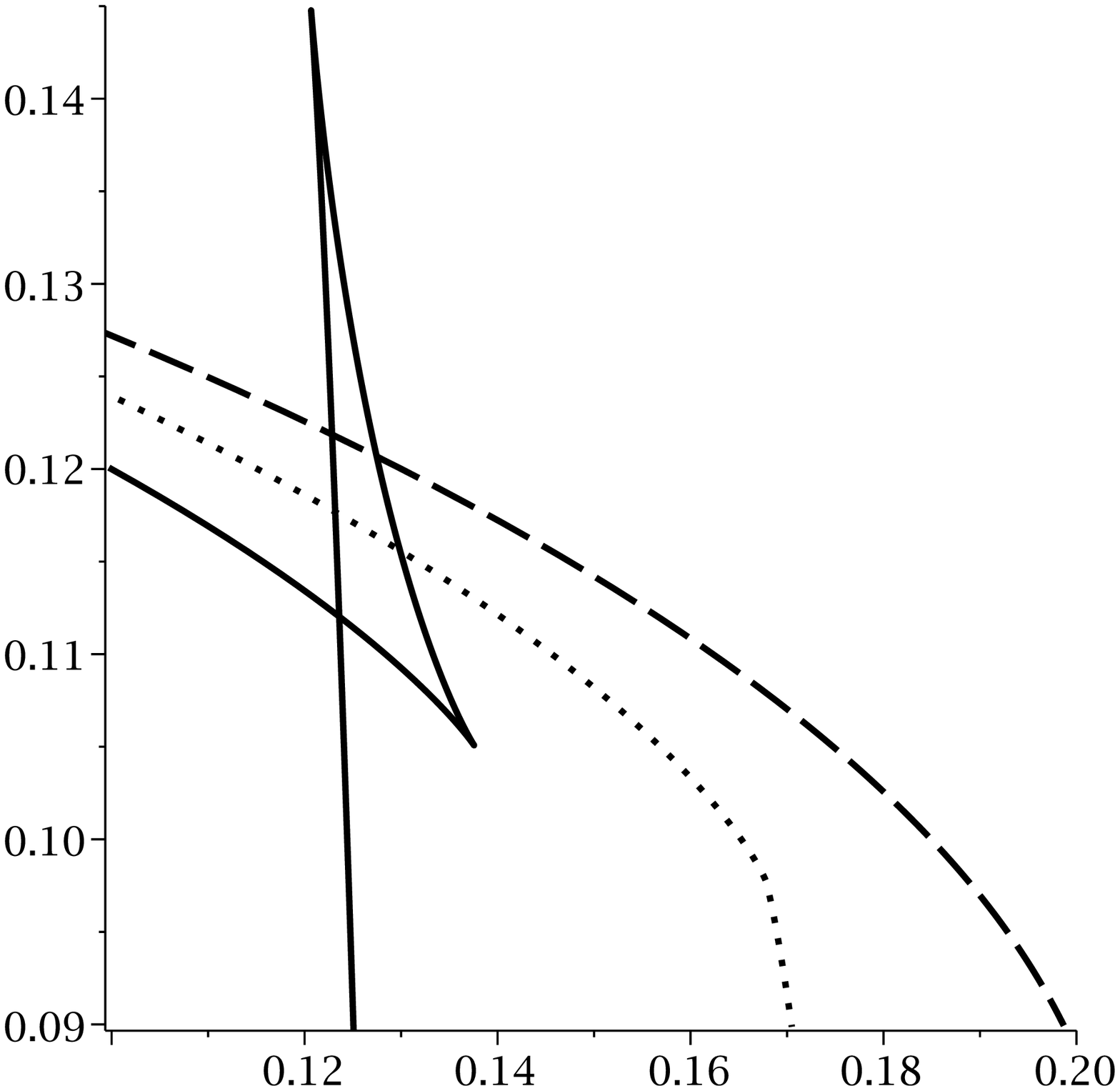}%
\end{array}
$%
\caption{ $P-r_{+}$ (left), $T-r_{+}$ (middle) and $G-T$ (right)
diagrams for $\protect\beta=2$, $q=1$, $m=0.1$,
$c=c_{1}=c_{2}=c_{3}=2$, $c_{4}=0$ and $d=5$.
\newline
$P-r_{+}$ diagram, from up to bottom $T=1.1T_{c}$, $T=T_{c}$ and $T=0.9T_{c}$%
, respectively. \newline
$T-r_{+}$ diagram, from up to bottom $P=1.1P_{c}$, $P=P_{c}$ and $P=0.9P_{c}$%
, respectively. \newline
$G-T$ diagram for $P=0.5P_{c}$ (continuous line), $P=P_{c}$ (dotted line)
and $P=1.5P_{c}$ (dashed line). }
\label{Fig8}
\end{figure}

\begin{figure}[tbp]
$%
\begin{array}{ccc}
\epsfxsize=5cm \epsffile{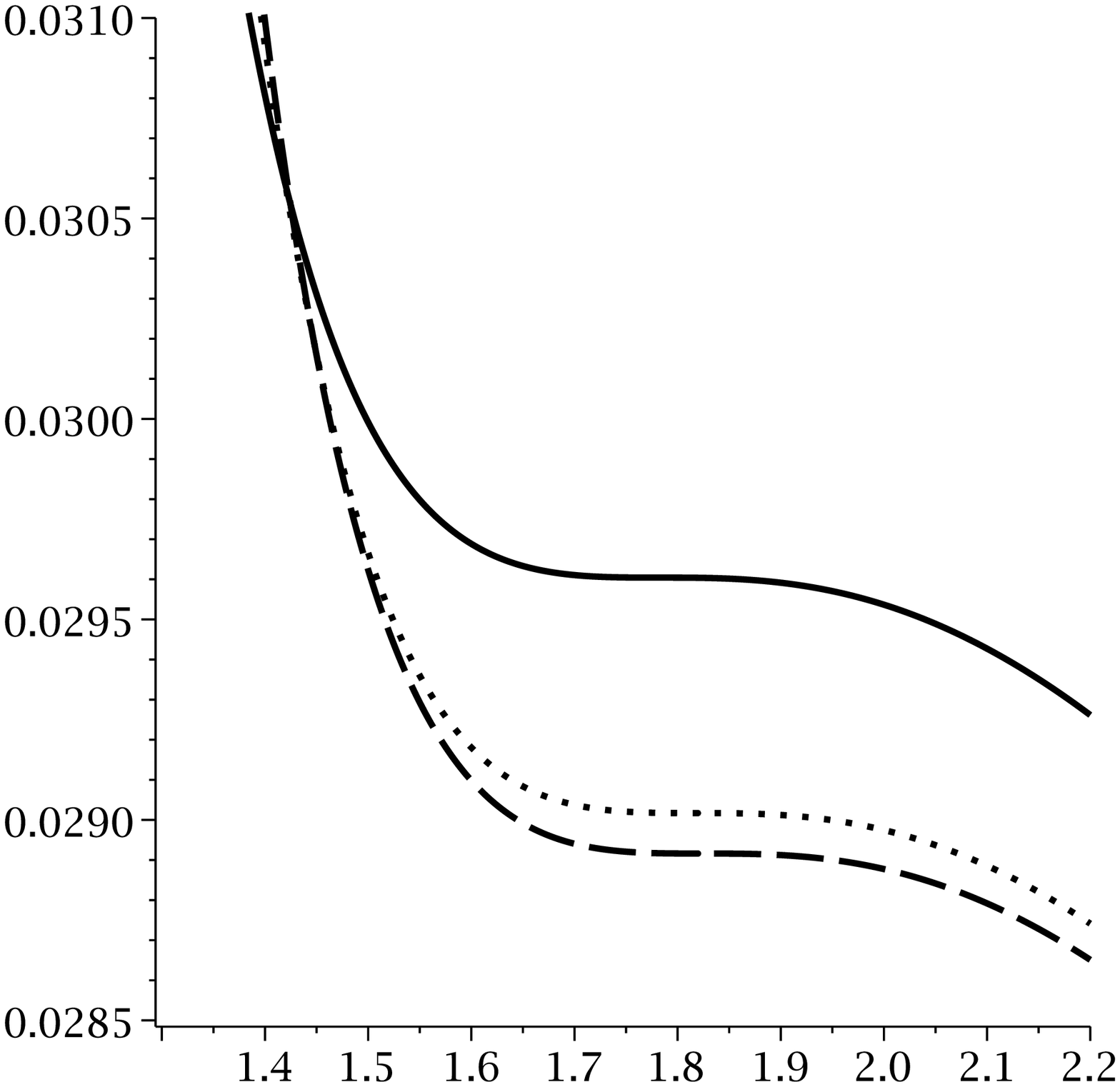} & \epsfxsize=5cm %
\epsffile{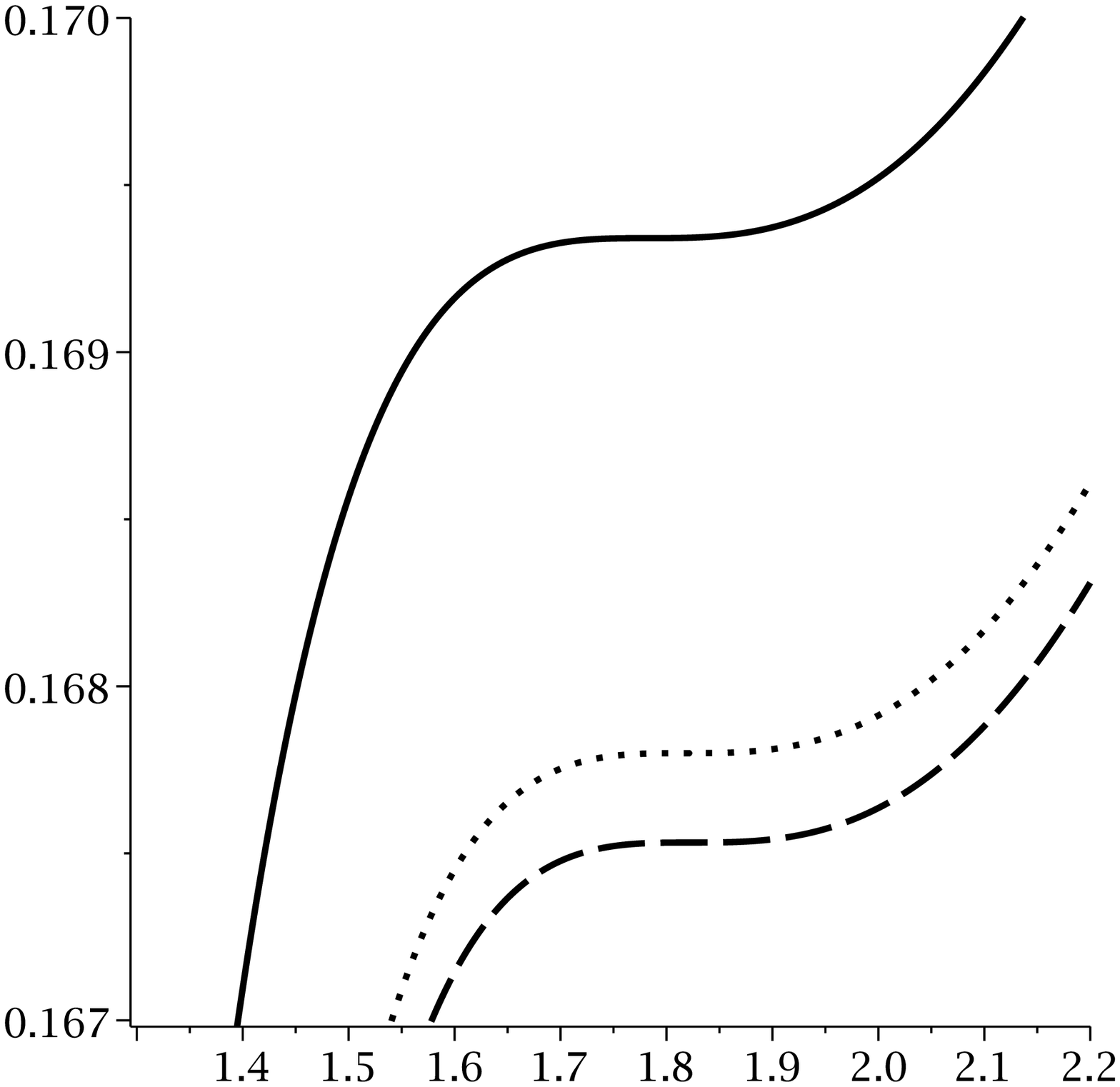} & \epsfxsize=5cm %
\epsffile{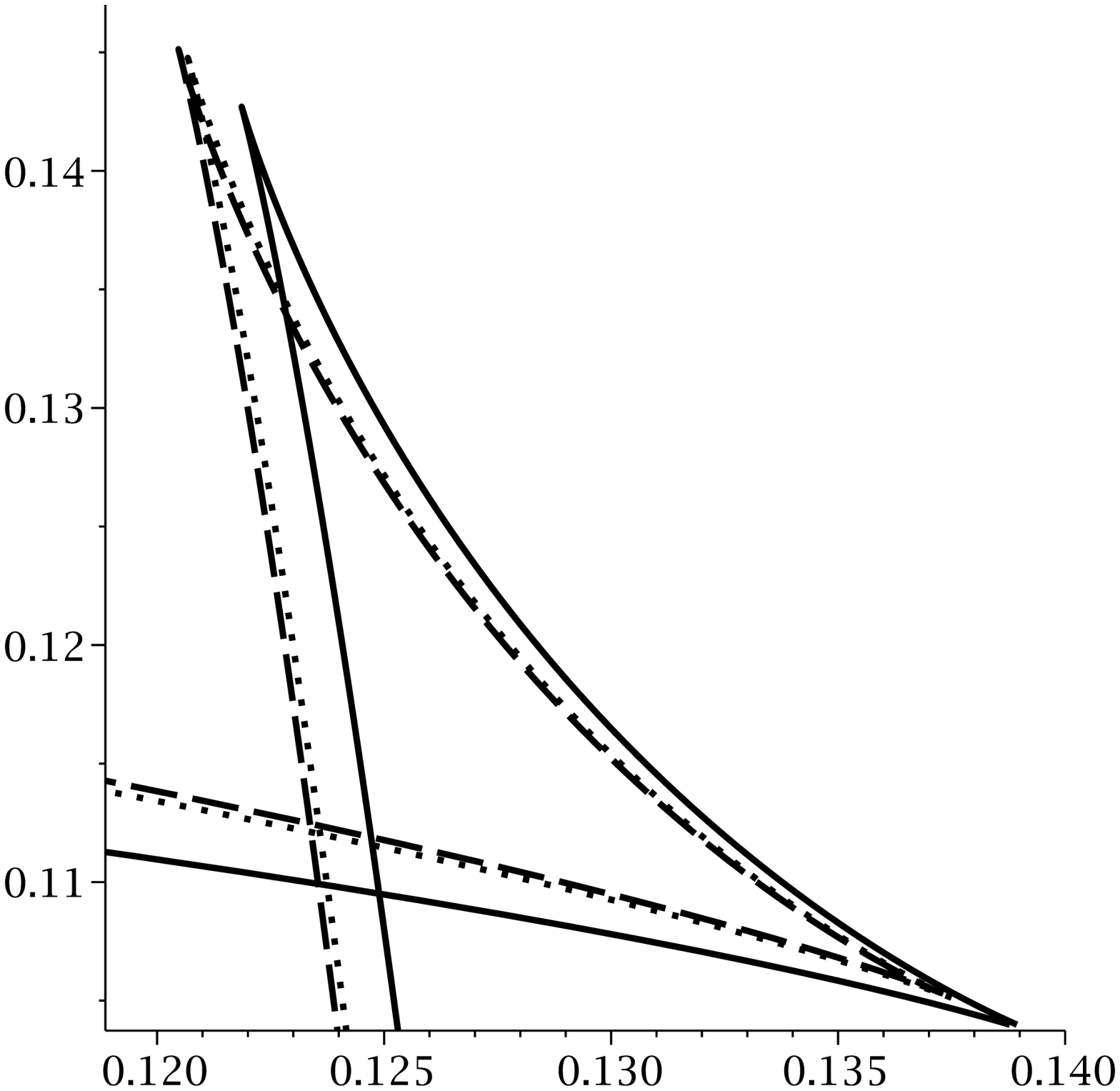}%
\end{array}
$%
\caption{ $P-r_{+}$ (left), $T-r_{+}$ (middle) and $G-T$ (right)
diagrams for $m=0.1$, $q=1$, $c=c_{1}=c_{2}=c_{3}=0.2$, $c_{4}=0$,
$d=5$, $\protect\beta=1$ (continuous line), $\protect\beta=2$
(dotted line) and $\protect\beta=3$ (dashed line). \newline
$P-r_{+}$ diagram for $T=T_{c}$, $T-r_{+}$ diagram for $P=P_{c}$
and $G-T$ diagram for $P=0.5P_{c}$. } \label{Fig9}
\end{figure}

In addition, we plot following diagrams (Figs. \ref{Fig6} - \ref{Fig9}) to
investigate that obtained values are the ones in which phase transition
takes place or not.

The formation of swallow tail in $G-T$ diagrams for pressure
smaller than critical pressure (Figs. \ref{Fig6} and \ref{Fig8}
right panels), subcritical isobars in $T-r_{+}$ diagrams for
critical pressure (Figs. \ref{Fig6} and \ref{Fig8} middle panel)
and isothermal diagrams in case of critical temperature in
$P-r_{+}$ diagrams (Figs. \ref{Fig6} and \ref{Fig8} left panels),
show that obtained values are critical ones in which phase
transition takes place.

It is evident that critical pressure (Fig. \ref{Fig7} left panel) and
temperature (Fig. \ref{Fig7} middle panel) are increasing functions of the
massive parameter, whereas the critical horizon (Fig. \ref{Fig7} left and
middle panels) and universal ratio of $\frac{P_{c}r_{c}}{T_{c}}$ are
decreasing functions of this parameter.

It is worthwhile to mention that length of subcritical isobars (which is
known as phase transition region) is a decreasing function of massive
parameter (Fig. \ref{Fig7} middle panel). In opposite, the size of swallow
tail and the energy of different phases are increasing functions of $m$
(Fig. \ref{Fig7} right panel).

Interestingly, the effects of variation of nonlinearity parameter is
opposite of massive parameter. In other words, critical pressure (Fig. \ref%
{Fig9} left panel), temperature (Fig. \ref{Fig9} middle panel) and the size
of swallow tail (Fig. \ref{Fig9} left panel) are decreasing functions of $%
\beta$, whereas, the critical horizon radius (Fig. \ref{Fig9} left and
middle panels), length of subcritical isobars (Fig. \ref{Fig9} middle panel)
and universal ration of $\frac{P_{c}r_{c}}{T_{c}}$ are increasing functions
of nonlinearity parameter.

It should be pointed that the length of subcritical isobars affects single
regions of different states which in our cases are smaller and larger black
holes. In other words, increasing the length of subcritical isobars (phase
transition region) decreases the single state regions.

\subsection{Neutral Massive black holes}

In this section, by cancelling the electric charge ($q=0$), we will study
the critical behavior of the system. Previously, it was shown that
Schwarzschild black holes does not have any phase transition in context of
extended phase space. Now, we are investigating the effects of massive
gravity in case of EN-massive gravity. Using obtained relation for
calculating critical horizon radius in previous part and setting $q=0$, one
can find following relation for calculating critical horizon radius
\begin{equation}
m^{2}\left(
6d_{4}d_{5}c_{4}c^{4}+3d_{4}c_{3}c^{3}r_{+}+c_{2}c^{2}r_{+}^{2}\right)
+r_{+}^{2}=0.
\end{equation}

It is a matter of calculation to show that this relation has following roots
which are critical horizon radii
\begin{equation}
r_{c}=-\frac{mc^{2}\left( 3d_{4}mc_{3}c\pm \sqrt{-3d_{4}\left[ 8d_{5}\left(
1+c_{2}c^{2}m^{2}\right) c_{4}-3d_{4}m^{2}c_{3}^{2}c^{2}\right] }\right) }{%
2\left( 1+c_{2}c^{2}m^{2}\right) }.  \label{root}
\end{equation}

Obtained relation shows that in absence of massive gravity, critical horizon
radius will be zero which is not of our interest. This result consistent
with Schwarzschild case. Now, for the simplicity, we consider the case of $%
c_{4}=0$. This leads into following critical horizon radius
\begin{equation}
r_{cc}=-\frac{3d_{4}mc_{3}c}{1+c_{2}c^{2}m^{2}}.  \label{rcc}
\end{equation}

It is evident that for the cases of $d=4$ and $d>4$ with vanishing
$c_{3}$, the critical horizon radius will be zero. Therefore,
there is no phase transition for these black holes. Interestingly
for case of $d>4$, the condition for having a positive critical
horizon radius will be $c_{3}<0$ and $1+c_{2}c^{2}m^{2}>0$. By
employing obtained value for critical horizon radius, one can find
critical temperature and pressure in the following forms
\begin{equation}
T_{cc}=\frac{\left( 3d_{4}c_{1}c_{3}-d_{3}c_{2}^{2}\right)
m^{4}c^{4}-d_{3}\left( 2c_{2}m^{2}c^{2}+1\right) }{12\pi d_{4}m^{2}c_{3}c^{3}%
},  \label{Tcc}
\end{equation}
\begin{equation}
P_{cc}=-\frac{\left( c_{2}m^{2}c^{2}+1\right) ^{3}d_{2}d_{3}}{432\pi
d_{4}^{2}m^{4}c_{3}^{2}c^{6}}.  \label{Pcc}
\end{equation}

\begin{figure}[tbp]
$%
\begin{array}{cc}
\epsfxsize=5cm \epsffile{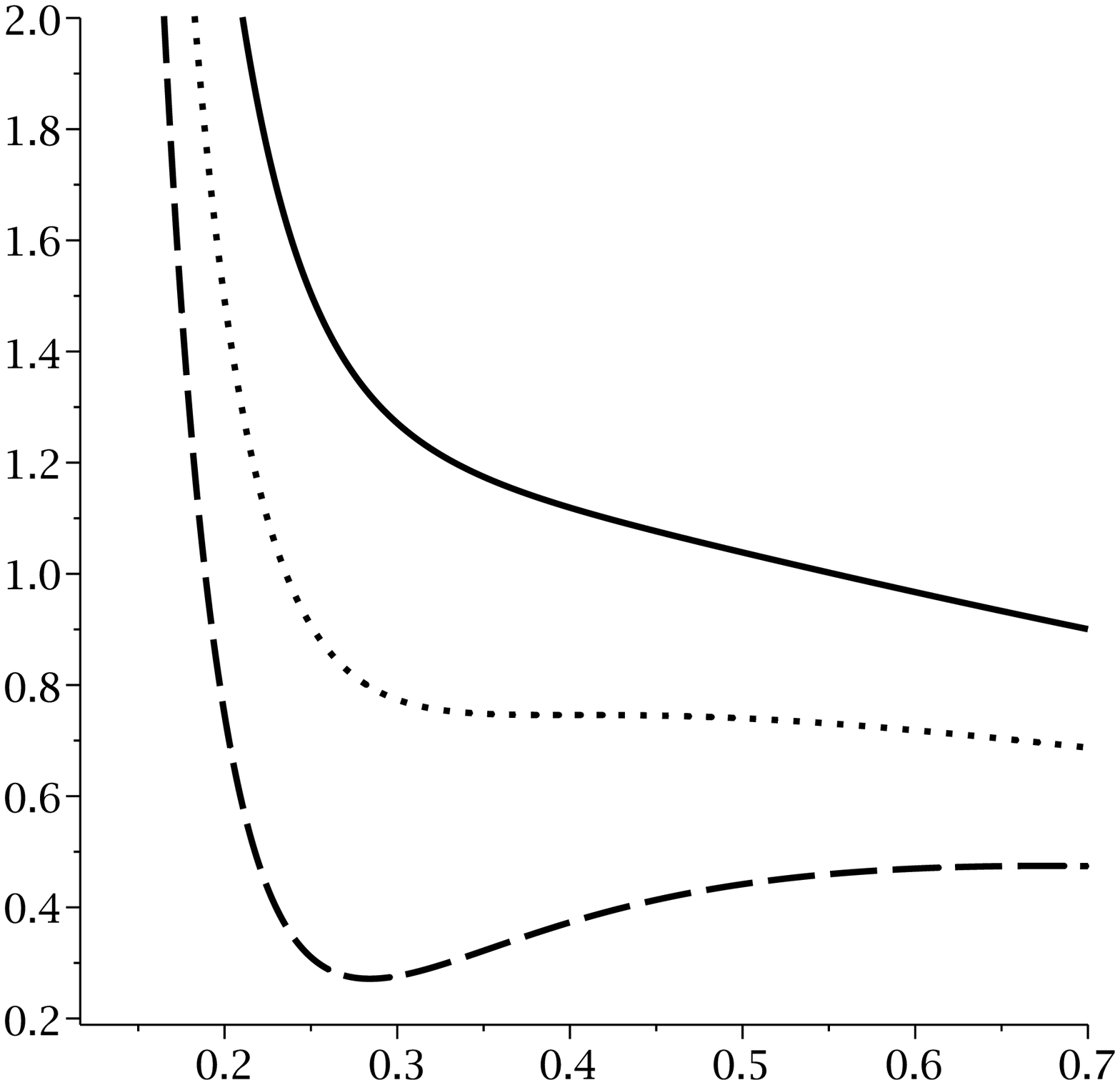} & \epsfxsize=5cm %
\epsffile{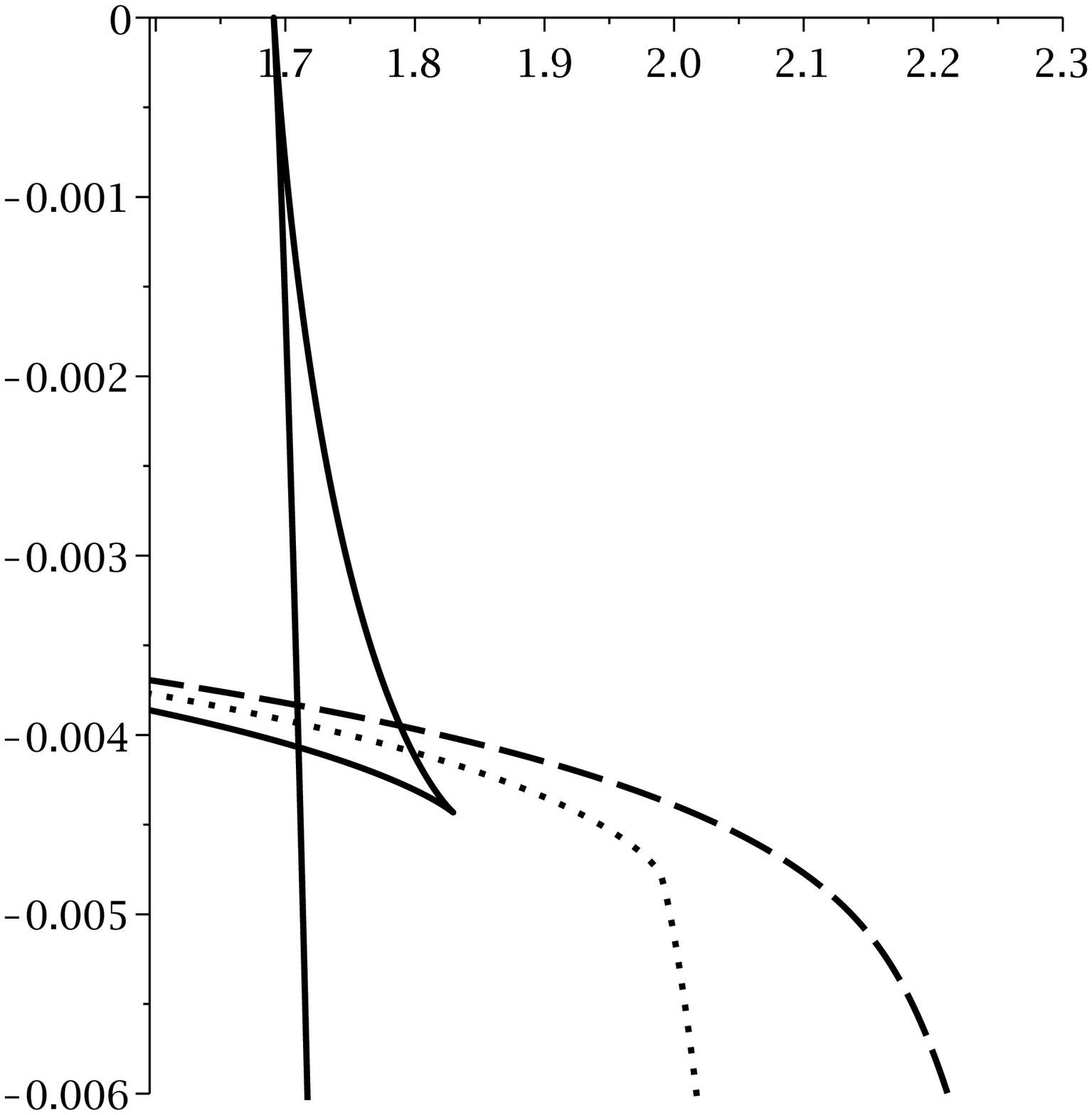}%
\end{array}
$%
\caption{ $P-r_{+}$ (left) and $G-T$ (right) diagrams for $m=5$, $c=0.2$, $%
c_{1}=c_{2}=2$, $c_{3}=-2$, $c_{4}=0$ and $d=5$. \newline
$P-r_{+}$ diagram, from up to bottom $T=1.1T_{c}$, $T=T_{c}$ and $T=0.9T_{c}$%
, respectively. \newline
$G-T$ diagram for $P=0.5P_{c}$ (continuous line), $P=P_{c}$ (dotted line)
and $P=1.5P_{c}$ (dashed line). }
\label{q0}
\end{figure}

\begin{figure}[tbp]
$%
\begin{array}{cc}
\epsfxsize=5cm \epsffile{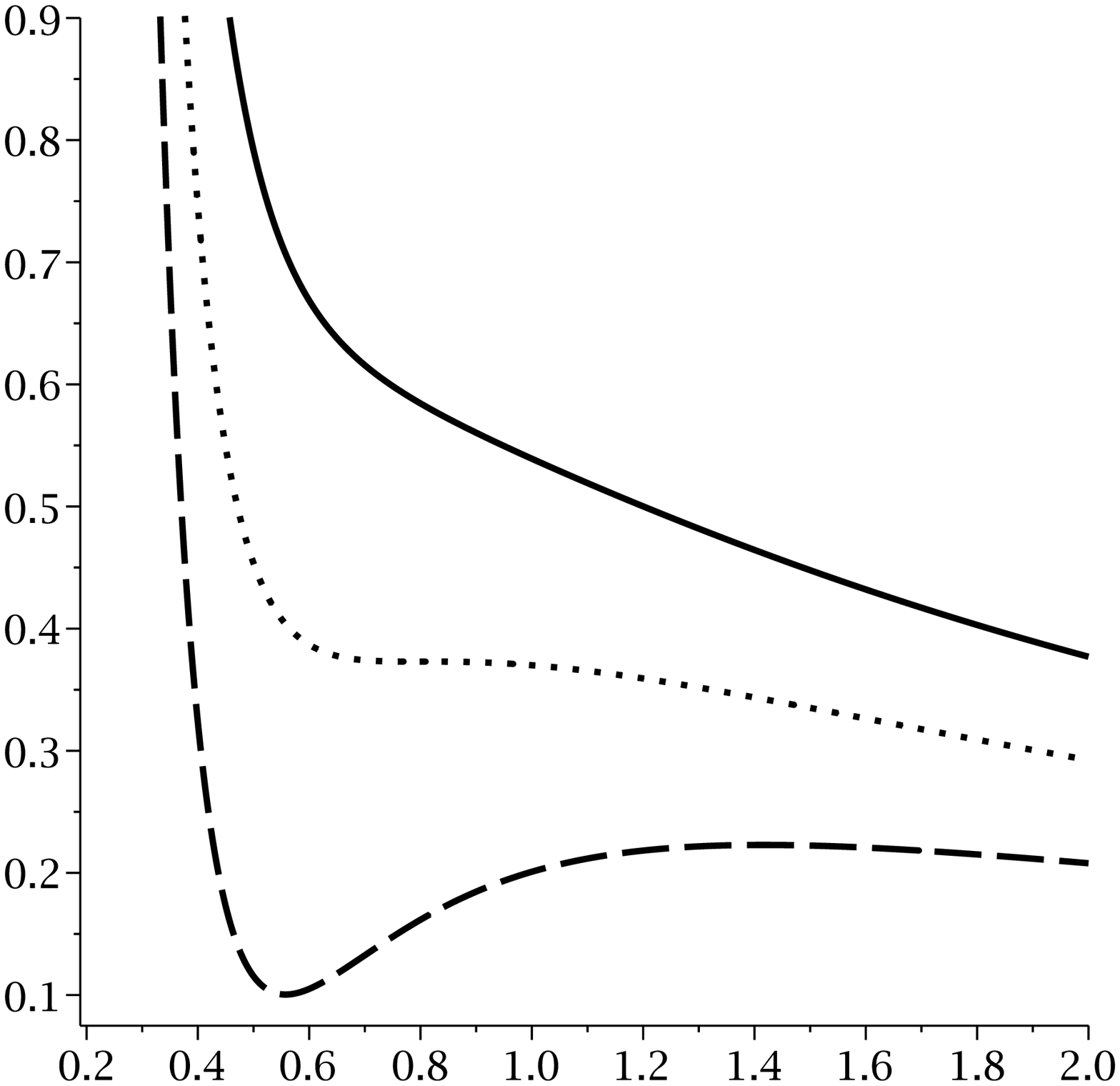} & \epsfxsize=5cm %
\epsffile{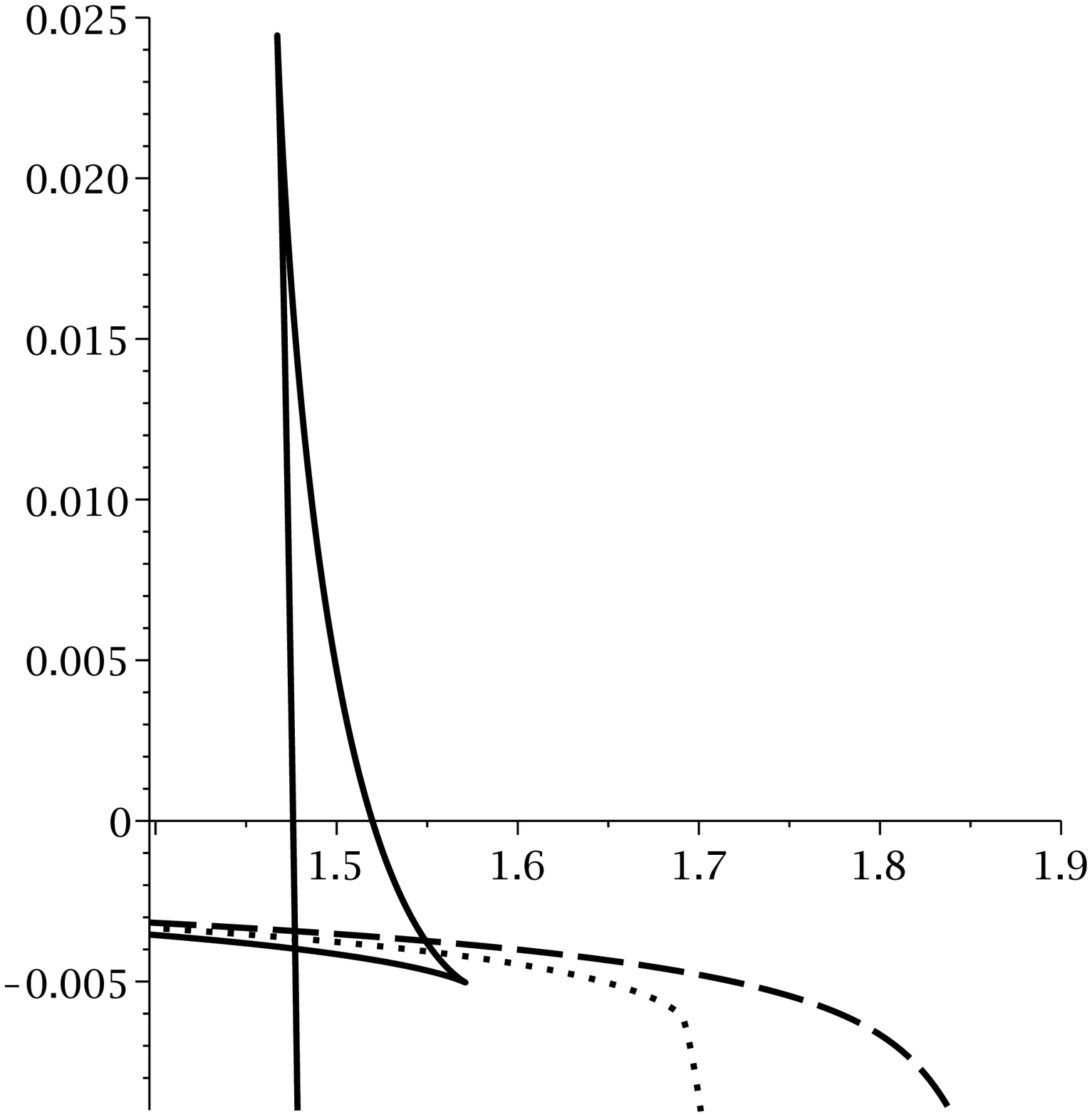}%
\end{array}
$%
\caption{ $P-r_{+}$ (left) and $G-T$ (right) diagrams for $m=5$, $c=0.2$, $%
c_{1}=c_{2}=2$, $c_{3}=-2$, $c_{4}=0$ and $d=6$. \newline
$P-r_{+}$ diagram, from up to bottom $T=1.1T_{c}$, $T=T_{c}$ and $T=0.9T_{c}$%
, respectively. \newline
$G-T$ diagram for $P=0.5P_{c}$ (continuous line), $P=P_{c}$ (dotted line)
and $P=1.5P_{c}$ (dashed line). }
\label{q06}
\end{figure}

Considering obtained values, one can show that following equality is hold
\begin{equation}
\frac{P_{cc}r_{cc}}{T_{cc}}=\frac{\left( c_{2}m^{2}c^{2}+1\right)
^{2}d_{2}d_{3}}{12\left( 3d_{4}c_{1}c_{3}-d_{3}c_{2}^{2}\right)
m^{4}c^{4}-12d_{3}\left( 2c_{2}m^{2}c^{2}+1\right) },  \label{pcrc/tc}
\end{equation}%
which shows that in this case, $\frac{P_{cc}r_{cc}}{T_{cc}}$ is a function
of massive parameter and coefficients. Using obtained critical values (Eqs. (%
\ref{rcc}), (\ref{Tcc}) and (\ref{Pcc})) with Eqs. (\ref{TotalTT}), (\ref{P}%
), (\ref{G}), (\ref{PP}) and setting $q=0$, we plot following diagrams for $%
5 $ and $6$ dimensions (Figs. \ref{q0} and \ref{q06}).


In Ref. \cite{GBCai}, it was shown that in context of neutral Gauss-Bonnet
black holes, no phase transition is observed in $6$-dimensions. Here, the
extension of the massive gravity enables the black holes to enjoy the
existence of second order phase transition in $6$-dimensions. Also, we
observed that contrary to Gauss-Bonnet case, $\frac{P_{cc}r_{cc}}{T_{cc}}$
is a function of massive gravity.

\section{Geometrical phase transition in context of heat capacity and
extended phase space \label{GTs}}

\begin{figure}[tbp]
$%
\begin{array}{ccc}
\epsfxsize=5cm \epsffile{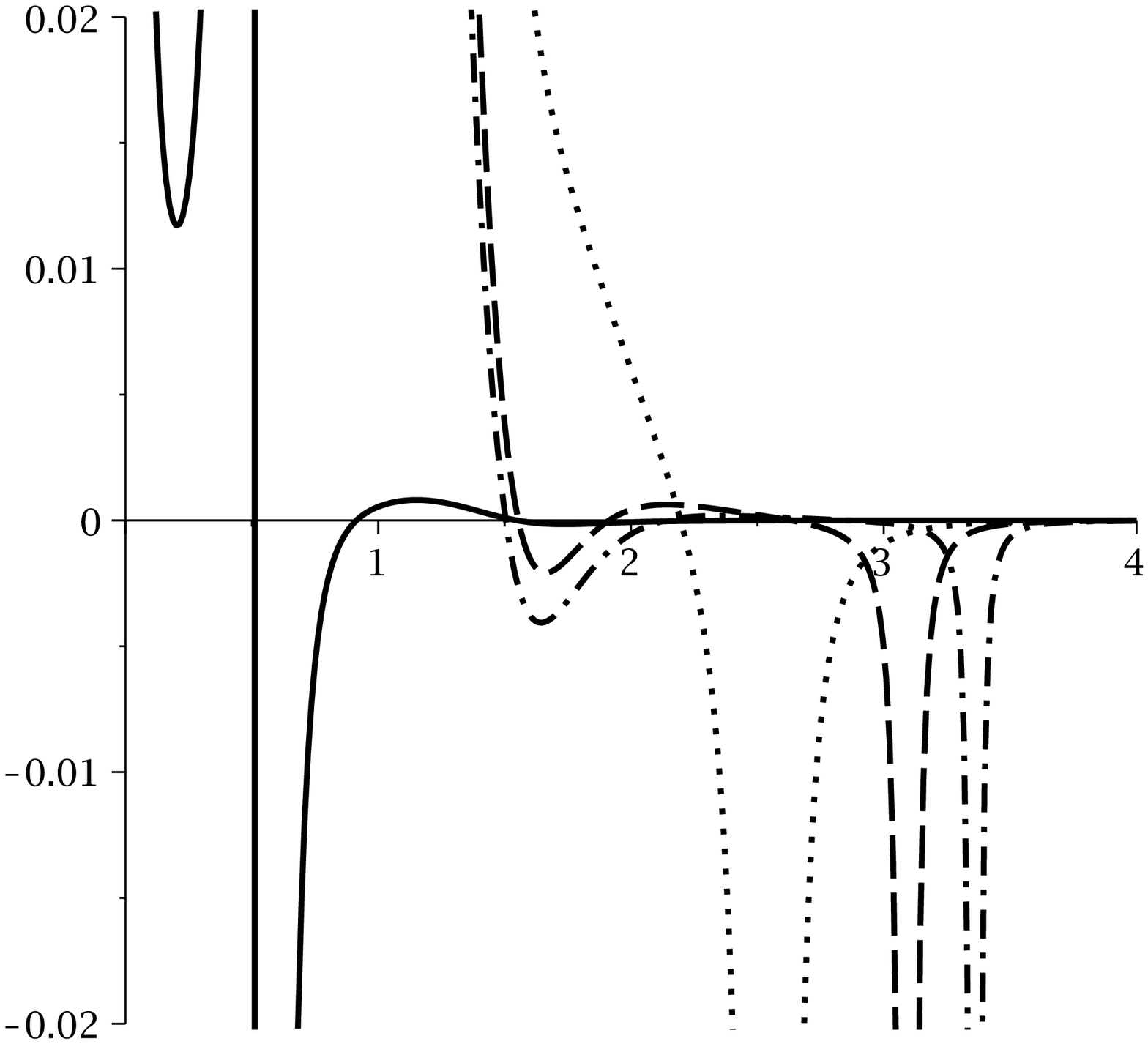} & \epsfxsize=5cm %
\epsffile{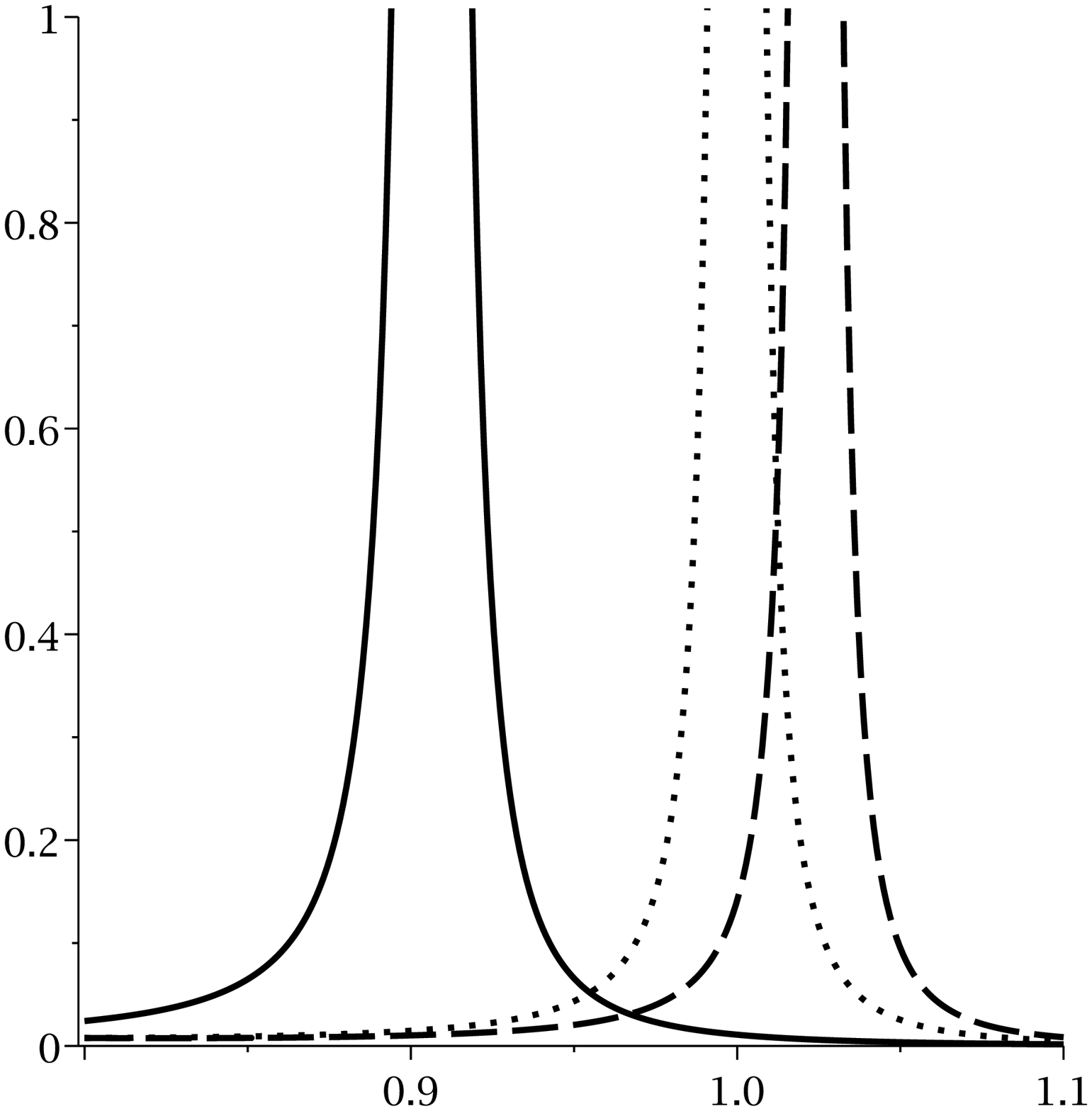} & \epsfxsize=5cm %
\epsffile{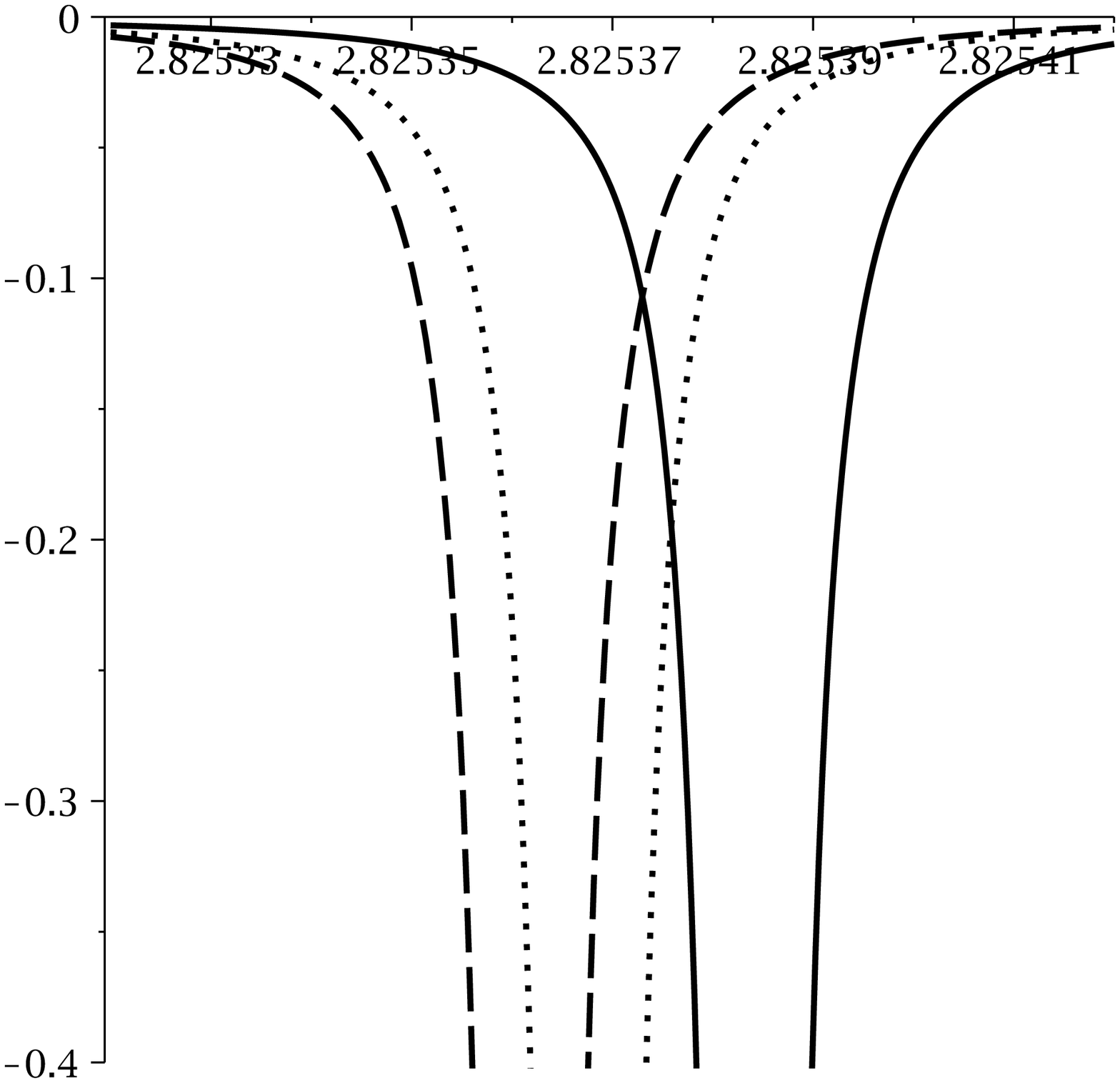}%
\end{array}
$%
\caption{For different scales: $\mathcal{R}$ versus $r_{+}$ diagrams for $%
q=1 $, $c=c_{1}=c_{2}=c_{3}=2$, $c_{4}=0$, $d=5$ and $k=1$.
\newline
left: $\protect\beta=0.5$, $m=0$ (continues line), $m=0.25$ (dotted line), $%
m=0.35$ (dashed line) and $m=0.40$ (dashed-dotted line). \newline
middle and right: $m=0.3$, $\protect\beta=2$ (continues line), $\protect\beta%
=3$ (dotted line) and $\protect\beta=4$ (dashed line). }
\label{Fig10}
\end{figure}

In this section, we employ the geometrical concept for studying
thermodynamical behavior of the obtained solutions. In order to do so, we
employ HPEM method. In this method, the thermodynamical phase space is
constructed by considering mass of the black holes as thermodynamical
potential. By doing so, the components of the phase space will be extensive
parameters such as electric charge, entropy and etc. The general form of
HPEM metric is \cite{HPEM}
\begin{equation}
ds_{New}^{2}=\frac{SM_{S}}{\left( \Pi _{i=2}^{n}\frac{\partial ^{2}M}{%
\partial \chi _{i}^{2}}\right) ^{3}}\left(
-M_{SS}dS^{2}+\sum_{i=2}^{n}\left( \frac{\partial ^{2}M}{\partial \chi
_{i}^{2}}\right) d\chi _{i}^{2}\right) ,  \label{HPEM}
\end{equation}%
where $M_{S}=\partial M/\partial S$, $M_{SS}=\partial ^{2}M/\partial S^{2}$
and $\chi _{i}$ ($\chi _{i}\neq S$) are extensive parameters which are
components of phase space. Now, we will investigate whether the phase
transition points that were obtained in section ($IV$) coincide with all
divergencies of the Ricci scalar of HPEM metric. For economical reasons, we
only plot diagrams correspond to variation of massive and nonlinearity
parameters. To do so, we use Eqs. (\ref{TotalQ}), (\ref{TotalS}) and (\ref%
{TotalM}) with HPEM metric (Eq. \ref{HPEM}). This leads into following
diagrams (Fig. \ref{Fig10}).

It is evident that employed metric has consistent results with what were
found in case of heat capacity (compare Figs. \ref{Fig1} and \ref{Fig2} with %
\ref{Fig10}). In other words, the divergencies of the Ricci scalar are
matched with phase transition points of the heat capacity. An interesting
characteristic behavior of the diagrams is the different divergencies for
different types of phase transition. In case of larger to smaller black
holes phase transition, the divergency of the Ricci scalar is toward $%
+\infty $ (compare Fig. \ref{Fig2} left panel with Fig. \ref{Fig10} middle
panel), whereas in case of smaller to larger phase transition, the
divergency is toward $-\infty$ (compare Fig. \ref{Fig2} middle panel with
Fig. \ref{Fig10} right panel). This specific behavior enables us to
recognize the type of phase transition independent of heat capacity.

\begin{figure}[tbp]
$%
\begin{array}{ccc}
\epsfxsize=5cm \epsffile{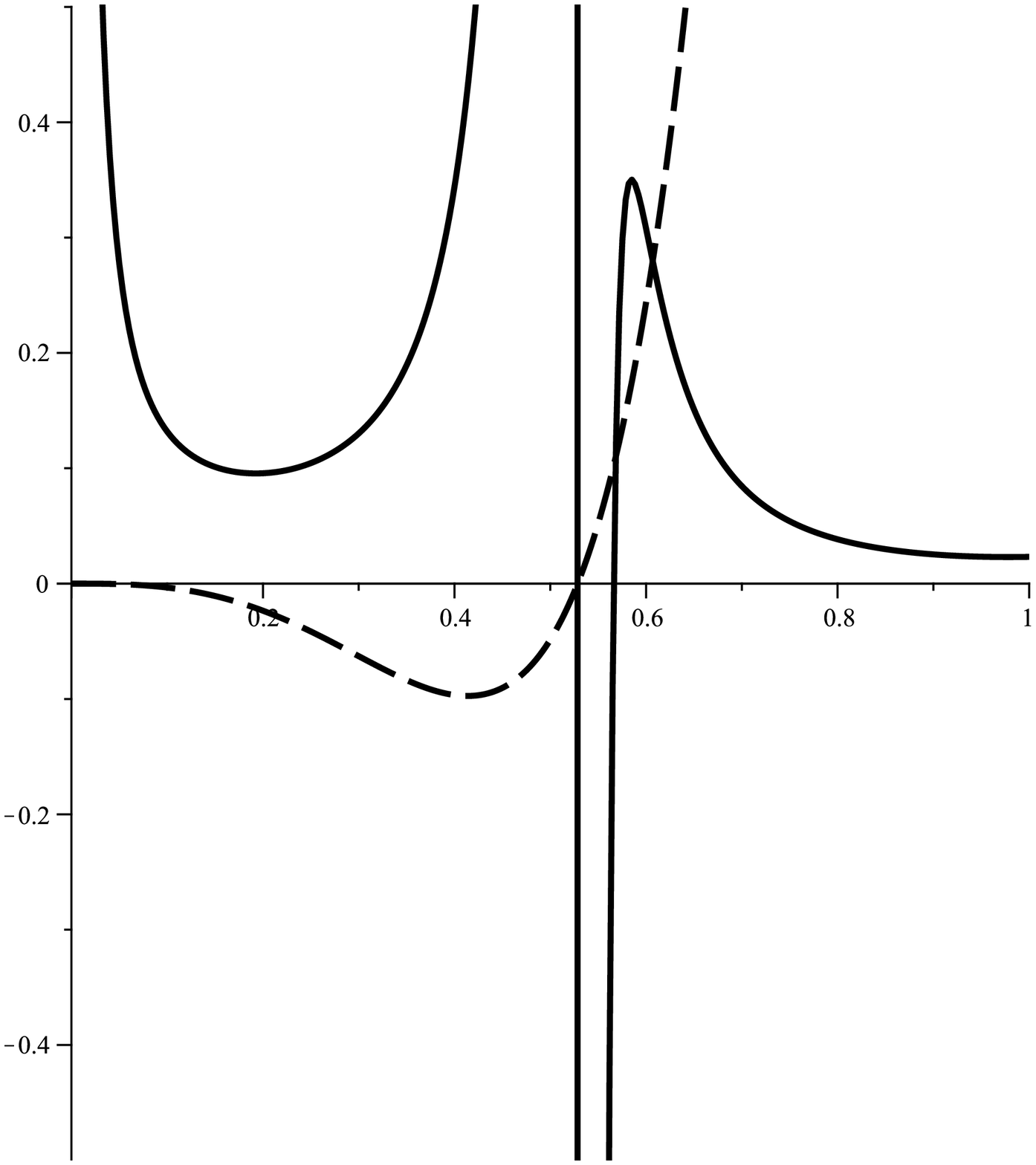} & \epsfxsize=5cm %
\epsffile{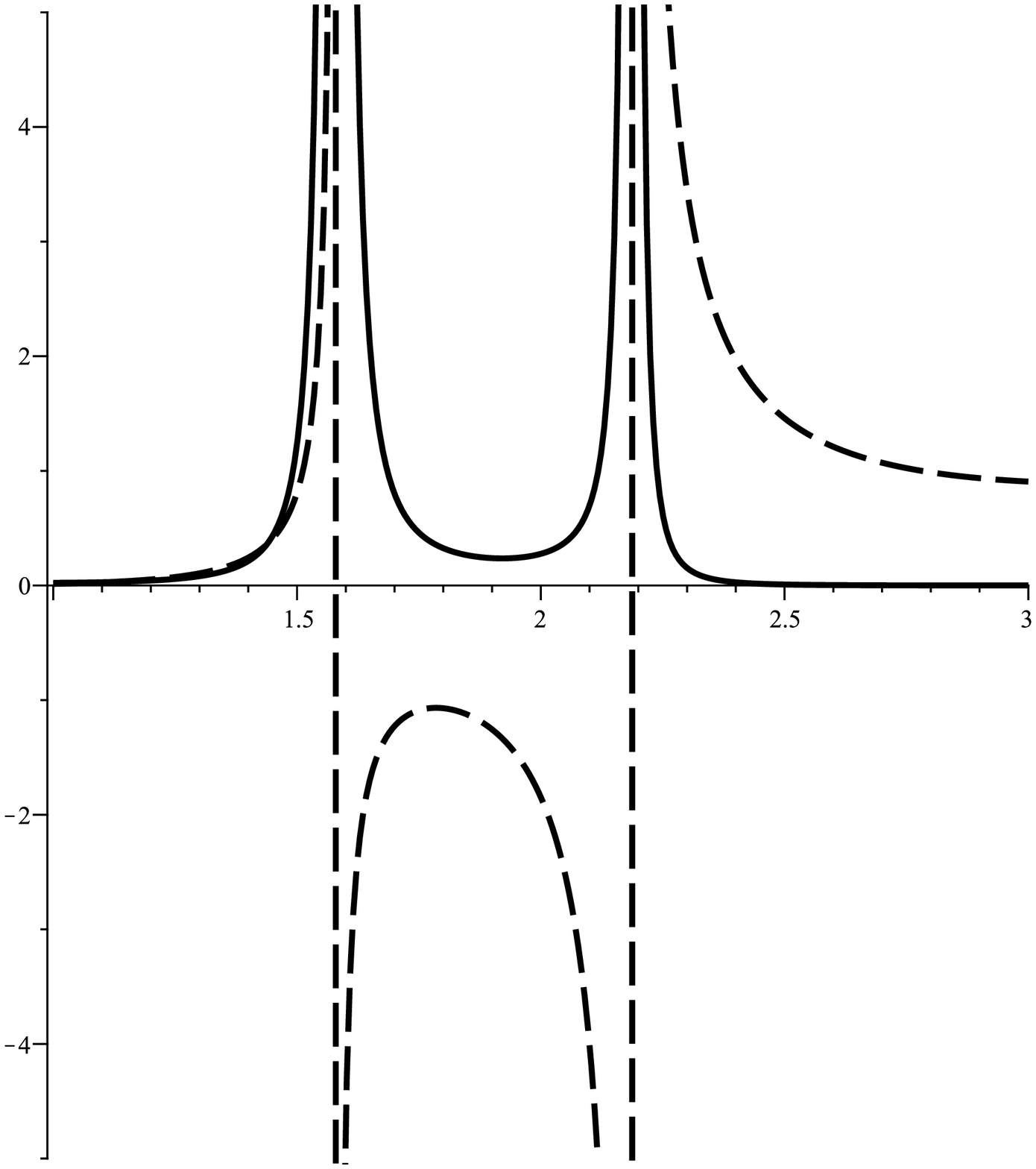} & \epsfxsize=5cm \epsffile{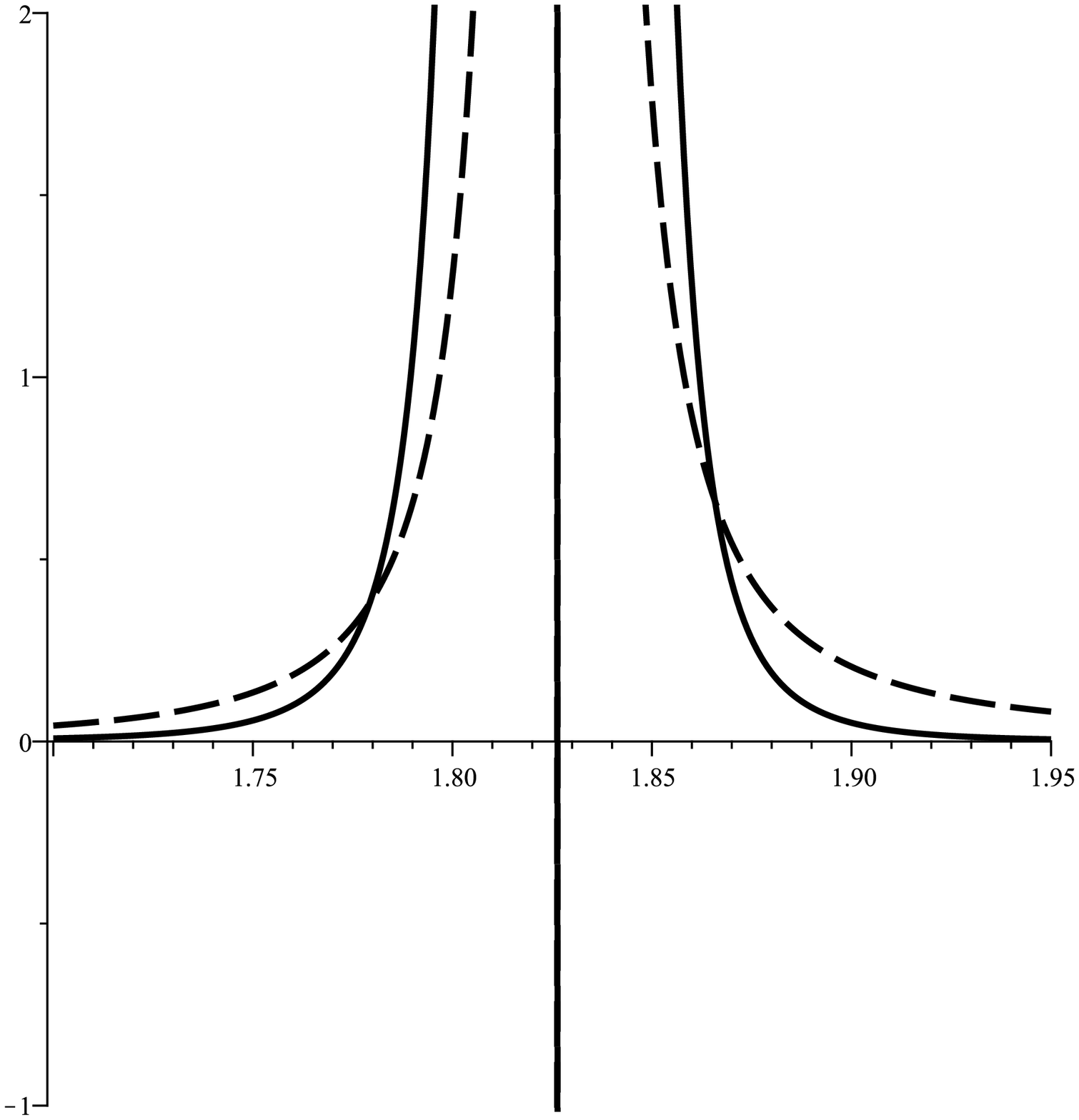}%
\end{array}
$%
\caption{For different scales: $\mathcal{R}$ (continuous line), $C_{Q}$
(dashed line) diagrams for $q=1$, $c=c_{1}=c_{2}=c_{3}=2$, $c_{4}=0$, $%
\protect\beta=0.5 $, $d=5$, $k=1$ and $m=0.1$. \newline
$P=0.9P_{c}$ left and middle panels, $P=P_{c}$ left and right panels and $%
P=1.1P_{c}$ left panel.}
\label{Fig11}
\end{figure}

Next, we employ another geometrical metric for studying the critical
behavior of the system in context of extended phase space. In this metric,
Due to consideration of the cosmological constant as thermodynamical
pressure, we have three extensive parameters; electric charge, entropy and
pressure. In order to construct phase space we employ following metric \cite%
{PVEinstein}
\begin{equation}
ds^{2}=
\begin{array}{cc}
S\frac{M_{S}}{M_{QQ}^{3}}\left(-M_{SS}dS^{2}+M_{QQ}dQ^{2}+dP^{2}\right) &
\end{array}%
.  \label{GTDPV}
\end{equation}
Considering Eqs. (\ref{TotalM}), (\ref{Heat}), (\ref{P}) and (\ref{GTDPV}),
we plot following diagram (Fig. \ref{Fig11}) with respect to Fig. \ref{Fig6}.

Due to existence of a root for heat capacity, in all plotted diagrams, a
divergency is observed (Fig. \ref{Fig11} left panel). It is evident that for
pressures smaller than critical pressure, system goes under two phase
transitions with different horizon radii (Fig. \ref{Fig11} middle panel).
This is consistent with what was observed in studying $T-r_{+}$ diagrams of
Fig. \ref{Fig6}. On the other hand, for critical pressure system goes under
a phase transition. The place of this divergency is exactly located at the
critical horizon which is obtainable through $T-r_{+}$ diagrams of Fig. \ref%
{Fig6} (Fig. \ref{Fig11} right panel). Finally, for pressures larger than
critical pressure no phase transition is observed and the behavior of Ricci
scalar will be what is plotted in Fig. \ref{Fig11} (left panel). These
results are consistent with ordinary thermodynamical concepts and indicates
that these three pictures (phase diagrams, heat capacity and geometrical
thermodynamics) are in agreement.

\section{Heat capacity and critical values in the extended phase space \label%
{HC}}

The final section of this paper is devoted to calculation of the critical
pressure in extended phase space by using denominator of the heat capacity.
It was shown that one can calculate critical pressures that were obtained in
section ($V$) by using denominator of the heat capacity \cite{PVEinstein}.
To do so, one should replace the cosmological constant in denominator of the
heat capacity (\ref{Heat}) with its corresponding pressure (\ref{P}). Then,
solve the denominator of the heat capacity with respect to pressure. This
will lead into following relation
\begin{eqnarray}
P &=&\frac{d_{2}d_{3}c_{4}m^{2}c^{2}}{16\pi r_{+}^{4}}\left(
3d_{4}d_{5}c_{4}c^{2}+2d_{4}c_{3}cr_{+}+c_{2}r_{+}^{2}\right) -\frac{%
d_{2}d_{3}^{2}q^{2}}{8\pi r_{+}^{2d_{2}}\sqrt{1+\Gamma _{+}}}  \notag \\
&&-\frac{\left( \sqrt{1+\Gamma _{+}}-1\right) \beta ^{2}}{4\pi \sqrt{%
1+\Gamma _{+}}}+\frac{d_{2}d_{3}}{16\pi r_{+}^{2}}  \label{PNEW}
\end{eqnarray}

Obtained relation for pressure is different from what was obtained through
use of temperature (\ref{PP}). In this relation, the maximum(s) of pressure
and its corresponding horizon radius are critical pressure and horizon
radius in which phase transition takes place. Now, by using indicated values
in table $1$ and Eq. (\ref{PNEW}), we plot following diagram (Fig. \ref%
{Fig12}).

\begin{figure}[tbp]
$%
\begin{array}{cc}
\epsfxsize=6cm \epsffile{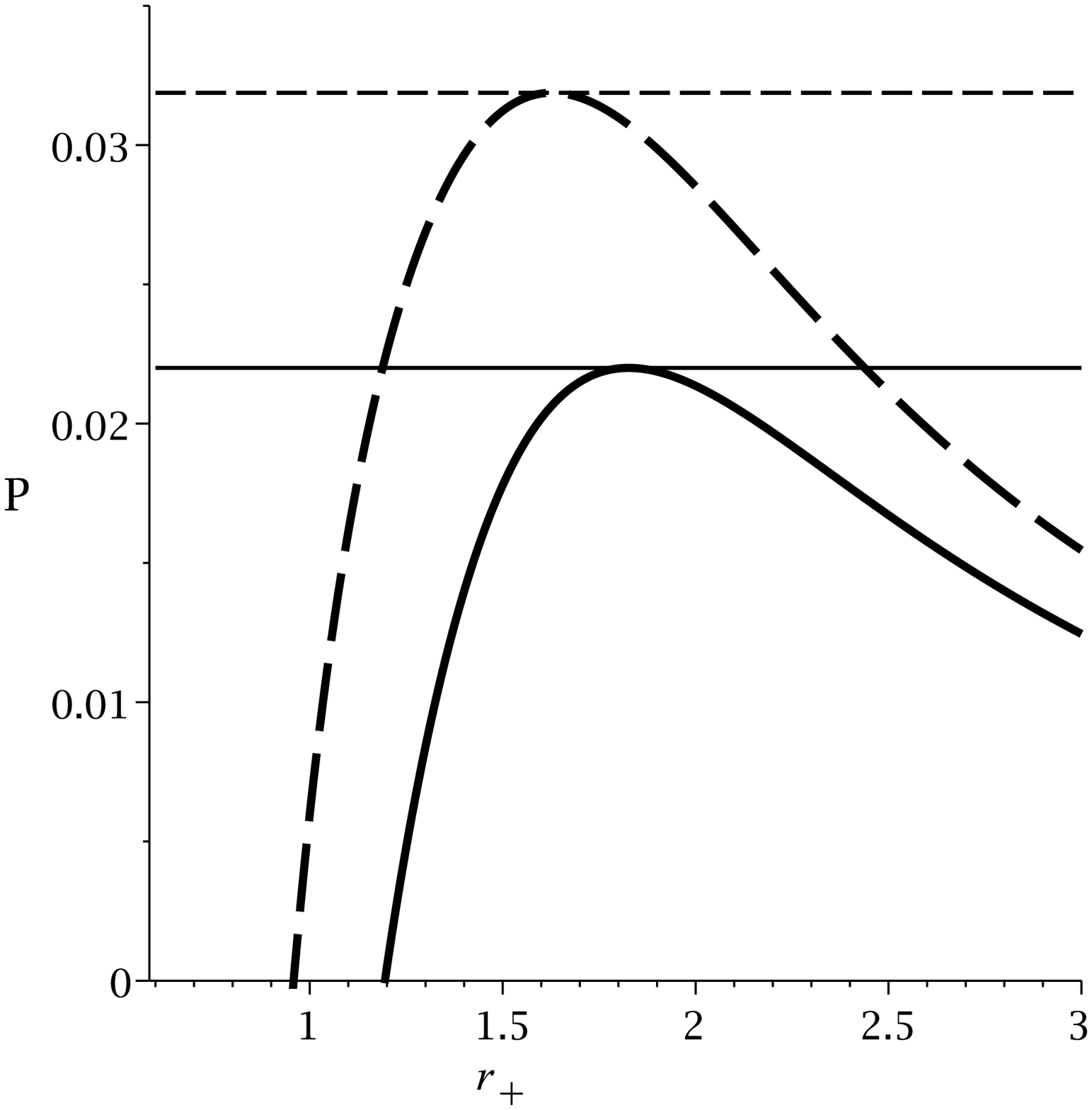} & \epsfxsize=6cm %
\epsffile{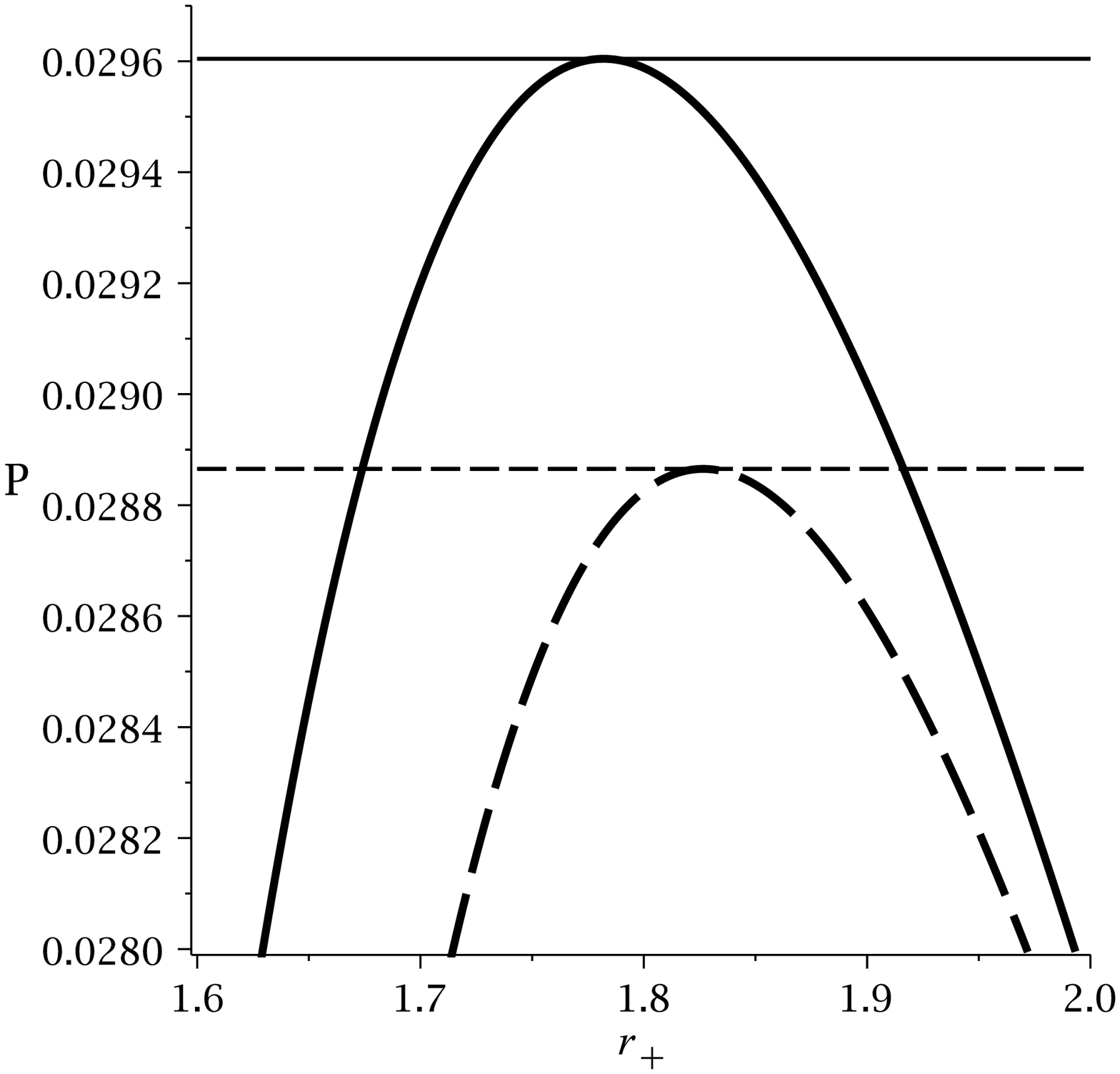}%
\end{array}
$%
\caption{$P$ versus $r_{+}$ diagrams for $q=1$, $c=c_{1}=c_{2}=c_{3}=2$, $c_{4}=0$, $%
d=5 $ and $k=1$. \newline left panel:$\protect\beta=0.5$ and $m=0$
(bold-continues line), $P=0.02200146 $ (continues line), $m=5$
(bold-dashed line) and $P=0.03187871$ (dotted line). \newline
right panel: $m =0.1$ and $\protect\beta =1$ (bold-continues line), $%
P=0.0296043$ (continues line), $\protect\beta=5$ (bold-dashed line) and $%
P=0.0288653$ (dashed line). }
\label{Fig12}
\end{figure}

It is evident that obtained maximums are critical pressures in which phase
transitions take place. The thermodynamical concept that was mentioned in
last section (pressure being smaller than critical pressure leads to
existence of two phase transitions and for pressures larger than critical
pressure no divergency is observed) is also hold in case of this approach.
In other words, this approach is an additional method for studying critical
behavior of the system and the results of this approach is consistent with
GTs, heat capacity and extended phase space.

\section{Conclusions}

In this paper, we have considered EN-massive gravity in presence of BI
nonlinear electromagnetic field. It was shown that considering this
configuration leads to modification of the number and place of horizons that
black holes can acquire. In other words, cases of multiple horizons were
observed with different phenomenologies. Next, conserved and thermodynamical
quantities were obtained and it was shown that first law of thermodynamics
hold for these black holes.

Next, we studied the thermodynamical behavior of the system. It was shown
that temperature in the presence and absence of massive gravity presents
different behaviors. Adding massive put limitations on values that
temperature can acquire, while, there was no limitation for temperature in
the absence of it. Interestingly, this behavior was modified in the presence
of strong nonlinearity parameter. In strong nonlinearity parameter, the
behavior of temperature returned to a massive-less like behavior but the
effects of the massive were observed in existences of extrema. It was also
seen, that in case of different dimensions, for each pair of dimensions, one
can find a point in which temperature for both dimensions are equal.

Regarding the stability, it was seen that in the presence of massive
gravity, black holes enjoy one phase transition of the smaller unstable to
larger stable. The phase transition was related to the divergency of heat
capacity. Then again, in strong nonlinearity parameter, this behavior was
modified. In this case, black holes had three phase transitions of smaller
non-physical unstable to larger physical stable (in place of root), larger
unstable to smaller stable (in place of smaller divergency) and smaller
unstable to larger stable (in place of larger divergency).

Clearly, one can conclude that nonlinear electromagnetic field has an
opposing effect comparing to massive gravity. Strong nonlinearity parameter
modifies the effects of the massive gravity and return the system to the
massive-less like behavior, although the effects of massive still observed
through extrema.

It was pointed out that at maximums of the temperature, larger unstable to
smaller stable and at minimums, smaller unstable to larger stable phase
transitions take place. Therefore, studying temperature provides an
independent picture for studying phase transitions and stability of the
solutions.

Next, we extended phase space by considering cosmological constant as
thermodynamical variable known as pressure. It was shown that volume of the
black holes is independent of generalization of the electromagnetic field
and extension of the massive gravity. Obtained values where critical points
in which phase transitions took place. It was shown that the effects of
variation of nonlinearity parameter was opposite of the massive parameter.
In other words, these two factors put restrictions on each others effects.

Interestingly, in Ref. \cite{HendiPEM} variation of massive gravity highly
modified the critical temperature and pressure. In case of obtained
solutions in this paper, the modification was not as considerable as what
was observed in case of Gauss-Bonnet-Maxwell-massive black holes. This shows
that generalization of electromagnetic field puts stronger restrictions on
the effects of the massive gravity. In other words, in order to have
stronger control over contributions of the massive gravity one should
increase the nonlinearity of the electromagnetic sector.

In addition, a study in context of neutral solutions was conducted. It was
shown that due to contribution of the massive gravity, the chargeless
solutions of this gravity also enjoy the existence of phase transition. In
other words, black holes in EN-massive gravity go under phase transitions in
extended phase space. Also, it was shown that ratio of $\frac{P_{c}r_{c}}{%
T_{c}}$ was a function of massive gravity. It was also pointed out
that in $6 $-dimensions, contrary to case of Gauss-Bonnet black
holes, these black holes enjoy second order phase transition. In
addition, it was shown that in case of $d=4$, no phase transition
for massive black holes is observed.

Next, geometrical approach was used for studying critical behavior of the
system in context of heat capacity and extended phase space. It was shown
that employed metrics for both cases have consistent results and follow the
concepts of ordinary thermodynamics. The characteristic behavior of
divergencies in Ricci scalar of the geometrical thermodynamical metrics,
enabled us to recognize the type of phase transition (smaller to larger or
larger to smaller).

Finally, another method which was based on denominator of the heat capacity
was used to calculate critical pressure and horizon radius. It was shown
that this method has consistent results with extended phase space and follow
the concepts of ordinary thermodynamics. In other words, this method
provides an independent approach for investigating critical behavior of the
system.

Due to generalization of Born-Infeld for electromagnetic sector of
the solutions, it will be worthwhile to study the effects of this
generalization on conductivity of these black holes and their
corresponding superconductors phase transition. Specially, it will
be interesting to see how this generalization will affect the
interpretation of graviton as lattice and the Drude like behavior.
Also, it will be worthwhile to study the metal-insulator
transition in context of these solutions.

\begin{acknowledgements}
We thank Shiraz University Research Council. This work has been
supported financially by the Research Institute for Astronomy and
Astrophysics of Maragha, Iran.
\end{acknowledgements}

\end{document}